\documentclass[12pt]{article}
\usepackage{amsmath}
\usepackage{amsthm}
\usepackage{amssymb}
\usepackage{amsfonts}
\usepackage{epic}
\usepackage{eepic}
 \usepackage{times}
\usepackage[matrix,arrow]{xy}
\usepackage{macros}
\usepackage{epsfig}

\theoremstyle{plain}

\theoremstyle{remark}
\newtheorem{rem}{Remark}

\newcommand{\sst}{\scriptscriptstyle}

\newcommand{\nn}{\nonumber}

\renewcommand{\1}{\one}

\newcommand{\beq}{\begin{equation}}
\newcommand{\eeq}{\end{equation}}

\newcommand{\pa}{\partial}
\newcommand{\ot}{\otimes}
\newcommand{\ra}{\to}

\newcommand{\fr}[2]{{\textstyle \frac{#1}{#2} }}

\newcommand{\fsl}{{\mathfrak s}{\mathfrak l}}
\newcommand{\hfsl}{\widehat{\fsl}}

\newcommand{\bra}{\langle}

\newcommand{\ket}{\rangle}

\newcommand{\al}{\alpha}

\newcommand{\be}{\beta}
\newcommand{\ga}{\gamma}
\newcommand{\Ga}{\Gamma}
\newcommand{\de}{\delta}
\newcommand{\De}{\Delta}
\newcommand{\ep}{\epsilon}

\newcommand{\Om}{\Omega}
\newcommand{\si}{\sigma}

\newcommand{\vf}{\varphi}

\newcommand{\bq}{\bar{q}}

\newcommand{\bx}{\bar{x}}

\newcommand{\bz}{\bar{z}}

\newcommand{\CA}{{\mathcal A}}
\newcommand{\CB}{{\mathcal B}}

\newcommand{\CC}{{\mathcal C}}
\newcommand{\CD}{{\mathcal D}}
\newcommand{\CE}{{\mathcal E}}
\newcommand{\CF}{{\mathcal F}}
\newcommand{\CG}{{\mathcal G}}
\newcommand{\CH}{{\mathcal H}}

\newcommand{\CI}{{\mathcal I}}
\newcommand{\CJ}{{\mathcal J}}

\newcommand{\CL}{{\mathcal L}}
\newcommand{\CM}{{\mathcal M}}

\newcommand{\CO}{{\mathcal O}}
\newcommand{\CP}{{\mathcal P}}
\newcommand{\CQ}{{\mathcal Q}}
\newcommand{\CR}{{\mathcal R}}
\newcommand{\CS}{{\mathcal S}}
\newcommand{\CT}{{\mathcal T}}

\newcommand{\CU}{{\mathcal U}}
\newcommand{\CV}{{\mathcal V}}
\newcommand{\CW}{{\mathcal W}}
\newcommand{\CX}{{\mathcal X}}
\newcommand{\CY}{{\mathcal Y}}
\newcommand{\CZ}{{\mathcal Z}}

\newcommand{\SA}{{\mathsf A}}
\newcommand{\SB}{{\mathsf B}}
\newcommand{\SC}{{\mathsf C}}

\newcommand{\SH}{{\mathsf H}}

\newcommand{\sk}{{\mathsf k}}
\newcommand{\SL}{{\mathsf L}}

\newcommand{\SM}{{\mathsf M}}
\newcommand{\SN}{{\mathsf N}}

\renewcommand{\SS}{{\mathsf S}}
\newcommand{\ST}{{\mathsf T}}
\newcommand{\SU}{{\mathsf U}}
\newcommand{\SV}{{\mathsf V}}
\newcommand{\SW}{{\mathsf W}}

\newcommand{\SZ}{{\mathsf Z}}

\newcommand{\fe}{{\mathfrak e}}
\newcommand{\fg}{{\mathfrak g}}
\newcommand{\FG}{{\mathbf G}}

\newcommand{\fv}{{\mathfrak v}}

\newcommand{\fw}{{\mathfrak w}}

\newcommand{\sa}{{\mathsf a}}

\newcommand{\sfc}{{\mathsf c}}
\newcommand{\sd}{{\mathsf d}}

\newcommand{\sh}{{\mathsf h}}

\newcommand{\so}{{\mathsf o}}
\newcommand{\sq}{{\mathsf q}}

\newcommand{\su}{{\mathsf u}}
\newcommand{\sv}{{\mathsf v}}

\newcommand{\sw}{{\mathsf w}}

\newcommand{\sz}{{\mathsf z}}

\newcommand{\BG}{{\mathbb G}}

\newcommand{\FV}{{\mathfrak V}}

\newcommand{\0}{{\mathfrak 0}}
\newcommand{\one}{{\mathfrak 1}}

\newcommand{\BR}{{\mathbb R}}

\newcommand{\BC}{{\mathbb C}}

\newcommand{\BP}{{\mathbb P}}

\newcommand{\BU}{{\mathbb U}}
\newcommand{\Lfg}{{{}^{\rm\sst L}\fg}}
\newcommand{\BZ}{{\mathbb Z}}

\newcommand{\rf}[1]{(\ref{#1})}

\newcommand{\aufz}
{\begin{list}{$\bullet$}{\topsep0cm \itemsep0cm \parsep0cm}}
\newcommand{\eaufz}{\end{list}}

\def \ignore#1 { {} }

\def \Fig#1#2#3 {
\begin{figure}
\begin{center}
\scalebox{.6}{\includegraphics{#1.eps}}
\label{#1}
\end{center}
\caption{#3}
\end{figure}
}

\def \half {\frac{1}{2}}

\def \hib {\frac{1}{2b}}

\def \p {\partial}
\def \pp#1 {{\frac{\p}{\p #1}}}
\def \ppd#1 {{\frac{\p^2}{\p #1 ^2}} }

\def \bg {\bar{\gamma}}
\def \vf {\varphi}

\def \a {\alpha}

\def \bz {\bar{z}}

\def \bx {\bar{x}}

\def \bq {\bar{q}}

\def \ra {\right\rangle}

\def \H      {H_3^+}

\def \vir { \ , \ }
\def \F32#1#2#3#4#5#6{{} _3F_2\left(\left.\begin{array}{c}#1 \vir #2 \vir
    #3 \\ #4\vir #5
    \end{array} \right| #6 \right) }

\def \bea {\begin{eqnarray}}
\def \eea {\end{eqnarray}}
\def \bee {\begin{eqnarray*}}
\def \eee {\end{eqnarray*}}
\def \nn  {\nonumber}
\def \sst {\scriptscriptstyle}

\renewcommand{\ra}{\rightarrow}

\newcommand{\Locsl}[1]{{\rm Loc}_{\rm PSL(2,\BC)}(#1)}
\newcommand{\Homsl}[1]{{\rm Hom}(\pi_1(#1),{\rm PSL(2,\BC)})}
\newcommand{\Homslr}[1]{{\rm Hom}(\pi_1(#1),{\rm PSL(2,\BR)})}

\newcommand{\pslc}{{\rm PSL(2,\BC)}}

\begin{document}

\title{Quantization of the Hitchin moduli spaces, Liouville theory,
and the geometric Langlands correspondence I}

\author{J. Teschner$^{(1)}$}

\address{$^{(1)}$ DESY Theory, Notkestr. 85,
22603 Hamburg, Germany}

\maketitle

\section{Introduction}

\subsection{Motivation}

It is known for a while that the low-energy theory of $N=2$ supersymmetric
gauge theories in four dimensions can be described in terms of the data
characterizing an algebraically integrable system which is
canonically associated to a given
gauge theory \cite{GKMMM,DW,Fd}. More recently it was found that
studying the gauge theories in finite volume \cite{Pe},
or in the presence of certain deformations like the
so-called Omega-deformation \cite{Ne}
is a useful tool to extract
some highly nontrivial non-perturbative information about such
gauge theories. It was in particular
recently argued in \cite{NS} that the gauge theory in the presence of
a certain one-parameter deformation can at low energies
effectively be described in terms
the quantization of the above-mentioned algebraically integrable
system. An amazing correspondence was furthermore observed in
\cite{Pe,AGT} between
the partition functions of a certain class of gauge theories
on $S^4$ and the correlation functions
in Liouville theory \cite{ZZ,T01,T09}.
Knowing  the modular transformation properties of
the Liouville conformal blocks \cite{PT,T01,T03a,T09} now allows
us to investigate and test the S-duality conjectures in these gauge theories,
as illustrated in \cite{AGGTV,DGOT}.

It seems, however, that the deeper reasons for this relationship between a two- and
a four-dimensional theory remain to be understood.
A clue in this direction may be seen in the fact that the instanton partition functions
which represent the building blocks of the partition functions studied in \cite{Pe,AGT}
are obtained by specializing the two-parameter family $\CZ(a,\ep_1,\ep_2;q)$ of instanton
partition functions introduced in \cite{MNS,LNS,Ne}. The functions $\CZ(a,\ep_1,\ep_2;q)$
not only allow one to obtain the Seiberg-Witten prepotential of the gauge theory on
$\BR^4$ in the limit where both $\ep_1$ and $\ep_2$ tend to zero \cite{Ne,NO}, but
also the Yang's potential determining the spectrum of the quantized integrable
model mentioned above in the limit where only one of the two parameters $\ep_1$ or
$\ep_2$ vanishes. This was observed in  \cite{NS} for a certain class of  examples,
and is expected to hold much more generally.
The functions $\CZ(a,\ep_1,\ep_2;q)$ were identified with the
conformal blocks of Liouville theory in \cite{AGT}.

This indicates that the relationship between certain gauge theories and Liouville
theory involves in particular a two-parametric deformation of the algebraically integrable
model associated to the gauge theories on $\BR^4$ which ultimately produces
Liouville theory as a result.
One of my intentions in this paper is to
clarify in which sense this point of view is correct. Such a study
may be seen as being complementary to the recent work of
Nekrasov and Witten \cite{NW}, where certain aspects of the correspondence
between Liouville theory and gauge theory were understood by studying
a certain two-parameter generalization of the set-up from \cite{NS}.
We'll make some comments on this relation in the conclusions.

Another piece of motivation comes from the relations discussed in \cite{KW} between
four-dimensional gauge theories and the geometric Langlands correspondence.
A puzzling  aspect of the resulting picture is the fact that the geometric
Langlands correspondence is also related to conformal field theory as shown in the
works of Beilinson, Drinfeld, Feigin, Frenkel and others, see \cite{Fr2} for a nice review
and further references. However, the relation between the gauge theory approach to the
geometric Langlands correspondence of \cite{KW} and the conformal field theory approach
has remained mostly unclear up to now. The author feels that the above-mentioned relations
between gauge theory and conformal field theory offer new clues in this regard.
It is therefore my second main aim to clarify the relations between the quantization
of the Hitchin system, the geometric Langlands correspondence and the
Liouville conformal field theory.

\subsection{From the Hitchin integrable system to Liouville theory}

One of my aims is to explain that it is possible to
understand the relation between Liouville and the Hitchin
system in two ways
as the result
of a two-step process which is
a combination of a one-parameter deformation and
quantization, schematically:
\begin{equation}\label{maindiag1}
\begin{array}{ccccc}
& & \text{Hitchin system} & & \\
& {{\rm (A)}_{\ep_2}}\swarrow & & \searrow{{\rm (B)}_{\ep_1}} & \\
\text{Fuchsian} & & & & \text{quantized}\\
\text{isomonodromic} & & & & \text{Hitchin} \\
\text{deformations} & & & & \text{systems} \\
& {{\rm (C)}_{\ep_1}}\searrow & & \swarrow{{\rm (D)}_{\ep_2}} & \\
& & \text{Liouville theory} & &
\end{array}
\end{equation}
where the arrows may be schematically characterized as follows:
\begin{itemize}
\item[(A)] 
Hyperk\"ahler rotation for the Hitchin moduli space $\CM_{\rm H}(C)$.
This is explained in Section \ref{sec:isomon}.
\item[(B)] Quantization of the Hitchin system in the sense discussed
in \cite{NS} and \cite{NW} with quantization conditions determined by
Yang's potential (Section 4).
\item[(C)] Quantization of the Hitchin moduli spaces $\CM_{\rm H}(C)$.
This is explained in
Section \ref{sec:quantlocsys}.
\item[(D)] This arrow will be referred to as {\it quantum} hyperk\"ahler
rotation. The motivation for this terminology come from the closure
of the diagram together with the observation that the quantized
Hitchin system can be recovered from Liouville theory
in suitable limits, as discussed in Section \ref{sec:Liou}.
\end{itemize}
The parameters $\ep_1$, $\ep_2$ that govern the different relations
will also be parameterized as
\begin{equation}\label{eppar}
\ep_1\,=\,\hbar b\,,\qquad \ep_2\,=\,\hbar/b\,,
\end{equation}
with $b$ being the parameter that is often used in the literature on Liouville theory.

Liouville theory
is known to be related to the quantum theory of Teichm\"uller spaces \cite{T,T05}.
The Teichm\"uller spaces can be identified with one of the connected components of the moduli spaces of
flat SL$(2,\BR)$-connections on Riemann surfaces. We will show that
the relations above can be understood as natural generalizations
of the relation between Liouville theory and the quantization of
Teichm\"uller spaces to the other components of the moduli spaces of
flat SL$(2,\BR)$-connections.

\subsection{Separation of Variables}

It is known for a while that the Hitchin system is related
to conformal field theory by a similar-looking two-step procedure of
deformations and formal quantization.
\begin{equation}\label{Ddiag}
\begin{array}{ccccc}
& & \text{Hitchin system} & & \\
& {\rm (a)}_{\ep}\swarrow & & \searrow {\rm (b)}_{\hbar}& \\
\text{Isomonodromic} & & & & \text{Beilinson-Drinfeld} \\
\text{deformations} & & & & \text{system} \\
& {\rm (c)}_{\check{k}}\searrow & & \swarrow {\rm (d)}_{{k}} & \\
& & \text{KZB-equations} & &
\end{array}
\end{equation}
\begin{itemize}
\item[(a)$_\ep$] The Hitchin system can be obtained as a limit of
the isomonodromic deformation system as shown in \cite{LO,kr2}.
\item[(b)$_\hbar$] The quantization of the Hitchin Hamiltonians constructed by
Beilinson-Drinfeld \cite{BD}.
\item[(c)$_k$] The KZB equations of the WZNW conformal field theory
can be obtained as a formal quantization of the ismonodromic deformation system, as
was observed for $g=0$
in \cite{Re,Ha} and shown for $g>0$ in \cite{BF}.
\item[${\rm (d)}_{\check{k}}$] The eigenvalue equation for the Gaudin Hamiltonians, which are
the $g=0$ cases of the quantized Hitchin Hamiltonians arise in
the critical level limit of the KZ equations as shown for $g=0$
in \cite{RV} and for $g>0$ in \cite{BF}.
\end{itemize}
The whole diamond of relations was discussed in \cite{LO,BF}.

The diagram \rf{Ddiag} is of course not unrelated to the previous one in \rf{maindiag1}.
On the classical level there are two natural representations for the Hitchin system,
one coming from the representation of an open dense subset of
the Hitchin moduli space $\CM_{\rm H}(C)$ as $T^*{\rm Bun}_{G}$,
the other is related to a natural map $\SS\so\sv$ from $\CM_{\rm H}(C)$
to a Hilbert scheme $(T^*C)^{[h]}$ of points on $T^*C$ which can also be
used to introduce a set of coordinates for $\CM_{\rm H}(C)$.
The change of variables $\SS\so\sv$ is closely related to what is called the
Separation of Variables (SOV) in the integrable systems literature \cite{Sk}.

And indeed, we are going to explain that the full set of relations
between the diagrams \rf{maindiag1} and \rf{Ddiag} originates from
the change of variables underlying the Separation of Variables method.
For the quantized Hitchin system the relation between
the diagrams  was found in \cite{Sk} for $g=0$ and for $g=1$
in \cite{EFR}. It is related to the quantum version of the
Separation of Variables method.
At the bottom of \rf{Ddiag} and \rf{maindiag1} one
finds on the level of systems of differential equations
a correspondence between the null-vector decoupling equations of
Belavin-Polyakov-Zamolodchikov (BPZ) and the KZ equations
discovered in \cite{Sto}.
The correspondence between the respective systems of differential equations can be extended
to a  correspondence  between Liouville theory and SL$(2)$-WZNW-model on the level of
the full correlation functions as was established in \cite{RT} for $g=0$ and extended to
higher genus in \cite{HS}. We will finally show that the relation between
the Fuchsian isomonodromic
deformation equations and the theory of isomonodromic deformations of flat holomorphic connections
can also be seen to follow from a
variant of the change of variables as used in the Separation of Variables
method.

The relations in \rf{Ddiag} were so far only discussed
on the level of system of differential equations.
The connection with Liouville theory allows us to go much further: It enables us
to construct and parameterize interesting spaces of solutions to the KZB
equations which are complete in the sense that all the monodromies can be
represented as linear transformations.

\subsection{Geometric Langlands correspondence}

Our second main aim in this article is to point out relations
to the geometric Langlands correspondence and a certain generalization thereof.
The geometric Langlands correspondence 
(see \cite{Fr2} for a nice review and further references)
is often schematically presented as a correspondence between
\begin{equation}\label{Lang}
\boxed{\quad\phantom{\Big|}
{}^{\rm\sst L}_{}{G}-\text{local systems}
\;\quad}\quad
\longrightarrow \quad
\boxed{\quad\phantom{\Big|}
\CD-\text{modules on ${\rm Bun}_G$}
\;\quad}
\end{equation}
It is connected to the quantization of the Hitchin Hamiltonians \cite{BD}
by noting that an important part of the  $\CD$-module structure on the
right hand side  of \rf{Lang} can be represented as the system of
eigenvalue equations
\begin{equation}
\SH_r\Psi\,=\,E_r\Psi\,,
\end{equation}
for the quantized Hitchin Hamiltonians $\SH_r$,
with $\SH_r$ being certain second order differential operators on a line bundle on
${\rm Bun}_G$.

We are going to propose that important aspects  of
the geometric Langlands correspondence can be understood as
arising in a suitable limit from a correspondence between the conformal blocks
of Liouville theory and those of
the SL$(2)$-WZNW model that will be described below. This correspondence
is based on the relations observed in \cite{Sto} between
the Belavin-Polyakov-Zamolodchikov (BPZ)
and Knizhnik-Zamolodchikov (KZ) systems of differential equations.
We are going to show that this correspondence opens the way to
{\it construct} the conformal blocks of the SL$(2)$-WZNW model from those of Liouville theory.
The possibility to reconstruct the SL$(2)$-WZNW model from Liouville theory,
\begin{equation}\label{q-Lang}
\boxed{\quad\phantom{\Big|}\SL{\mathsf i}\so\su_b\;
\quad}
\quad
\longrightarrow \quad
\boxed{\quad\phantom{\Big|}{\SW\SZ\SN\SW}_{k}(\fsl_2)\;
\quad}
\end{equation}
may be seen as a kind of inversion of the Drinfeld-Sokolov reduction.
The correspondence \rf{q-Lang} will be shown to reproduce important aspects of the geometric
Langlands correspondence in the limit $k\ra -2$, which is called the critical level limit.
The KZ equations yield the eigenvalue equations for the Hitchin
Hamiltonians representing the right hand side of \rf{Lang}.
This limit is related to the limit $b\ra\infty$ in Liouville theory.
Liouville theory has the profound property to be self-dual under
inversion of the parameter $b$,
which means that almost\footnote{The only exception being the dependence on the cosmological
constant, the parameter in front of the interaction term $e^{2b\vf}$ in the Liouville action.}
all characteristic quantities of Liouville theory
like in particular the conformal blocks
are unchanged if one replaces $b$ by $1/b$.
This phenomenon will be referred to as the modular duality of Liouville theory.
The modular duality
of Liouville theory implies that the critical level
limit is equivalent to the classical limit in Liouville theory. Fuchsian differential
equations of the second order arise naturally in this limit. The monodromies of the
solutions to these Fuchsian differential equations are the local systems on the left hand side
of \rf{Lang}.

On the level of the representation theory of chiral algebras a related way
to explain the local geometric Langlands correspondence was developed in
\cite{FF1}, see also \cite{Fr05} and in particular \cite[Section 8.6]{Fr2}
for a nice discussion. Relations between the geometric Langlands correspondence
and the Separation of Variables method have first
been discussed in \cite{Fr},
which was an important source of inspiration for this work.

There are two
elements that the relationship with Liouville theory
adds to the story. First, it allows one to lift certain aspects of
the geometric
Langlands correspondence from the local level (opers on a disc
vs. representations of the current algebra at the critical level)
to the global level where both sides of \rf{Lang}
are associated to Riemann surfaces.
Even more interesting appears to be the possibility to
extend the geometric Langlands correspondence from the level of
$\CD$-modules
to the level of the multivalued holomorphic solutions of the
differential equations
coming from the $\CD$-module structure.

\subsection{Modular duality vs. Langlands duality}

The modular duality of Liouville theory offers another way to
construct a
SL$(2)$-WZNW model from Liouville theory \cite{GNN}, obtained from the
first by the exhange $b\ra b^{-1}$, schematically
\begin{equation}\label{q-Langl}
\boxed{\quad\phantom{\Big|}{\SW\SZ\SN\SW}_{\check{k}}(\fsl_2)\;
\quad}\quad
\longleftarrow \quad
\boxed{\quad\phantom{\Big|}\SL{\mathsf i}\so\su_b\;
\quad}
\quad
\longrightarrow \quad
\boxed{\quad\phantom{\Big|}{\SW\SZ\SN\SW}_{k}(\fsl_2)\;
\quad}
\end{equation}
The level $\check{k}$ of the SL$(2)$-WZNW model on the left is
determined by
\begin{equation}
\check{k}+2\,=\,(k+2)^{-1}\,=\,-b^2\,.
\end{equation}
We are going to show that the corresponding relations between
spaces of conformal blocks lead to another
approach to the geometric Langlands correspondence in which
both sides of \rf{Lang} are obtained in the limit $b\ra\infty$.
The same limit which reduces the KZB equations to the eigenvalue
equations of the quantized Hitchin Hamiltonians is now observed to be
the classical limit $\check{k}\ra\infty$
for the dual WZNW model
${\SW\SZ\SN\SW}_{\check{k}}(\fsl_2)$.
Local systems will be found to arise very naturally in the
classical limit $\check{k}\ra\infty$ of the WZNW model.
This means that the somewhat asymmetric looking geometric
Langlands correspondence \rf{Lang} is obtained in the limit $b\ra\infty$
from a much more symmetric looking duality between
two WZNW-models at different levels,
\begin{equation}\label{dmaindiag1}
\begin{array}{ccccc}
\boxed{\phantom{\Big|}
{\rm PSL}(2)-\text{local systems}
\;} &  & \leftrightarrow & & \boxed{\phantom{\Big|}
\CD\text{-modules on ${\rm Bun}_{{\rm SL}(2)}$}
\;} \\
\uparrow & & & & \uparrow \\
\boxed{\quad\phantom{\Big|}{\SW\SZ\SN\SW}_{\check{k}}(\fsl_2)\;
\quad} &
\longleftarrow &
\boxed{\quad\phantom{\Big|}\SL{\mathsf i}\so\su_b\;
\quad}
&
\longrightarrow &
\boxed{\quad\phantom{\Big|}{\SW\SZ\SN\SW}_{k}(\fsl_2)\;
\quad}
\end{array}
\end{equation}
It seems natural to call the relations schematically represented
at the bottom of
\rf{dmaindiag1} a
{\it quantum geometric Langlands correspondence}.
Other approaches to defining ``quantum'' versions
of the geometric Langlands correspondence have been discussed
in \cite{Fr2,St06,Ga,Kap}.

The author views this
paper as a first look on a huge iceberg, most of which remains
invisible. It is hoped
that this look stimulates further investigations
of this subject.

In the present Part I of our paper
we will mostly illustrate the picture
proposed above by examples related to
Riemann surfaces of genus 0. The forthcoming second part of
the paper \cite{TII} will discuss the cases of higher genus in more detail.
Nevertheless, whenever easily possible we'll
present the relevant background and the main claims in full
generality already in this paper.

{\small {\bf Acknowledgements.} The author is grateful to
D. Ben-Zvi, D. Gaiotto, E. Frenkel, A. Neitzke, N. Nekrasov, S. Shatashvili
and E. Witten for discussions and remarks on topics
related to this work.

Some preliminary results were presented at
the miniprogram ``Gauge Theory and Langlands Duality'' in 08/2008,
the mini-conference ``Langlands-Duality week'' in 03/2009,
both at the KITP, UCSB Santa Barbara, as well as the
Simons Center Workshop ``Perspectives, Open Problems and
Applications of Quantum Liouville Theory'', Stony Brook, Spring 2010.
The author would like to thank
the organizers for the invitation to these inspiring conferences.

The author furthermore gratefully acknowledges support
from the EC by the Marie Curie Excellence
Grant MEXT-CT-2006-042695.
}

\section{The classical Hitchin system}

\setcounter{equation}{0}
\renewcommand{\sh}{{\rm\sst H}}

The following is a (rather incomplete) reminder of some basic
definitions and results about the Hitchin system.

\subsection{Self-duality equations vs. Higgs pairs}

The Hitchin moduli space $\CM_{\rm H}(C)$ on a Riemann surface $C$
is the space of solutions $(A,\theta)$ of the
SU$(2)$ self-duality equations
 \begin{equation}\label{Hitchin}
F_A+R^2\,[\,\theta\,,\,\bar\theta\,]\,=0\,,\qquad\quad\begin{aligned}
&\bar\pa_A\theta+\theta\,\bar\pa_A\,=\,0\,,\\
&\pa_A\bar\theta+\bar\theta\,\pa_A\,=\,0\,,
\end{aligned}
\end{equation}
where $d_A=d+A$ is an SU$(2)$-connection on a vector bundle $V$, and $\theta$ is a holomorphic one-form with
values in ${\rm End}(V)$, modulo
SU$(2)$ gauge transformations.
$\CM_{\rm H}(C)$
is a space of complex dimension $6g-6+2n$ if $C=C_{g,n}$ is
a Riemann surface of genus $g$ with $n$ marked points.

Decomposing $d_A$ into the
$(1,0)$ and $(0,1)$ parts $\pa_A$ and $\bar{\pa}_A$, respectively,
we may associate to each solution a holomorphic vector bundle $\CE$ with
holomorphic structure being defined by $\bar\pa_A=\bar\pa+A^{0,1}$.
The equations \rf{Hitchin} imply in particular
that $\theta$ is holomorphic with respect to the holomorphic
structure defined by $\bar{\pa}_A$. This means that each solution of the
self-duality equations \rf{Hitchin} defines a Higgs pair $(\CE,\theta)$,
which is a pair $(\CE,\theta)$ of objects,
with $\CE$ being  a holomorphic vector bundle, and
$\theta\in H^0(C,{\rm End}(\CE)\ot\Omega^1_C)$. Conversely, Higgs pairs
come from solutions of the self-duality equations iff they are stable,
which means that any $\theta$-invariant sub-bundle of $V$ must have
a degree that is smaller than half of the degree of $V$ \cite{Hi}.

We will allow for a finite number of regular singularities on $C$.
Introducing a local coordinate $y_r$ near the singular point $z_r$,
$r=1,\dots,n$, we will require that the singular behavior is of the
form
\begin{equation}
\begin{aligned}
&A\,=\,\frac{1}{2i}A_r\bigg(\frac{dy_r}{y_r}-\frac{d\bar{y}_r}
{\bar{y}_r}\bigg)+{\rm regular}\,,\qquad
&\theta\,=\,\frac{1}{2}\theta_r\frac{dy_r}{y_r}+{\rm regular}\,,
\end{aligned}
\end{equation}
with $\theta_r$ and $A_r$ being simultaneously diagonalizable matrices,
and $A_r$ skew-hermitian.

There is a natural slice within $\CM_{\rm H}(C)$ defined by the condition
$\theta=0$. It is clearly isomorphic to ${\rm Bun}_G(C)$, the moduli space of
holomorphic bundles on $C$. Sections $\theta$
of $H^0(C,{\rm End}(\CE)\ot K_C)$, where $K_C$ is the canonical line bundle, naturally represent vectors in the
cotangent space of ${\rm Bun}_G(C)$. It follows that
an open dense subset of $\CM_{\rm H}(C)$ is naturally isomorphic to
the cotangent bundle $T^*{\rm Bun}_G(C)$.

\subsection{The Hitchin integrable system}

To begin with, let us consider an ${\rm SL}(2)$ Higgs pair $(\CE,\theta)$.
Associate to it the quadratic differential
\begin{equation}\label{qfromphi}
\vartheta\,=\,{\rm tr}(\theta^2)\,.
\end{equation}
Expanding $\vartheta$ with respect to a basis $\{\vartheta_1,
\dots,\vartheta_{3g-3+n}\}$
of the $3g-3+n$-dimensional
space of quadratic differentials,
\begin{equation}\label{H_rdef}
\vartheta\,=\,\sum_{r=1}^{3g-3+n}H_r\,\vartheta_r\,,
\end{equation}
defines functions $H_r$, $r=1,\dots,3g-3+n$
on $\CM_{\rm H}(C)$ which are called Hitchin's Hamiltonians.
The subspaces $\Theta_E\subset\CM_{\rm H}(C)$
defined by the equations
$H_r=E_r$ for $E=(E_1,\dots,E_{3g-3+n})$
are abelian varieties (complex tori) for generic $E$.
This means that
$\CM_{\rm H}(C)$ can be described as a torus fibration with base
$\CB$ which can be identified with the space $\CQ(C)$
of quadratic differentials on the underlying
Riemann surface $C$.

There is a complex structure $I$ on $\CM_{\rm H}(C)$ for which both
$E$ and complex analytic coordinates for the fibres $\Theta_E$
are holomorphic. Associated with the
complex structure $I$ is the holomorphic
symplectic structure $\Omega_I$ which can be defined as
 \begin{equation}\label{OmegaI}
\Omega_I\,=\,2iR\int_C{\rm tr}(\de\theta\wedge\de A^{0,1})\,,
\end{equation}
where $\bar\pa_A=\bar\pa+A^{0,1}$.
The functions $H_r$ are Poisson-commuting with respect to the
Poisson structure coming from the symplectic structure $\Omega_I$.

The assertions above can be summarized in the
statement that $\CM_{\rm H}(C)$
is an algebraically
completely integrable system in complex structure $I$.
It is useful to encode the values of $E$ into the
definition of the spectral curve
\begin{equation}
\Sigma\,=\,\big\{ \,(v,y)\,|\,{\rm det}(v-\theta(y))\,=\,0\,\big\}\,,
\end{equation}
which defines a double cover $\Sigma$ of the surface $C$.

Certain generalizations of this set-up will become relevant for us later.
Instead of considering holomorphic $G={\rm SL}(2)$-bundles one may consider bundles in
$G={\rm GL}(2)$. One may furthermore consider
Higgs fields $\theta$ in $H^0(C,{\rm End}(\CE)\ot L)$, with $L$ being a line bundle different from the
canonical line bundle $K_c$.
In this case one gets additional degrees of freedom and additional Hamiltonians from ${\rm tr}(\theta)$.
This will be discussed in more detail in Part II of this paper.

\subsection{Separation of variables}\label{SOV}

In the Separation of Variables method \cite{Sk,Sk2}
one maps the dynamics of an integrable
system to the motion of a divisor on the spectral curve. It furnishes
a set of canonically conjugate variables which can be used as a
starting point for the quantization of the model.

Let ${\rm Bun}_G$ be the moduli space of holomorphic vector bundles $\CE$ on $V$.
In the case of SL$(2)$-bundles on $C_{g,n}$, for example, we have
\begin{equation}
d:=\,{\rm dim}_{\BC}(\CM_k)\,=\,3g-3+n\,.
\end{equation}
The separation of variables amounts to the existence of a birational map
\[ {\SS\so\sv}\;:\;T^*\CM \ra (T^*C)^{[d]}\,,\]
from $T^*\CM$ to the Hilbert scheme of points on $T^*C$ which is a
symplectomorphism
on open dense subsets.
The open dense subset of $(T^*C)^{[d]}$ which is relevant here
is the set
\[ \CY\,\equiv\,((T^*C)^d-\Delta)/S_d\,,
\]
with $\Delta$ being union of all diagonals
and $S_d$ is the symmetric group. On this subset one may choose coordinates
$(y,v)\equiv[(y_1,v_1),\dots,(y_d,v_d)]$ such that the symplectic form $\Omega_I$ becomes
\begin{equation}\label{symsepvar}
\omega\,=\,\sum_{r=1}^d dv_r \wedge dy_r\,.
\end{equation}
The main idea behind
the definition of the coordinates $(y,v)$ can be described most
easily in the
case of $g=0$ with $n$ marked points  corresponding to the Gaudin model.
Choosing a gauge where ${A}^{0,1}=0$,
the Higgs pair $(V,\theta)$ is characterized
by Higgs fields of the form
\begin{equation}\label{Higgsform}
\theta\,=\,\bigg(\,
\begin{matrix} \theta^0 & \phantom{-}\theta^+ \\ \theta^- & -\theta^0\end{matrix}\,\bigg)
,\qquad
\theta^a\,=\,
\sum_{r=1}^n\frac{\theta_r^a}{y-z_r}\,,
\end{equation}
subject to the global $\fsl_2$-invariance
constraints $\sum_{r=1}^n\theta_r^a=0$ for $a=-,0,+$.
$\vartheta(y)$ is the form
\begin{equation}
\vartheta(y)\,=\,
\sum_{r=1}^n\left(\frac{\de_r}{(y-z_r)^2}+\frac{H_r}{y-z_r}\right)\,,
\end{equation}
where $\de_r$ are central elements,
and the $H_r$ are the Hitchin Hamiltonians. In the following we will
mostly consider a slightly simpler version of this model obtained
by sending $z_n\ra\infty$, $\theta_n^-\ra 0$ and
imposing $\sum_{r=1}^{n-1}\theta_r^a=\de_{a,0}\sqrt{\de_n}$ for $a=-,0$.
The difference is in the treatment of the global $\fsl_2$-invariance,
and will turn out to be inessential even on the quantum level. 

The coordinates $y_r$ are then found as the zeros of $\theta^-(y)$,
\begin{equation}
\theta^-(y)\,=\,u\frac{\prod_{j=1}^{n-3}(y-y_j)}{\prod_{i=1}^{n-1} (y-z_i)}\,,\qquad u=\sum_{i=1}^{n-1}
\mu_i z_i\,,
\end{equation}
where $\mu_r={\rm Res}_{y=z_r}\theta^-(y)$. The conjugate variables
$v_r$ can be found from the condition that
the point
$(y_r,v_r)$ of $T^*C$ lies on the curve $\Sigma$,
\begin{equation}\label{speccurve}
v^2_r\,=\,\vartheta(y_r)\,=\,{\rm tr}(\theta^2(y_r))\,.
\end{equation}
Given the tuple $(y,v)$ one recovers the spectral curve $\Sigma$
as the curve that goes through
all points $(y_r,v_r)$, while for fixed values of the conserved quantities one
may view the equations $v_r^2=\vartheta(y_r)$
as equations determining the "momenta" $v_r$
in terms of the variables $y_r$ and the values of the conserved Hamiltonians.

The separation of variables for $g>0$ was discussed in \cite{GNR,kr1}.
It can be recast in a form more similar to the $g=0$ case
as will be discussed in \cite{TII}.

\subsection{Special geometry of the base of the Hitchin fibration}

It is known that the base of any algebraically completely integrable system canonically
has special geometry \cite{Fd}. In the case at hand it can be described as follows.
The spectral curve $\Sigma$ is a double covering of the surface $C$. On $\Sigma$ let
us introduce the differential
\begin{equation}\label{SWdiff}
dS\,=\,vdy\,.
\end{equation}
We then get the special coordinates $a_r$, $a_s^{\sst D}$ as the periods of $S$ along the homology
cycles $\al_r$, $\be_s$, $r,s=1,\dots,h$,
respectively,
\begin{equation}\label{periods1}
a_r\,=\,\int_{\al_r}dS\,,\qquad a_r^{\sst D}\,=\,\int_{\be_r}dS\,.
\end{equation}
Both  $a=(a_1,\dots,a_{h})$ and
$a^{\sst D}=(a_1^{\sst D},\dots,a_{h}^{\sst D})$
represent systems of coordinates for the base $\CB$. The change of coordinates
can be described in terms of a holomorphic
function $\CF(a)$ called prepotential such that
\begin{equation}
a_r^{\sst D}\,=\,\frac{\pa \CF}{\pa a_r}\,.
\end{equation}
There are coordinates $\tau=(\tau_1,\dots,\tau_{h})$
on the torus fibres $\Theta_{E(a)}$ which are
Poisson-conjugate to the variables
$a$. The coordinates $(a,\tau)$ are action-angle variables for the
Hitchin system.

\section{Isomonodromic deformations as a deformation of the Hitchin system}\label{sec:isomon}
\setcounter{equation}{0}

\subsection{Hitchin moduli space as space flat connections}

There is a useful description of the Hitchin moduli space $\CM_{\rm H}$
as a moduli space of flat complex connections.
To each solution $(A,\theta)$ to the self-duality equations \rf{Hitchin}, we may associate
the connection
\begin{equation}\label{nonHodge}
\nabla\,=\,\nabla'+\nabla''\,,\qquad\quad\begin{aligned}
&\nabla'\,=\,\pa_A+R\,\theta\,, \\
&\nabla''\,=\,\bar\pa_{A}+R\,\bar\theta\,.
\end{aligned}
\end{equation}
The connection is flat thanks to the equations \rf{Hitchin}.
In \rf{nonHodge} we have introduced a
parameter $R$ which can be eliminated by a rescaling of $\theta$,
$\bar\theta$, but which is sometimes useful.

Conversely, given a flat connection $\nabla$ on a vector bundle $V$ on $C$, there is a canonical
way to associate to it a solution to the self-duality equations. For given
connection $\nabla$,
let $\rho:\pi(X)\ra{\rm PSL}(2,\BC)$ be its monodromy representation.
The key result \cite{Do,Co,Si90} to be used is the existence of a {\it canonical}
hermitian metric $h$ on the fibres of $V$, which may be represented as
a smooth $\rho$-equivariant {\it harmonic}
map from the universal cover $\tilde{C}$
of $C$ to $H=G/K$, with $K$ being
the maximal compact subgroup of $G={\rm PSL}(2,\BC)$.
The metric $h$ allows us to decompose the connection $\nabla$ into the
component $\nabla_K=d+A$ preserving the subgroup $K$, and the component $\Theta$ orthogonal to Lie algebra of $K$.
Decomposing further into the $(1,0)$ and $(0,1)$ parts $\nabla'=\pa_A+R\theta$
and $\nabla''=\bar\pa_A+R\bar\theta$
yields a solution to the self-duality equations, as is reviewed
in \cite[Section 2]{Si97}.

\subsection{Flat connection vs. local systems}

Using the complex structure of the underlying surface, it is
possible to represent the connections $\nabla$ in
holomorphic terms. To this aim one may note that
$\nabla''=\bar\pa_A+R\bar\theta$ is an
integrable holomorphic structure and $\nabla'$ is an integrable
holomorphic connection on $\CE=(V,\nabla'')$. We may
introduce local trivializations such that
$\nabla''=\bar\pa$. The connection
$\nabla$ is then locally described by
holomorphic differential operators
of the form
\begin{equation}\label{holoconn}
\nabla'\,=\,(\pa_y+M(y))\,dy\,.
\end{equation}
One may furthermore
trivialize the bundle by means of a basis of local solutions
of $\nabla's=0$. The transition functions between the patches of such a
trivialization must then be {\it constant}. This means that a flat connection
$\nabla$ on a surface $C$ canonically defines a {\it local system},
a vector bundle defined by
a local trivialization with {\it constant} transition functions between
the patches.

Let ${\rm Loc}_{G}(C)$
be the moduli space of $G$-local systems for
a complex group $G$. The space
${\rm Loc}_{\rm PSL(2,\BC)}(C)$ is also known as the space of projective
structures on $C$.

Two alternative realizations of local systems
will be used. First, each local system canonically defines
a representation
\begin{equation}
\rho: \pi_1(C)\;\;\rightarrow\;\;{\rm PSL}(2,\BC)\,.
\end{equation}
Conversely, each such representation $\rho$ canonically defines
a local system, see e.g. \cite[Section 3.1]{Fr2} for the three-line proof.
The space of local systems is therefore isomorphic
to the space $\Homsl{C}$ of representations of the fundamental group
$\pi_1(C)$ in $\pslc$. In the following we will often identify
the representations $\rho$ of $\pi_1(C)$ with the corresponding local
systems.

Alternatively, one may associate to each local system a pair of objects
$(\CE,\nabla')$, where $\CE$ is a holomorphic vector bundle on $C$, and
$\nabla'$ is a holomorphic connection, which may be locally represented in the
form
\begin{equation}\label{nablaM}
\nabla'\,=\,\frac{\pa}{\pa y}+M(y)\,,
\end{equation}
where $M(y)$ is a matrix-valued holomorphic function.
The correspondence between
local systems and pairs $(\CE,\nabla')$ is called the
Riemann-Hilbert correspondence. 

It may also be useful to consider holomorphic 
vector bundles $\tilde{\CE}$ with
{\it meromorphic} connections $\tilde\nabla'$. 
As illustrated later, we may then have pairs $(\tilde\CE,\tilde\nabla')$ 
which have the same monodromy representation 
$\rho:\pi(X)\ra{\rm PSL}(2,\BC)$ as a given local system $(\CE,\nabla')$.

\subsection{Hyperk\"ahler structure}

For a given Higgs bundle $(\CE,\theta)$ one may introduce, generalizing
the decomposition \rf{nonHodge}, a one-parameter family of flat connections
as
\begin{equation}\label{nonHodgedef}
\nabla_\zeta^{}\,=\,\nabla'_\zeta+\nabla''_\zeta\,,\qquad\quad\begin{aligned}
&\nabla'_\zeta\,=\,\pa_A+\frac{1}{\zeta} R\,\theta\,, \\
&\nabla''_\zeta\,=\,\bar\pa_{A}+\zeta R\,\bar\theta\,.
\end{aligned}
\end{equation}
Associated with this one-parameter family of flat connections
are a one-parameter family of natural complex structures $J^{(\zeta)}$ and
holomorphic symplectic forms $\varpi_\zeta$
on the Hitchin moduli space $\CM_{\rm H}(C)$ \cite{Hi}.
The complex structures $J^{(\zeta)}$ can be characterized by the
property that holomorphic functions of the flat connection $\nabla_\zeta$
like the traces of monodromies of $\nabla_\zeta$ are holomorphic
in complex structure $I_\zeta$. The holomorphic symplectic forms
$\varpi_\zeta$ can be defined as
\begin{equation}\label{holosym}
\varpi_\zeta\,=\,\frac{1}{2}\int_C{\rm tr}(\de\CA_\zeta\wedge
\de\CA_\zeta)\,,
\end{equation}
where $\CA$ is defined by $\nabla_\zeta=d+\CA_\zeta$. The form $\varpi_\zeta$
can be expanded as
\begin{equation}
\varpi_\zeta\,=\,-\frac{i}{2\zeta}\,\omega_++\omega_3-\frac{i}{2}\zeta\,
\omega_-\,,
\end{equation}
where, in particular, $\omega_+\equiv \Omega_I$,
the natural holomorphic symplectic form associated with the Higgs
bundle picture for $\CM_{\rm H}(C)$ defined in \rf{OmegaI}.

In order to describe the situation in purely holomorphic terms, let
$\CE_{\zeta R}$ be the holomorphic structure on the vector bundle $V$ defined by
$\nabla''_\zeta=\bar\pa_A+\zeta R\theta$.
On $\CE_{\zeta R}$ let us, following \cite[Section 4]{Si97}, consider
the holomorphic ${\ep}$-connection which locally is obtained from $\nabla'_\zeta$
by
\begin{equation}
\pa_{\ep}\,\equiv\,{\ep}\nabla'_\zeta\,=\,
{\ep}\pa+I(y)\,,\qquad {\ep}\,=\,\frac{\zeta}{R}\,.
\end{equation}
$I(y)$ transforms under gauge transformations as
$I\rightarrow g^{-1}I g+{\ep} g^{-1}\pa g$.

\subsection{Drinfeld-Sokolov reduction}\label{Fuchsian}

Important for us will be a special class of local systems called
{\it opers} \cite{BDop}, which in the case $\fg=\fsl_2$ may be
described as bundles admitting a connection that locally looks
as in \rf{nablaM} with
\begin{equation}\label{operconn}
M(y)\,=\,\bigg(\,\begin{matrix} 0 & t(y) \\ 1 & 0 \end{matrix}\,\bigg)\,.
\end{equation}
The equation $(\pa_y+M(y))\phi=0$ now implies that the component $\chi$ of
$\phi=(\eta,\chi)$ solves a second order differential equation of the form
\begin{equation}\label{DFuchsian}
(\pa_y^2+t(y))\chi=0\,.
\end{equation}
Under holomorphic
changes of the local coordinates on $C$,
the differential operator $\pa_y^2+t(y)$ transforms as
\begin{equation}\label{opertrsf}
t(y)\;\mapsto\;(y'(w))^2t(y(w))-\frac{1}{2}\{y,w\}\,,
\qquad\{y,w\}\,\equiv\,\frac{y'''}{y'}-
\frac{3}{2}\left(\frac{y''}{y'}\right)^2\,,
\end{equation}
which is the transformation law characteristic for a {\it projective}
connection. The transformation law \rf{opertrsf} follows from the
transformation law for a connection if one takes into account
that a compensating gauge transformation is generically needed in order
to recover the form \rf{operconn} of the connection after having
changed the local coordinate.

It is useful to note that
any local system can be represented in the form \rf{operconn} away from
finitely many points on $C$, as discussed in \cite[Section 9.6]{Fr2}.
In order to see this for $G={\rm SL}(2)$ in a simple way,
let us represent the elements of the
connection matrices $M(y)$ as
\begin{equation}
M(y)\,=\,\bigg(\,\begin{matrix} \al(y) & \beta(y) \\ \gamma(y) & -\al(y)
\end{matrix}\, \bigg)\,.
\end{equation}
$\gamma(y)$ may be set to one by a singular gauge transformation
\begin{equation}\label{gauge}
\pa_y+M'\,\equiv\,g\cdot(\pa_y+M)\cdot g^{-1}\,,\quad g\,=
\bigg(\,\begin{matrix} h & 0 \\ 0 & h^{-1}\end{matrix}\,\bigg)\,,
\end{equation}
where $h(y)=\sqrt{\gamma(y)}$.
The gauge transformation $g$ is singular at the zeros $w_1,\dots,w_d$
of $\gamma(y)$.
This is where $M'(y)$ has additional singularities.
By means of a further gauge transformation one may set the
diagonal elements of $M'(y)$ to zero, leading to
\begin{equation}
M'(y)\,=\,\bigg(\,\begin{matrix} 0 & t(y) \\ 1 & 0
\end{matrix}\, \bigg)\,.
\end{equation}
The corresponding equation $(\pa_y^2+t(y))\chi=0$ has regular singular points
$z_1,\dots,z_n$ {\it and} $w_1,\dots,w_d$.
The behavior near the
singular points is of the form
\begin{equation}\label{locfuchs}
\begin{aligned}
&t(y)\,\sim\,\frac{\de_r}{(y-z_r)^2}+\frac{H_r}{y-z_r}\qquad\text{near}\;\;y=z_r\,,\\
&t(y)\,\sim\,\frac{-3}{4(y-w_k)^2}+
\frac{\kappa_k}{y-w_k}\qquad\text{near}\;\;y=w_k\,.
\end{aligned}
\end{equation}
However, the additional
singularities as $w_1,\dots,w_d$ are
gauge artefacts, and the monodromy of $\pa_y+M'(y)$
is the same as the one of $\pa_y+M(y)$.
The singular points $w_1,\dots,w_d$ of $t(y)$ are called apparent
singularities which expresses the fact that the monodromy around these
singular points is trivial in $\pslc$. It can be shown
\cite[Section 3.9]{Fr} that this implies the equations
\begin{equation}\label{nullclass}
t_{k,2}+t_{k,1}^2\,=\,0\,,\quad {\rm where}\quad
t(y)=\sum_{i=0}t_{k,i}(y-w_k)^{i-2}\,.
\end{equation}
These equations give relations between the parameters
$w_k$, $\kappa_{k'}$ and $H_r$ of the projective
connection $\pa_y^2+t(y)$.

\subsection{Space of opers}\label{opers}

Of particular importance for us will be the cases where $d=0$,
where there are no apparent singularities.
Let ${\rm Op}_{\fsl_2}(C)$ the space of $\fsl_2$-opers on a
Riemann surface $C$.
Two opers ${{P}}$ and ${{P}'}$ differ by a holomorphic
quadratic differentials $\vartheta={{P}}-{{P}'}$.
This implies that the space ${\rm Op}_{\fsl_2}(C_{g,n})$ of $\fsl_2$-opers
on a fixed surface $C_{g,n}$ of genus $g$ with $n$ marked points
is $3g-3+n$-dimensional.
Complex analytic coordinates for ${\rm Op}_{\fsl_2}(C_{g,n})$ are
obtained by picking a reference projective connection ${P}_0$,
a basis ${\vartheta}_1,\dots,{\vartheta}_{3g-3+n}$ for the vector
space of quadratic differentials,
and writing any other projective connection ${P}$ as
\begin{equation}\label{accessdef}
{P}\,=\,{P}_0+\sum_{r=1}^{3g-3+n}H_r\,{\vartheta}_r\,.
\end{equation}
The parameters $H_r$ are sometimes called accessory parameters.

The monodromy representations $\rho_{{P}}:\pi_1(C_{g,n})\ra {\rm SL}(2,\BC)$
of the differential operators ${{P}}$ will generate a
$3g-3+n$-dimensional subspace
of the space ${\rm Loc}_{\rm PSL(2,\BC)}(C_{g,n})$  of local systems.
Varying the complex structure of the underlying surface $C$, too,
we get a subspace of ${\rm Loc}_{\rm PSL(2,\BC)}(C)$ of complex dimension
$6g-6+2n$. The space of opers forms an affine bundle $\CP$ over the Teichm\"uller space
of deformations of the complex structure of $C$.
Standard Teich\-m\"uller theory identifies the space of
quadratic differentials with the holomorphic cotangent space
of the Teichm\"uller space of deformations of the complex structure
of $C$.  It follows that $\CP$ is canonically isomorphic to
the cotangent bundle $T^*\CT(C)$ over the Teichm\"uller
space $\CT(C)$.
It is important that the mapping $\CP\ra \Homsl{C}$
defined by the associating to the projective connection $P$ its monodromy
representation $\rho_P^{}$ is locally biholomorphic, and
that the corresponding mapping $T^*\CT\ra \Homsl{C}$ is
symplectic in the sense that the canonical cotangent
bundle symplectic structure
is mapped to the natural symplectic structure $\Omega_J\equiv \varpi_\zeta\big|_{\zeta=1}$ on the
space of flat complex connections, see \cite{Ka} and references therein.
We may, therefore, choose a set of local coordinates ${q}=({q}_1,\dots,{q}_{3g-3+n})$
on $\CT(C_{g,n})$
which are conjugate to the coordinates $H_r$ defined above in the sense that
the Poisson brackets coming from this symplectic structure are
\begin{equation}\label{Poissopers}
\{\,{q}_r\,,{q}_s\,\}\,=\,0\,,\qquad \{\,H_r\,,{q}_s\,\}\,=\,\de_{r,s}\,,\qquad
\{\,H_r\,,H_s\,\}\,=\,0\,.
\end{equation}
Other useful sets of coordinates for the space of opers can be defined
in terms of the monodromy map ${\mathbf M}: {\rm Op}_{\fsl_2}(C_{g,n})\ra
\Homsl{C_{g,n}}$ as follows. Let $\CC$ be a pants decomposition
of $C_{g,n}$ defined by a collection $\{\ga_1,\dots,\ga_{3g-3+n}\}$ of
simple mutually non-intersecting closed curves. To each curve $\ga_r$
there corresponds a unique generator $\ga_r$ of the fundamental group
$\pi_1(C_{g,n})$. For given oper, let
\begin{equation}\label{monofix}
L_{r}:=\,2\cosh\frac{{l}_r}{2}\,:=\,{\rm tr}(\rho_{{P}}^{}(\ga_r))\,.
\end{equation}
The tuple ${l}_{{P}}^{}
=({l}_1,\dots,{l}_{3g-3+n})$ can be used to parameterize
${\rm Op}_{\fsl_2}(C_{g,n})$ at least locally.
For given ${l}
=({l}_1,\dots,{l}_{3g-3+n})$
one may generically find
accessory parameters $H_r=H_r({l},{q})$ such that \rf{monofix} is satisfied
(Riemann-Hilbert correspondence).

\subsection{Isomonodromic deformations}\label{isomon}

The representation of the connection $\nabla$ in terms of
holomorphic data $(\CE,\nabla')$ was
using the complex structure on $C$.
It is natural to ask how $(\CE,\nabla')$ vary
if we consider variations of the complex structure of $C$ for
fixed monodromy of the connection $\nabla$. This defines families
of compatible flows on the space of pairs $(\CE,\nabla')$
\cite{Si97,kr2,BF}. The differential
equations characterizing these flows are called the
isomonodromic deformation equations. For $g=0$
one gets well-known systems of partial differential equations,
and more explicit forms of the resulting equations
for $g>0$ were obtained in \cite{kr2}.

\subsubsection{Example: The Schlesinger system}

In the case of $g=0$ with n punctures
we can describe ${\rm Loc}_{\rm SL(2,\BC)}$ as the space of all meromorphic connections of the form
\begin{equation}\label{holoconn2}
\pa_y+M(y)\,=\,\pa_y+\sum_{r=1}^{n}\frac{M_r}{y-z_r}\,,\qquad M_r\in\fsl(2,\BR)_\BC\,,
\end{equation}
with fixed conjugacy class of $M_r$.
The equations
\begin{equation}\label{Schlesinger}
\begin{aligned}
&\frac{\pa}{\pa z_s}M_r\,=\,\frac{[M_r,M_s]}{z_r-z_s}\,,\qquad r\neq s\,,\\
&\frac{\pa}{\pa z_r}M_r\,=\,\sum_{s\neq r}\frac{[M_r,M_s]}{z_r-z_s}\,,
\end{aligned}
\end{equation}
ensure that the monodromy of $\pa_y+M(y)$ stays constant under variations of the complex structure
of $\CC$. The equation \rf{Schlesinger} are integrable, and define
what is called the Schlesinger system.

\subsubsection{Separation of variables for the Schlesinger system}

For the case at hand ($g=0$)
it is particularly easy to see that
the relation between the holomorphic connection \rf{holoconn2}
and the second order differential operator $\pa_y^2+t(y)$
is based on a change of variables very similar to the one that was
giving the separation of variables for the Gaudin model in Subsection
\ref{SOV}. Following the discussion in Subsection \ref{Fuchsian}
leads to the differential equation $(\pa_y^2+t(y))\chi=0$ with
$t(y)$ of the form
\begin{equation}\label{tgen}
t(y)\,=\,\sum_{r=1}^{n}\left(\frac{\de_r}{(y-z_r)^2}+\frac{H_r}{y-z_r}\right)-
\sum_{k=1}^{l}\left(\frac{3}{4(y-w_k)^2}-\frac{\kappa_k}{y-w_k}\right)\,.
\end{equation}
In order to eliminate the constraints following from projective
invariance, let us choose $z_n=0$, $z_{n-1}=1$, $z_{n-2}=\infty$.
$t(y)$ may then be written in  the form
\begin{equation}\label{tred}\begin{aligned}
t(y)\,=\,\frac{\de_{n}}{y^2}+\frac{\de_{n-1}}{(1-y)^2}+\frac{\upsilon}{y(y-1)}
&+\sum_{r=1}^{n-3}\left(\frac{\de_r}{(y-z_r)^2}+\frac{z_r(z_r-1)}{y(y-1)}\frac{H_r}{y-z_r}\right)\\ &-
\sum_{k=1}^{d}\left(\frac{3}{4(y-w_k)^2}-\frac{w_k(w_k-1)}{y(y-1)}\frac{\kappa_k}{y-w_k}\right)\,.
\end{aligned}\end{equation}
In the case where $d=n-3$, the equations \rf{nullclass}
can be written explicitly as
\begin{equation}\label{BPZclass}
\kappa_k^2+
\sum_{r=1}^{n}\left(\frac{\De_r}{(w_k-z_r)^2}+\frac{H_r}{w_k-z_r}\right)
-\sum_{\substack{k=1\\k'\neq k}}^d
\left(\frac{3}{4(w_k-w_{k'})^2}-\frac{\kappa_k}{w_k-w_{k'}}
\right)=0\,,
\end{equation}
The equation \rf{BPZclass} can be solved
to express
$H_r$ in \rf{tred} in terms of variables $w_k$ and $\kappa_k$. The
resulting expression is a quadratic polynomial $H_r(\kappa,w)$ in the
variables $\kappa_k$.
This is precisely the form of an projective connection considered in the
theory of the Garnier systems.
The monodromy of the projective connection $\pa_y^2+t(y)$
stays constant under a variation of the variables $z_r$ provided that
$\kappa_k$, $w_k$ are varied according to
\begin{equation}\label{garnier}
\frac{\pa w_k}{\pa z_r}\,=\,\frac{\pa H_r}{\pa \kappa_k}\,,\qquad
\frac{\pa \kappa_k}{\pa z_r}\,=\,-\frac{\pa H_r}{\pa w_k}\,.
\end{equation}
These equations are nothing but the rewriting of the Schlesinger-equations
\rf{Schlesinger} in terms of the separated variables $y_k$ defined by the condition
of $\gamma(y_k)=0$.

\subsubsection{Symplectic structure}

The Hamiltonian form
of the isomonodromic deformation equations \rf{garnier}
naturally suggests the Poisson structure
\begin{equation}\label{Poissonisomon}
\{\,w_r\,,w_s\,\}\,=\,0\,,\qquad
\{\,\kappa_r\,,w_s\,\}\,=\,\de_{r,s}\,,\qquad
\{\,\kappa_r\,,\kappa_s\,\}\,=\,0\,.
\end{equation}
In the generalization to higher genus \cite{Iw2} it is natural to set $d=3g-3+n$. The
positions $w_k$ of the apparent singularities together with the residues
$\kappa_k$ introduced in \rf{locfuchs}
then form a local set of coordinates for the subset of $\Homsl{C}$ given
by the monodromies of the Fuchsian differential operators $\pa_y^2+t(y)$.
This Poisson structure \rf{Poissonisomon} coincides with the one coming from
the holomorphic symplectic form
$\Omega_J$ on Hitchin moduli space \cite{Iw2}.

\subsection{Real slices}

A real slice in the space $\Homsl{C}$ is naturally defined by the
requirement that the representation $\rho\in\Homsl{C}$ is conjugate
to a subgroup of ${\rm PSL}(2,\BR)$. The space $\Homslr{C}$
has finitely many connected components, as will be described
in the following.

\subsubsection{Representation in terms of Higgs pairs}

Let us first describe
how these components are represented in terms of Higgs pairs $(\CE,\theta)$.
For a given effective divisor $D$ of degree $d$, and chosen
square-root $K^{\frac{1}{2}}_C$ of the canonical line bundle
let us consider holomorphic bundles $\CE$ of the
form
\begin{equation}
\CE\,=\,L_1\oplus L_2\,,\quad
\begin{aligned} & L_2:=K_C^{+\frac{1}{2}}\,,\\
& L_1:=K_C^{-\frac{1}{2}}\ot D\,.
\end{aligned}\end{equation}
Let us then consider Higgs fields of the form
\begin{equation}
\theta\,=\,\bigg(\,\begin{matrix} 0 & \vartheta \\ \gamma & 0
\end{matrix}\, \bigg)\,,
\end{equation}
where $\gamma$ is a holomorphic section of the line bundle
corresponding to $D$, and ${\vartheta}$ is a quadratic differential.

\subsubsection{Representation in terms of flat connections}

It can be shown (see \cite{Hi} for details) that the
flat connection $\nabla=d+\CA$ associated to such Higgs pairs may then
be represented in the form
\begin{equation}\label{SU11conn}
\CA\,=\,\left(\,\begin{matrix} -\frac{1}{2}\pa\vf & R\,\frac{1}{\zeta}\,\vartheta\, e^{-\vf} \\
  R\,\frac{1}{\zeta}\,\gamma \,e^{\vf} &  +\frac{1}{2}
\pa\vf \end{matrix}\right)dz
+\left(\,\begin{matrix} +\frac{1}{2}\bar\pa\vf &  R\,\zeta\,\bar{\gamma}\,e^{\vf} \\
 R\,\zeta\,\bar{\vartheta} \,e^{-\vf} & -
\frac{1}{2}\bar\pa\vf\end{matrix}\right)d\bz\,.
\end{equation}
It is manifest that this is a SU$(1,1)$-connection when $\zeta=1$.
The flatness is equivalent to
\begin{equation}\label{zerocurv}
\pa\bar\pa \vf\,=\,R^2(\gamma\bar{\gamma}
\,e^{2\vf}-{\vartheta}\bar{\vartheta}\,e^{-2\vf})\,.
\end{equation}
This is a variant of the Sinh-Gordon equation. It reduces to a variant
of the Liouville-equation for ${\vartheta}=0$.
In this case, equation \rf{zerocurv}
implies that
\begin{equation}
ds^2\,=\,\gamma\bar{\gamma}\,e^{2\vf}\,dzd\bz
\end{equation}
is a metric of constant negative curvature on $C$. This metric has
conical singularities with excess angle $2\pi$ at the zeros of $\gamma$.

\subsubsection{Complex structures on the real slices}\label{cplxreal}

Let  $\Homslr{C}^d$ be connected component in $\Homslr{C}$ that
is described in this way. It will be important for us to note
that there is a convenient description of $\Homslr{C}^d$ as a complex
analytic manifold associated with this description.
It is proven in \cite[Section 10]{Hi}, see also
\cite[Section 6.2]{Go}, that
$\Homslr{C}^d$ has the structure of a holomorphic vector bundle
over the symmetric power ${\rm Sym}^d(C)$, with fiber over
$D\in {\rm Sym}^d(C)$ being the vector space
\begin{equation}
\{\,\vartheta\in H^0(C,K_c^2)\,|\,{\rm div}(\vartheta)\geq D\,\}\,\simeq
\BC^{3g-3+n-d}\,.
\end{equation}
The relation to the representation in terms of Fuchsian differential
equations described in Subsection \ref{Fuchsian} is easy to see:
The divisor $D$ is the collection $(w_1,\dots,w_d)$ of apparent
singularities. The equations \rf{nullclass} imply that
there are $d$ relations among the $3g-3+n$ holomorphic
quadratic differentials.

\subsubsection{Teichm\"uller component}

Of particular interest and importance is the case where
$d=0$. The representations $\rho\in \Homslr{C}^0$
are then {\it Fuchsian}, which means that quotient of the
upper half plane by the representation $\rho$ produces
a Riemann surface $C$ with natural constant curvature metric
induced from the hyperbolic metric of upper half plane \cite{Go}.
The component $\Homslr{C}^0$ is therefore
called the Teichm\"uller component.

The relation to the discussion above can easily be seen as follows.
Setting $d=0$ implies that $\ga$ can be set to unity in above equations.
Each point in $\Homslr{C}^0$ can be represented by a flat
connection of the form
\rf{SU11conn}. It is shown in \cite[Theorem (11.2)]{Hi} that
any solution of the flatness condition \rf{zerocurv} defines
a metric of constant negative curvature via
\begin{equation}\label{defhyp}
ds^2\,=\,e^{2\vf}\,\big(dz+e^{-2\vf}\bar\vartheta \,d\bz\big)
\big(d\bz+e^{-2\vf}\vartheta \,dz\big)\,.
\end{equation}
We see that the quadratic differentials $\vartheta$
parameterize deformations of the constant negative
curvature metric associated to the complex structure of $C$.
The natural
complex structure on the Teichm\"uller space $\CT(C)$ of such
deformations coincides with
the complex structure on the Teichm\"uller component $\Homslr{C}^0$
introduced in Subsection \ref{cplxreal}, as follows from the
fact that $e^{-2\vf}\bar\vartheta$ is the so-called harmonic
Beltrami-differential associated to the quadratic
differential $\vartheta$ from Teichm\"uller theory.

One should note, however, that
in order to get the corresponding Fuchsian representative
$\pa_y^2-t(y)$, we need to set $\vartheta=0$, as was observed above. It is not hard to show that $t(y)$ is then equal
to the so-called energy-momentum tensor associated to the metric of constant
negative curvature,
\begin{equation}\label{enmom}
t(y)\,=\,-(\pa_z^{}\vf)^2+\pa^2_z\vf\,.
\end{equation}
This means that the space of opers ${\rm Op}_{\fsl_2}(C)$ is another
slice in $\Homsl{C}$ which
intersects the real slice $\Homslr{C}^0$ transversally.
This fits naturally to our earlier observation that the
space of opers is naturally isomorphic to the holomorphic
cotangent space of $\CT(C)$: While $\Homslr{C}^0$ is naturally isomorphic
to the Teichm\"uller space, the space ${\rm Op}_{\fsl_2}(C)$ represents the
cotangent space of $\CT(C)$. Both spaces are naturally isomorphic
to each other, but this isomorphism is not holomorphic, as it involves the
constant curvature metric $e^{2\vf}dzd\bz$.

\subsection{K\"ahler potential on the real slices}\label{Kahlerreal}

The symplectic structure $\Omega_J$ on $\Homsl{C}$, restricted to the real slices $\Homslr{C}^d$
gives us the natural symplectic structure $\Omega_J^{\BR}$ we will consider. We have seen, on the other
hand, that the real slices $\Homslr{C}^d$ have a natural complex structure
related to the the complex structure from Teichm\"uller theory.

For $d=0$ it is known
that the symplectic structure $\Homslr{C}^d$ is {\it K\"ahler} w.r.t.
this symplectic structure, with K\"ahler potential given by the
Liouville action.
The Liouville action functional $S^{\rm cl}\big[\vf]$ is defined as
\begin{equation}
S_L[\vf]\;=\; \frac{1}{2\pi}
\int_{C_{g,n}}d^2z \,\bigg(\,\frac{1}{2}(\pa_a\vf)^2+8\pi\mu b^2e^{2\vf}\,\bigg)
+[{\rm boundary\;\, terms}]\,,
\end{equation}
with a suitable choice of boundary terms which was determined in \cite{TZ85,TZ87a,TZ87b}.
The Liouville action defines a natural symplectic form on $\CT(C)$ as
\begin{equation}\label{omdef}
\Omega_\CT\;=\;2 i\,\pa\bar{\pa}S^{\rm cl},
\end{equation}
where $\pa$, $\bar{\pa}$ are the holomorphic and anti-holomorphic
components of the de Rham differential on $\CT_{g,n}$ respectively.
It was shown in \cite{TZ85,TZ87a,TZ87b} that $\Omega_\CT$ coincides with the Weil-Petersson symplectic form
from Teichm\"uller theory, which in turn is known to coincide \cite{Go0,Hi}
with the symplectic structure $\Omega_J^{\BR}$ on $\Homslr{C}^0$.
This implies that the Poisson structure on the real slices is still of the form
\rf{Poissopers}, but the variables $H$
are no longer independent, but rather given as functions of the variables ${q}$.
A convenient reference projective connection $P_S$ is e.g. given by the Schottky
uniformization, and
\begin{equation}\label{HfromSLiou}
P-P_S\,=\,\frac{1}{2}\pa S_L[\vf]\,,
\end{equation}
for a suitable choice of boundary terms in the definition of the Liouville action functional
\cite{TZ85,TZ87a,TZ87b}.

For $g=0$, $C_{0,n}=\BP^1\setminus\{z_1,\dots,z_n\}$
one can represent ${{P}}$ in the form
 \begin{equation}\label{tg0}
{{P}}=\pa_y^2+t(y)\,,
\qquad t(y)\,=\,\sum_{r=1}^{n}\left(\frac{\de_r}{(y-z_r)^2}+
\frac{H_r}{y-z_r}\right)\,.
\end{equation}
The parameters $H_r$ are restricted by the relations
\begin{equation}
\sum_{r=1}^{n}H_r=0\,,\qquad
\sum_{r=1}^{n}(z_rH_r+\de_r)=0\,,\qquad
\sum_{r=1}^{n}(z_r^2H_r+2\de_rz_r)=0\,.
\end{equation}
The coordinates ${q}_r$ conjugate to the $H_r$ may be chosen as cross-ratios of the
coordinates $z_1,\dots,z_n$. Alternatively, one may set $z_n=\infty$, $z_{n-1}=1$ and $z_{n-2}=0$,
and identify the remaining coordinates $q_1\equiv z_1,\dots, q_{n-3}\equiv z_{n-3}$ as the conjugates of $H_1,\dots,H_{n-3}$,
respectively.

For $d>0$ one needs to take into account the fact that the quadratic differentials that are
holomorphic on $C$ are constrained by the relations \rf{nullclass}. For general $g$  let us pick a set
$\{\vartheta_1,\dots,\vartheta_{3g-3+n-d}\}$
of linearly independent holomorphic quadratic differentials.
In order to generate all
Teichm\"uller deformations one has to add $d$ meromorphic quadratic differentials
$\{\vartheta_1^{\rm\sst mer},\dots,\vartheta_d^{\rm\sst mer}\}$,
where $\vartheta_k^{\rm \sst mer}$
has a pole at
the point $w_k$. Expanding
\begin{equation}
P-P_0\,=\,\sum_{r=1}^{3g-3+n-d}H_r\,\vartheta_r+\sum_{k=1}^d \kappa_k \,
\vartheta_k^{\rm\sst mer}\,,
\end{equation}
The quadratic differentials define $(1,0)$-forms on $\CT(C)$. There are corresponding local coordinates
${q}_1,\dots,{q}_{3g-3+n-d}$ and $w_1,\dots,w_d$ such that these $(1,0)$-forms are representable as $d{q}_r$
and $dw_k$, respectively.
The only non-vanishing Poisson brackets are then
\begin{equation}
\{\,\kappa_k\,,w_l\,\}\,=\,\de_{r,s}\,,\qquad\{\,H_r\,,{q}_s\,\}\,=\,\de_{r,s}\,.
\end{equation}
The coordinates $w_k$ will parameterize the variations of the
positions of the apparent singularities.

\subsection{Limit to the Hitchin system}

Let us now consider the limit $\zeta\ra 0$, $R\ra\infty$ 
such that $R\zeta$ stays
constant. This implies in particular
that the
integrable holomorphic structure
$\nabla''_\zeta=\bar\pa_A+\zeta R\theta$
is kept fixed in the limit.
The ${\ep}$-connection $\pa_{\ep}={\ep}\pa-I$ becomes
the Higgs field $\theta$.
In terms of opers, one may take the limit by
rescaling $t(y)={\ep}^{-2}\vartheta_{\ep}^{}(y)$. The transformation of
$\vartheta_{\ep}^{}(y)$ is then
\begin{equation}\label{opertrsf'}
\vartheta_{\ep}^{}(y)\;\mapsto\;(y'(w))^2\,\vartheta_{\ep}^{}(y(w))-\frac{{\ep}^2}{2}\{y,w\}\,.
\end{equation}
For ${\ep}\ra 0$
we get the transformation law of quadratic differentials.
We may in this sense regard the space ${\rm Op}_{\fsl_2}(C)$
as a deformation $\CB_{\ep}$ of the base $\CB$ of the
Hitchin fibration.

The complex structure $J^{(\zeta)}$ turns into
the complex structure $I$ characteristic for the Hitchin integrable system,
and the symplectic structure $\varpi_\zeta$ becomes the symplectic structure $\Om_I$
of $T^*{\rm Bun}_G$ in the sense that
\[
\Om_I\,=\,\underset{\zeta=0}{\rm Res}(\varpi_\zeta)\,.
\]
One may furthermore study the isomonodromic deformation equations
in this limit. This is slightly delicate, but the upshot is
that isomonodromic deformation equations indeed reduce to the
equations of motion of the Hitchin system in this limit \cite{kr2}.

We may naturally distinguish two types of observables,
the Hamiltonians $H_r$ on the one hand, and the traces
$L_r=2\cosh\frac{l_r}{2}={\rm tr}(\rho_P(\ga_r))$
of monodromies on the other hand. It seems natural to refer to them
as local and non-local observables, respectively.
The former are clearly related to the Hitchin Hamiltonians $H_r$
in the limit under consideration. In order to study the asymptotics
for ${\ep}\ra 0$ of the latter, let us note that
the leading WKB approximation to the
solutions of the equation $(\ep^2\pa_y^2+\vartheta(y))\chi(y)=0$ can be
constructed in terms of the differential $dS$ introduced in \rf{SWdiff}
as
\begin{equation}\label{WKBchi}
\chi_\pm(y)\,=\,\exp\left(\frac{i}{{\ep}}\int^y dz\; v_\pm\right)\,,
\end{equation}
where $v_\pm$ are two choices of a branch for the solution of the equation
$v^2=\vartheta(z)$.
It follows easily from \rf{WKBchi} that the
parameters ${l}_r$ are related to the action variables $a_r$
introduced in \rf{periods1}
in the limit ${\ep}\ra 0$,
\begin{equation}\label{lfroma}
{l}_r\,=\,\frac{4\pi}{{\ep}}a_r\,,\qquad r=1,\dots,3g-3+n\,,
\end{equation}
as follows from \rf{WKBchi}. We may therefore regard the nonlocal
observables ${L}_r$
parameterizing
the monodromies of the flat connections as deformations of the
action variables $a_r$
associated to the special geometry of the Hitchin fibration.

These remarks are supposed to clarify the meaning of the arrow marked
(A)$_{\ep_2}$ in \rf{maindiag1}.
In this regard let us note in particular that
the relation between the
isomonodromic deformations and the Hitchin
system involves a hyperk\"ahler rotation in the parameter $\zeta$.

\section{Quantization of the Hitchin system}\label{sec:quantHitchin}
\setcounter{equation}{0}
\subsection{Quantization scheme}\label{quantscheme}

The quantization of an algebraically integrable system like the Hitchin system can roughly
be approached in the following way.
\begin{itemize}
\item[a)] Deform the space of (algebraic) functions on the
phase space to a noncommutative algebra $\CA$, whose elements are supposed to become the
observables of the quantum theory. Of particular interest
are the Hamiltonians whose proper definition  will typically
involve ordering issues. Integrability means that $\CA$ should contain
a commutative subalgebra $\CI$ of "sufficient size" generated by the
quantized Hamiltonians.
\item[b)] Choose a Lagrangian subspace $\CL$ of the phase space, here $\CM_{\rm\sst H}$, and
represent the quantized algebra of observables as algebraic differential or difference
operators on $\CL$.
\item[c)] Choose a *-structure on the algebra of observables and find a scalar product on the
space of functions that realizes the *-structure via hermitian conjugation.
\end{itemize}
For any given value of $E$ there is typically a finite-dimensional space of
solutions to the
eigenvalue equations $\SH\Psi=E\Psi$ of the
Hamiltonian $\SH\in\CA$ that have suitable analytic properties.
Normalizability of the solutions w.r.t.
the scalar product introduced in step c) then selects in many cases
a discrete subset of the possible
values of $E$ and thereby yields the quantization conditions.

In the case of the Hitchin system it is in most cases difficult to implement
step c) explicitly since the complex structure on the phase space typically depends
on the complex structure of the underlying surface $C$, and is hard to describe explicitly,
making the definition of a suitable scalar product difficult. The only known examples of Hitchin-type
systems where it is know how to implement step c) explicitly are the Calogero systems.

In the following we shall describe basic elements of steps a) and b), but instead of implementing c)
we shall discuss another approach. Integrability means that the phase space in question has the
structure of a torus fibration with base $\CB$. We will (inspired by \cite{NS}) propose to replace step c) by
\begin{itemize}
\item[c')]
Define  a suitable  "deformation" $\CB_{\ep}$ of the base $\CB$, and
a function $\CW:\CB_{{\ep}}\ra\BC$ called Yang's potential whose
critical points define the eigenvalues.
\end{itemize}
Such a procedure appears to be well-motivated in the case
of algebraically integrables systems for the following reason.
In some prototypical examples like the quantum Toda chain
it is possible to prove that the quantization conditions
obtained in step c) can indeed be recast in the form c')
for a suitable choice of the Yang's potential $\CW$ \cite{KT}.
As $\CW$ depends analytically on all parameters, one may
use the characterization of the spectrum in terms of the
Yang's potential even in cases when step c) is hard
to implement.

In the case of the Hitchin system, the space $\CB_{\ep}$
will be identified with the moduli space of opers.
Our proposal will be to identify the Yang's potential with
the semiclassical Liouville conformal blocks, which
leads to a precise definition in terms
of the theory of ordinary differential equations.

\subsection{Semiclassical quantization of the separated variables}
\label{Semiquant}

One possible approach to the quantization of the Hitchin system
can be based on using the separated variables
$(y,t)\equiv[(y_1,t_1),\dots,(y_h,t_h)]$ with symplectic form
\rf{symsepvar} as a starting point.
In view of \rf{symsepvar} it seems natural to
regard the variables $t_r$
as momenta, the $y_r$ as coordinates.
The quantization of the equation \rf{speccurve} defining the spectral
curve of the Hitchin system
would then naturally lead to the
differential equation
\begin{equation}\label{BAX}
({\ep}^2\pa_{y_r}^2+\vartheta(y_r))\chi(y)\,=\,0\,.
\end{equation}
These equation will in the following be referred to as the Baxter equations.

The leading WKB approximation to the solutions of the Baxter equation \rf{BAX} can be
constructed in terms of the differential $dS$ introduced in \rf{SWdiff},
\begin{equation}\label{WKB}
\chi_\pm(y)\,=\,\exp\left(\frac{i}{{\ep}}\int^y dz\; v_\pm\right)\,,
\end{equation}
where $t_\pm$ are two choices of a branch for the solution of the equation
$v^2=\vartheta(z)$.

There are a few natural possibilities one could discuss for the definition of
quantization conditions.

\subsubsection{Real quantization}

It may happen that the integrable system of physical interest is actually a real slice of the algebraically
integrable system under mathematical study. This is the case e.g. in the Calogero model, which is
a special case of the Hitchin system, see e.g. the discussion in \cite{NS}. In this case one
needs to impose a reality condition on the coordinate functions $a_r$ (or $a_r^{\rm D}$) of the base
of the torus fibration. Combined with the
Bohr-Sommerfeld quantization conditions one arrives at the conditions
\begin{equation}\label{BS1}
a_r^{}\,=\,2\pi {\ep}\, n_r^{}\,,\qquad n_r\in\BZ\,,
\end{equation}
where $a_s^{}$ are the periods of the Seiberg-Witten
differential $dS =dy\, v$ w.r.t. the cycles generating a canonical
basis for $H_1(\Sigma,\BZ)$,
\begin{equation}\label{periods}
a_r\,=\,\int_{\al_r}dS,\qquad
a_s^{\sst D}\,=\,\int_{\be_s}dS\,.
\end{equation}
The concrete choice of a basis $(\al_1,\dots,\al_{3g-3+n};
\be_1,\dots,\be_{3g-3+n})$ may be tricky. For later convenience
we'll henceforth assume that the $\al_r$ coincide with a maximal
set of simple closed curves defining a pants decomposition of $C_{g,n}$.
Reference \cite{NS} contains a discussion of the
quantization conditions \rf{BS1} in a  related context.

\subsubsection{Complex quantization}

In the present case there is an interesting alternative one can discuss. The phase space in question
has a complex structure, allowing one to require that the *-structure on the algebra of observables
acts as complex conjugation.
One may simply choose the Lagrangian subspace $\CL$ to be a complex
subspace, and assume that the algebra $\CA$ of observables is realized
both by holomorphic and
anti-holomorphic differential operators. For $g=0$ one thereby gets the
${\rm SL}(2,\BC)$-Gaudin model.

One of the basic requirements that an eigenfunctions of the Hitchin Hamiltonians should satisfy
is single-valuedness.
In order to find a single-valued solution of the eigenvalue equations we need to form linear
combinations of the form
\begin{equation}
\phi(y,\bar{y})\,=\,
(\chi_+(y),\chi_-(y))\cdot\,K\, \cdot\bigg(\,
\begin{matrix}\bar{\chi}_+(\bar y)\\\bar{\chi}_-(\bar y)\end{matrix}\,\bigg)\,.
\end{equation}
Single-valuedness of $\phi$  leads
to the Bohr-Sommerfeld quantization conditions $K={\rm diag}(1,-1)$ and
\begin{equation}\label{BS}
{\rm Re}(a_r^{})\,=\,\pi {\ep}\, n_r^{}\,,\qquad
{\rm Re}(a_s^{\sst D})\,=\,\pi {\ep}\,n_s^{\sst D}\,,
\end{equation}
where $a_s^{}$ and $a_s^{\sst D}$ are the periods of the Seiberg-Witten
differential $dS =dy\, v$ as above. The derivation of \rf{BS}
is discussed in detail for the closely
related SL$(2,\BC)$-XXX-model in \cite{DKM}.
\begin{rem}
It is interesting to note\footnote{This fact 
has independently been remarked 
by S. Shatashvili, who had discussed it in various 
lectures long before this paper has appeared.}
that the conditions \rf{BS} coincide
with the so-called attractor
equations.
\end{rem}

\subsection{Quantization of the Hitchin Hamiltonians}

A dense open subset of $\CM_{\rm H}$ is isomorphic to  $T^*{\rm Bun}_G$,
the moduli space of stable $G$-bundles (here $G=SL(2)$) on $C$.
This forms the basis to an alternative approach to the quantization of
$\CM_{\rm H}$,
in which the Lagrangian subspace $\CL$ taken to be (possibly a real slice of)
${\rm Bun}_G$.
States in the quantum theory can then be described
in terms of functions (or sections of some line bundle) on ${\rm Bun}_G$.
Linear coordinates
on the fibres of $T^*{\rm Bun}_G$ play the role of momenta and would
consequently be realized as differential operators. The complex structure
on ${\rm Bun}_G$ (which is coming from the complex structure on $C_{g,n}$)
allows us
to distinguish holomorphic and anti-holomorphic coordinates and the
corresponding differential operators.

Hitchin's Hamiltonians are constructed from ${\rm tr}(\theta^2)$.
As $\theta$ is holomorphic
in complex structure $I$, they
should become holomorphic differential operators
on ${\rm Bun}_G$ after quantization.
Beilinson and Drinfeld constructed
such differential operators
from the representation theory of affine Kac-Moody algebras
at the critical
level \cite{BD}, as will be reviewed in Section \ref{geolangsect} below.
Here we will discuss the example of $g=0$ where
Hitchin's Hamiltonians can be quantized in an elementary way.

\subsubsection{Example: The $SL(2,\BC)$-Gaudin model}\label{SLCgaudin}

In the case of $g=0$,
$\CM$ parameterizes the choices of parabolic structures
at the marked points $z_n$. On an open dense subspace one may
use the collection of complex numbers $(x_1,\dots,x_n)$ modulo
Moebius-transformations as coordinates for $\CM$. The
complex number $x_r$ parameterizes a point in the flag manifold
$G/B$ ``attached'' to marked point $z_r$.

We will consider the tensor product of $n$ principal series
representations $\CP_j$ of
$SL(2,\BC)$. It corresponds to the tensor product
of representations of the Lie algebra $\fsl(2,\BC)$ generated
by differential operators ${\CJ}^a_r$
acting on functions $\Psi(x_1,\bx_1,\dots,x_n,\bx_n)$ as
\begin{equation}\label{CJdef}
{\CJ}^-_r=\pa_{x_r} ,\quad
{\CJ}^0_r=x_r\pa_{x_r}-j_r,\quad
{\CJ}^+_r=-x^2_r\pa_{x_r}+2j_rx_r,
\end{equation}
and the complex conjugate operators $\bar{\CJ}^a_r$.
The Casimir of the representation $\CP_{j_r}$
is parameterized via $j_r$ as $j_r(j_r+1)$.
The Gaudin Hamiltonians are defined as
\begin{equation}\label{Hdef}
{\mathsf H}_r^{}\,\equiv\,\sum_{s\neq r}\frac{
{\CJ}_{rs}}{{z}_r-{z}_s},\qquad
\bar{\mathsf H}_r^{}\,\equiv\,\sum_{s\neq r}\frac{
\bar{\CJ}_{rs}}{\bar{z}_r-\bar{z}_s},
\end{equation}
where the differential operator ${\CJ}_{rs}$ is defined as
\begin{align}\label{kzmu}
{\CJ}_{rs}:=\eta_{aa'}{\CJ_r^a\CJ_s^{a'}}:=
{\CJ}^0_r {\CJ}^0_s + & \half \left( {\CJ}^+_r {\CJ}^-_s
+{\CJ}^-_r {\CJ}^+_s \right)\,,
\end{align}
while $\bar{\CJ}_{rs}$ is the complex conjugate of ${\CJ}_{rs}$.
The Gaudin Hamiltonians are mutually commuting,
\begin{equation}
[\,{\mathsf H}_r\,,\,{\mathsf H}_s\,]\,=\,0\,,\qquad
[\,{\mathsf H}_r\,,\,\bar{\mathsf H}_s\,]\,=\,0\,,\qquad
[\,\bar{\mathsf H}_r\,,\,\bar{\mathsf H}_s\,]\,=\,0\,.
\end{equation}
It is therefore natural to look for joint eigenfunctions of the
Gaudin Hamiltonians in the space of
wave-functions $\Psi(x_1,\bx_1,\dots,x_n,\bx_n)$ which satisfy the conditions
\begin{equation}\label{globalSL2}
\sum_{r=1}^n\CJ_r^a\,\Psi\,=\,0\,,\qquad
\sum_{r=1}^n\bar\CJ_r^a\,\Psi\,=\,0\,,
\end{equation}
for $a=-,0,+$. It will be convenient for us to observe that the 
solutions of this problem are in one-to-one correspondence to the 
solutions of the slightly simplified problem which is found in the
limit when $z_n\ra\infty$, $x_n\ra\infty$. The simplified problem
can be defined directly by dropping the terms with $s=n$ from the 
expression \rf{Hdef}
for the Hamiltonians $\SH_r$, $r=1,\dots,n-1$. 
The eigenvalue equations for the resulting Hamiltonians 
are supplemented by the
equations $\sum_{r=1}^{n-1}\CJ_r^a\Psi=\de_{a,0}j_n$ and
$\sum_{r=1}^{n-1}\bar\CJ_r^a\Psi=\de_{a,0}j_n$
for $a=-,0$. The equivalence of the two problems is seen by 
expressing the solutions to \rf{globalSL2} in terms of 
functions $\psi$ that depend only on the cross-ratios
formed out of the variables $z_1,\dots,z_n$ and 
$x_1,\dots,x_n$. The same functions can be used to express
the solutions of the simplified problem.

\subsubsection{Eigenvalue problems?}\label{singrem}

It is not trivial to define a reasonable eigenvalue problem in the
case of the Gaudin model. In order to illustrate the point,
let us consider the $SL(2,\BR)$-Gaudin model, in which case
the variables $x_r$ are assumed to be real.
Let us look at the simplest case $n=4$ in some detail.
In this case one may
reduce the dependence on $x_1,\dots,x_4$
to the cross-ratio $x$.
There is only a single operator $\SH$ to consider, which   reduces to
a second order differential
operator $\CD_x^{(2)}$ in $x$ of the form
\begin{equation}
\CD^{(2)}_x\,=\,\frac{\CD^{(2)}_{21}}{z}+\frac{\CD^{(2)}_{32}}{1-z}\,,
\end{equation}
with ${\CD^{(2)}_{21}}$ and ${\CD^{(2)}_{32}}$
being second order differential operators that do not depend on $z$.
One natural quantization problem to consider would be
to assume $z\in\BR$, and to look for a measure $d\nu(x)$ making
$\CD_x^{(2)}$ self-adjoint in $L^2(\BR,d\nu(x))$.
The problem is that the definition of self-adjoint extensions
of $\CD_x^{(2)}$ may require careful choice of boundary conditions
at the singular points of $\CD_x^{(2)}$. In this regard let
us note that $\CD_x^{(2)}$ has regular singular points at
$x=0,z,1,\infty$, respectively, as follows from
\begin{equation}\label{CDrep}
\CD_x^{(2)}\,=\,x(x-1)(x-z)\frac{\pa^2}{\pa x^2}+\dots\,,
\end{equation}
up to terms with less derivatives with respect to $x$.
Of particular interest is the singularity at $x=z$.
We will return to this point later.

\subsection{Quantum separation of variables}\label{qSOV}

It is known that the quantization of the the separated variables and
the quantization of the Hitchin Hamiltonians are equivalent
even on the quantum level \cite{Sk}, as we shall now briefly recall.
We will use the simplified formulation obtained by sending $z_n\ra\infty$,
$x_n\ra\infty$ in the following, as was introduced at the end of 
Subsection \ref{SLCgaudin}.

The first step is to diagonalize ${\rm J}^-$ by means of the Fourier
transformation
\begin{equation}
\tilde\Psi(\mu_1,\dots,\mu_{n-1})\,=\,\frac{1}{\pi^{n-1}}
\int d^2 x_1 \dots \int d^2 x_{n-1} \;\prod_{r=1}^{n-1}
|\mu_r|^{2j_r+2} e^{\mu_rx_r-\bar\mu_r\bx_r}
\Psi(x_1,\dots,x_{n-1})\,.
\end{equation}
The generators ${\rm J}_r^a$ are mapped to the differential operators
${\rm D}_r^a$,
\begin{equation}\label{Dmu}
{\rm D}_r^-=\mu_r ,\quad
{\rm D}_r^0=\mu_r \p_{\mu_r} ,\quad
{\rm D}_r^+=\mu_r \p^2_{\mu_r}-\frac{j_r(j_r+1)}{\mu_r}\,,
\end{equation}
so that the Gaudin Hamiltonians get represented by
\begin{equation}\label{Hdef'}
{\mathsf H}_r^{}\,\equiv\,\sum_{s\neq r}\frac{
{\rm D}_{rs}}{{z}_r-{z}_s},\qquad {\rm D}_{rs}:=\eta_{aa'}
{\rm D}^a_r {\rm D}^{a'}_s \,,
\end{equation}
and their complex conjugates.
Let us then define variables
to $y_1,\dots,y_{n-3},u$ related to
the variables $\mu_1,\dots,\mu_{n-1}$ via
\begin{equation}\label{keyrel}
{\quad
\sum_{i=1}^{n-1} \frac{\mu_i}{t-z_i} =
u\frac{\prod_{j=1}^{n-3}(t-y_j)}{\prod_{i=1}^{n-1} (t-z_i)}
\, .\quad}
\end{equation}
Note that the constraints \rf{globalSL2} imply $\sum_{r=1}^{n-1} \mu_r=0$.

It was shown by Sklyanin \cite{Sk}
that the system of eigenvalue equations
$\SH_r\Psi=E_r\Psi$ is  transformed by the change of variables
$\mu_1,\dots,\mu_{n-1}\ra y_1,\dots,y_{n-3},u$ into the set of equations
\begin{equation}\label{BAX1}
(\pa_{y_k}^2+t(y_k))\chi(y_k)\,=\,0\,,\qquad
t(y)\,\equiv\,-\sum_{r=1}^{n-1}
\left(\frac{j_r(j_r+1)}{(y_k-z_r)^2}-\frac{E_{r}}{y_k-z_r}\right)\,.
\end{equation}
The dependence with respect to the variables $y_k$ has completely
separated.
Solutions to the Gaudin-eigenvalue equations $\SH_r\Psi=E_r\Psi$
can therefore be constructed from
solutions $\chi_k(y_k)$ of \rf{BAX1} by means of the ansatz
\begin{equation}
\Psi\,=\,
\prod_{k=1}^{n-3}
\chi_k^{}(y_k;{q})\,.
\end{equation}
Note, in particular that
the Baxter equations \rf{BAX1} reproduce \rf{BAX} if
$\de_r=\CO({\ep}^{-2})$ so that $t(y)={\ep}^{-2}\vartheta(y)$.

%

\subsection{Quantization from single-valuedness}\label{singleval}

In the case of the complex quantization as discussed above one may find
strong constraints on the eigenvalues already from the condition of single-valuedness.

Sklyanin's observation allows us to write $\Psi$ as a linear combination
of solutions to the Fuchsian differential equations (\ref{BAX1})
which have the factorized form
\begin{equation}\label{factorrep}
\Psi(y_1,\bar y_1,\dots,y_{n-3},\bar y_{n-3})\,=\,\prod_{a=1}^{n-3}\chi(y_a,\bar y_a)\,.
\end{equation}
We want to impose the condition of single-valuedness.
\newcommand{\by}{{\bar{y}}}
Let us focus on the dependence of $\Psi$ w.r.t. some $y\in \{y_1\dots
y_{n-3}\}$. $\chi(y,\by)$ can be represented as a linear combination of the
linearly independent solutions to the equation $(\pa_y^2+t(y))\chi_i=0$ and
its complex-conjugate counterpart in the form,
\begin{equation}\label{chians}
\chi(y,\bar{y})\,\equiv\,\chi(\,y,\bar{y}\,|\,z_1\dots z_{n-1})\,=\,
\left(\,\bar{\chi}_1(\bar{y})\,,\,\bar{\chi}_2(\bar{y})\, \right)
\cdot K\cdot
\biggl(\begin{matrix} \chi_1(y)\\
\chi_2(y)\end{matrix}\biggl)\,,
\end{equation}
where $K$ is a $2\times 2$ matrix which is constrained by the
condition of
single-valuedness,
\begin{equation}\label{monod}
M^{\dagger}_r\cdot K \cdot
M_r^{\phantom{\dagger}}=K~~{\rm for ~all}~~r=1,\dots,n.
\end{equation}
This is a
highly overdetermined
system of equations for the matrix $K$,
which can not be satisfied for arbitrary monodromy matrices $M_r$.


We claim that it is necessary and sufficient that the representation
of the fundamental group $\pi_1(\Sigma)$ which is generated by the matrices
$M_r$ is conjugate to a discrete subgroup of$SU(1,1)\subset
SL(2,{\mathbb C})$.  We may then use $K={\rm diag}(1,-1)$ to solve
(\ref{monod}).

Indeed, given a single-valued solution $\chi(y,\by)$ of $(\pa_y^2+t(y))\chi=0$
and the complex conjugate equation we may construct the metric
$e^{2\vf}dyd\by$ where $e^{-\vf}=\chi$.
This metric has negative constant curvature
since $\vf=-\log\chi$ satisfies the Liouville equation $\pa\bar\pa \vf=e^{2\vf}$.
The uniformization theorem ensures existence
and uniqueness of such a metric, which implies uniqueness
of the problem to find single-valued solutions to $(\pa_y^2+t(y))\chi_i=0$ and
its complex-conjugate counterpart.

We conclude that there exists a distinguished "state" $|q\rangle$ in the SL$(2,\BC)$ Gaudin model
corresponding to the metric of negative constant curvature on $C_{g,n}$.

\section{Liouville theory}\label{sec:Liou}
\setcounter{equation}{0}

Liouville theory is a field theory with conformal symmetry
generated by the energy-momentum tensor with central
charge $c$ that will be parameterized in terms of a parameter $b$ as
\begin{equation}
c\,=\,1+6Q^2\,,\qquad Q:=b+b^{-1}\,.
\end{equation}
It is characterized by
the correlation functions of $n$ primary fields $e^{2\al_r\phi(z_r,\bz_r)}$
denoted as
\begin{equation}\label{LcorrI}
 \big\langle\!\big\langle
\;e^{2\al_n\phi(z_n,\bz_n)}\cdots e^{2\al_1\phi(z_1,\bz_1)}\;
\big\rangle\!\big\rangle_{C_q}\,.
\end{equation}
$C_q$ is a family of Riemann surfaces parameterized by a collection $q=(q_1,\dots,q_{3g-3+n})$ of
complex-analytic local coordinates for the moduli space $\CM_{g,n}$ of Riemann surfaces.
The conformal dimension $\De_r$ of the primary field
$e^{2\al_r\phi(z_r,\bz_r)}$ is given as $\De_r\equiv\De_{\al_r}:=\al_r(Q-\al_r)$.
The correlation functions \rf{LcorrI}
can be represented in a holomorphically factorized form
\begin{equation}\label{holofact}
\big\langle\!
\big\langle \;e^{2\al_n\phi(z_n,\bz_n)}\cdots e^{2\al_1\phi(z_1,\bz_1)}\;
\big\rangle\!\big\rangle_{C_q}\,=\,\int d\mu(p)\;|\CF_{\al,C_q}^\si(p)|^2\,.
\end{equation}
The conformal blocks $\CF_{\al,C_q}^\si({q})$ are objects that are defined
from the representation theory of the Virasoro
algebra, as will be recalled in the following
two subsections.


\subsection{Virasoro conformal blocks}\label{Vircfbl}

\renewcommand{\fw}{w}
\renewcommand{\fv}{v}
\renewcommand{\fe}{e}

\subsubsection{Definition of the conformal blocks}

\renewcommand{\vir}{{\rm Vir}_c}
\newcommand{\CFB}{{\SC\SB}}

Let ${\rm Vir}_c$ be the Virasoro algebra with generators $L_n$,  $n\in\BZ$,
and relations
\begin{equation}\label{Vir}
[\,L_n\,,L_m\,]\,=\,(n-m)L_{n+m}+\frac{c}{12}n(n^2-1)\de_{n+m,0}\,,
\end{equation}
For given set of highest weight representations $\CV_r$, $r=1,\dots,n$
of $\vir$ with highest weights $\De_r$, and Riemann surface $C$ with $n$ marked points
at positions $z_1,\dots,z_n$
one defines the conformal blocks as linear
functionals $\CF_C:\CV_{[n]}\equiv\otimes_{r=1}^n\CV_r\ra\BC$ that
satisfy the invariance property
\begin{equation}\label{cfblvir}
\CF_C(T[\chi]\cdot v)\,=\,0\qquad \forall v\in\CR_{[n]},\;\;\forall\chi\in
\FV_{\rm out}^{}
\end{equation}
where $\FV_{\rm out}^{}$ is the Lie algebra of meromorphic differential operators on $C$ with poles only at $z_1,\dots,z_n$.
The action of $T[\chi]$ on
$\otimes_{r=1}^n\CR_r\ra\BC$ is defined
by taking the Laurent expansions of $\chi$ at the points $z_1,\dots z_n$,
w.r.t. local coordinates $t_r$ which vanish at $z_r$,
\begin{equation}
\chi(t_r)\,=\,\sum_{k\in\BZ} \chi_k^{(r)}\,t_r^{k+1} \,\pa_{t_r}^{}\;\;\in\;\;
\BC(\!(t_r)\!)\pa_{t_r}\,,
\end{equation}
to which we may associate the operator
\begin{equation}\label{Tdef}
T[\chi]\,=\,\sum_{r=1}^n {\rm id}\ot\dots\ot\underset{(\rm r-th)}{L[\chi^{(r)}]}\ot\dots\ot{\rm id}\,,\qquad
L[\chi^{(r)}]\,:=\,\sum_{k\in\BZ} L_k \,\chi_k^{(r)}\;\;\in\;\;\vir\,.
\end{equation}
It can be shown that the central extension vanishes
on the image of the Lie algebra $\FV_{\rm out}^{}$ in
$\bigoplus_{r=1}^n \vir$, making the definition consistent.
The defining invariance condition \rf{cfblvir} has generically many solutions.
We will denote the vector space of conformal blocks associated to the Riemann surface
$C$ with representations $\CV_r$ associated to the marked points $z_r$, $r=1,\dots,n$ by $\CFB(\CV_{[n]},C)$.

Physicists may be more familiar with conformal blocks as
expectation values of chiral vertex operators
associated to the representations $\CV_r$. State-operator correspondence
associates a chiral vertex operator $\Phi(\fv_r|z_r)$ to each
vector $\fv_r$ in $\CV_r$. The chiral vertex operators
associated to highest weight vectors $\fe_r$ in $\CV_r$ are
called primary fields, all other chiral vertex operators $\Phi(\fv_r|z_r)$
descendants. The functionals $\CF_C$ represent the expectation values
of a product of chiral vertex operators as
\begin{equation}\label{cfblnotation}
\CF_C\big(\fv_n\otimes\dots\otimes\fv_1)\,=\,\bigg\langle\,\prod_{r=1}^n
\Phi(\fv_r|z_r)\,\bigg\rangle_{\BG}\,.
\end{equation}
The subscript $\BG$ indicates the parameters
for the different
ways to compose the chiral vertex operators, as will
be made more explicit below.
The defining invariance property \rf{cfblvir} is a consequence of the
Virasoro Ward identities that $\big\langle\,\prod_{r=1}^n
\Phi(\fv_r|z_r)\,\big\rangle_\BG$ is required to satisfy. We shall often
use the notation on the right hand side of \rf{cfblnotation}
as it may be more appealing to the intuition of physicists.

\subsubsection{Propagation of vacua}

The vacuum representation $\CV_0$ which corresponds to $\De_r=0$
plays a distinguished role.
If $\Phi_0(\fv_0|z_0)$ is the
vertex operator associated to the vacuum representation,
we have
\begin{equation}\label{psi0}
\Phi_0(\fe_0|z_0)\,=\,{\rm id}\,,\qquad
\Phi_0(L_{-2}\fe_0|z_0)\,=\,T(z_0)\,,
\end{equation}
where $T(z)$ is the energy-momentum tensor.
It can be shown that
the spaces of conformal blocks with and without
insertions of the vacuum representation are canonically isomorphic.
The isomorphism between $\CFB(\CV_0\ot\CV_{[n]},C_{g,n+1})$ and $\CFB(\CV_{[n]},C_{g,n})$
is simply given by evaluation at the vacuum vector $\fe_0\in\CV_0$
\begin{equation}
\CF'_{C_{g,n+1}}\big(\fe_0\ot\fv_n\otimes\dots\otimes\fv_1)\,\equiv\,
\CF_{C_{g,n}}\big(\fv_n\otimes\dots\otimes\fv_1)\,,
\end{equation}
as is also obvious from \rf{psi0}.
This fact is often referred to as the "propagation of vacua".

\subsubsection{Deformations of the complex structure of $X$}

A key point that needs to be understood about spaces of conformal
blocks is the dependence on the complex structure of $C$.
There is a canonical way to represent infinitesimal variations
of the complex structure on the spaces of conformal
blocks.
By combining the definition of
conformal blocks with
the so-called ``Virasoro uniformization''
of the moduli space ${\CM}_{g,n}$ of complex structures on $C=C_{g,n}$
one may construct
a representation of
infinitesimal motions on ${\CM}_{g,n}$ on
the space of conformal blocks.

The ``Virasoro uniformization''
of the moduli space ${\CM}_{g,n}$  may be formulated as
the statement that the tangent space $T{\CM}_{g,n}$ to
${\CM}_{g,n}$ at $C$ can be identified with the
double quotient
\begin{equation}\label{VirUni}
T{\CM}_{g,n}\;=\;\Ga\big(C\setminus\{x_1,\dots,x_n\},\Theta_C\big) \;
\bigg\backslash \;
\bigoplus_{k=1}^n \BC(\!(t_k)\!)\pa_k \,
\bigg/\, \bigoplus_{k=1}^n \BC[[t_k]]\pa_k\,,
\end{equation}
where $\Ga(C\setminus\{x_1,\dots,x_n\},\Theta_C)$ is the set of vector
fields that are holomorphic on $C\setminus\{x_1,\dots,x_n\}$,
while $\BC(\!(t_k)\!)$ and $\BC[[t_k]]$
are formal Laurent and Taylor series
respectively.

Let us then consider $\CF_C(T[\eta]\cdot v)$ with
$T[\eta]$ being defined in \rf{Tdef} in the case that $\eta$ is an arbitrary element of
$\bigoplus_{k=1}^n \BC(\!(t_k)\!)\pa_k$ and $L_r v_k=0$ for
all $r>0$ and $k=1,\dots,n$.
The defining invariance property \rf{cfblvir} together
with $L_r v_k=0$ allow us to define
\begin{equation}\label{Viract}
\de_\vartheta \CF_C(v)\,=\,\CF_C(T[\eta_\vartheta^{}]\cdot v)\,,
\end{equation}
where $\de_{\vartheta}$ is the derivative corresponding to
a tangent
vector $\vartheta\in T{\CM}_{g,n}$ and $\eta_\vartheta^{}$
is any element of $\bigoplus_{k=1}^n \BC(\!(t_k)\!)\pa_k$
which represents $\vartheta$ via \rf{VirUni}.
Generalizing these observations one is led to the
conclusion that derivatives w.r.t. to the moduli parameters
of ${\CM}_{g,n}$ are (projectively) represented on the
space of conformal blocks, the central extension coming from the
central extension of the Virasoro algebra \rf{Vir}.

In the case of $g=0$, and $v_r$ being equal to the highest weight
vector $e_r$ of $\CV_r$ for $r=1,\dots,n$,
formula \rf{Viract} is closely related to  the familiar formula
\begin{equation}\label{cfWard}
\big\bra \,T(x)\, \Phi_{n}(z_n)\dots \Phi_{1}(z_1)\,\big\ket=
\;\sum_{i=1}^{n}\left(\frac{\De_{\al_i}}{(x-z_i)^2}+
\frac{1}{x-z_i}
\frac{\pa}{\pa z_i}\right)
\big\bra\,\Phi_{n}(z_n)\dots \Phi_{1}(z_1)\,\big\ket,
\end{equation}
where we have abbreviated the primary fields $\Phi(e_r|z_r)$ as
$\Phi_r(z_r)$.

\subsubsection{Conformal blocks vs. $\CD$-modules}

It may be worth noting the two possible ways to read \rf{Viract}.
Having defined the action of the Virasoro algebra on $\CV_{[n]}$,
\rf{Viract} tells us how the ring of holomorphic
differential operators on $\CM_{g,n}$
acts on the spaces of the conformal blocks. This makes the
spaces of conformal blocks a (twisted) $\CD$-module over $\CM_{g,n}$.

On the other hand,
given {\it any} holomorphic function $\CF$ defined in an
open subset $\CU\subset\CM_{g,n}$
one may use \rf{Viract} recursively in order to construct the values
of $\CF(v)$ on arbitrary vectors $v\in\CV_{[n]}$.
The Virasoro uniformization \rf{VirUni}
of $T\CM_{g,n}$ describes the local structure of $\CM_{g,n}$
in terms of the Lie algebra $\BC(\!(t)\!)\pa_t$ of
infinitesimal diffeomophisms of the circle,
and \rf{Viract} can be read as a
description of the space of local holomorphic sections of a projective
line-bundle over $\CM_{g,n}$ in terms of the representation theory
of the central extension
of $\BC(\!(t_k)\!)\pa_k$.


\subsection{Gluing construction of conformal blocks}\label{glue}

\subsubsection{Gluing two boundary components}

Let $C$ be a (possibly disconnected)
Riemann surface with marked points and choices of coordinates around the marked points.
We can construct a new Riemann surface $C'$
by picking two marked points $z_0$ and $z_0'$ with non-intersecting annuli $A$ and $A'$ embedded in
coordinate neighborhoods around the two points, choosing a bi-holomorphic mapping $I:A\ra A'$, and  by identifying the points
that are mapped to each other under $I$, see e.g. \cite{T09} for more details.

Let us in particular
consider a Riemann surface $C_{21}$ that was obtained by gluing two surfaces $C_2$ and $C_1$
with $n_2+1$ and $n_1+1$ boundary components, respectively.
Given an integer $n$, let sets $I_1$ and $I_2$ be such that
$I_1\cup I_2=\{1,\dots,n\}$. Let us consider
conformal blocks $\CF_{C_i}^{}\in\CFB(\CV^{[n_i]}_i,C_i)$
where  $\CV^{[n_2]}_2=(\otimes_{r\in I_2}\CV_r)\otimes\CV_0$
and $\CV^{[n_1]}_1=\CV_0\otimes(\otimes_{r\in I_1}\CV_r)$ with
the same representation $\CV_0$
assigned to $z_{0,1}$ and $z_{0,2}$, respectively.
Let
$\langle.,.\rangle_{\CV_0}$ be the invariant bilinear form on
$\CV_0$.
For given $v_2\in\otimes_{r\in I_2}\CV_r$ let
$W_{\fv_2}$ be the linear form on $\CV_0$ defined
by
\begin{equation}
W_{\fv_2}(\fw):=\CF_{C_2}^{}(\fv_2\ot\fw)\,,\qquad\forall
\fw\in\CV_0\,,
\end{equation} and let $\SC_\1(q)$ be the family of linear operators
$\CV_{1}^{[n_1]}\ra \CV_0$ defined as
\begin{equation}\label{Vprefactor}
\SC_\1(q)\cdot v_1\,:=\,
\sum_{e\in B(\CV_0)}
\,q^{L_0}{e}\;\CF_{C_1}^{}(\check{e}\ot v_1),
\end{equation}
where we have used the notation
$B(\CV_0)$ for a basis of the representation $\CV_0$ and $\check{e}$ for the dual
of an element $e$ of $B(\CV_0)$ defined by
$\langle \check{e},e'\rangle_{\CV_0}^{}
=\de_{e,e'}$.
We may then consider the expression
\begin{equation}\label{Vfactor}
\CF_{C_{21}}(\fv_2\ot\fv_1)\,{:=}\,
W_{\fv_2}(\SC_\1(q)\cdot\fv_1)\,.
\end{equation}
We have thereby defined a new conformal block associated to the
glued surface $C_{21}$, see \cite{T09} for more discussion.
The insertion of the
operator $q^{L_0}$ plays the role of
a regularization. It is not a priori clear that the linear form $W_{\fv_2}$
is defined on
infinite linear combinations such as $\SC_1(q)\cdot\fv_1$.
Assuming $|q|<1$, the factor $q^{L_0}$ will produce an
suppression of the contributions with large $L_0$-eigenvalue, which
renders the infinite series produced by the definitions \rf{Vfactor}
and \rf{Vprefactor}
convergent.

\subsubsection{Gluing from pairs of pants}\label{gluingpants}

One can produce any Riemann surface $C$ by gluing pairs of pants.
The different ways to obtain $C$ in this way are labeled by
cut systems $\CC$, a collection of mutually non-intersecting
simple closed curves on $C$.
Using the gluing construction recursively leads to the definition of a
family of conformal blocks denoted
\begin{equation}\label{confblnot}
\CF_{\be,C_q}^{\si}(p)
\equiv\, \big\langle\, e^{2\al_n\phi(z_r)}\dots e^{2\al_1(z_1)}\,\big\rangle_{C_q,\mathbb G}
\end{equation}
depending on the following set of data:
\begin{itemize}
\item $\si$ is a marking: A pants decomposition defined by
a cut system $\CC_\si$
together with three-valent graphs on the pairs of pants
glued together to form a
connected graph $\Gamma_\si$ on $C$.
\item $q$ is an assignments $q:\ga\mapsto q_\ga\in \BU$, defined
for all curves $\ga\in\CC_\si$. $q_\ga$ are the gluing parameters
$q_\ga$ entering the gluing construction
from three-punctured spheres. They parameterize the
complex structure of the family $C_q$ of Riemann surfaces obtained
in the gluing construction.
\item $p$ is an assignment $p:\ga\mapsto p_\ga\in \BR$, defined
for all curves $\ga\in\CC_\si$.
The parameters $p_\ga$ determine the
Virasoro representations $\CV_{\Delta_\ga}$ to be used in the gluing
construction of the conformal blocks
from pairs of pants via
\begin{equation}\label{Dedef}
\De_\ga\,=\,\frac{Q^2}{4}+\frac{p^2_\ga}{\hbar^2}\,.
\end{equation}
\item $\be=(\be_1,\dots,\be_n)$ is taken to parameterize the
external representations $\CV_{1},\dots,\CV_{n}$ via
\begin{equation}
\al_r\,=\,\frac{\be_r}{\hbar}.
\end{equation}
\end{itemize}
The pair of data $(\si,p)$ is condensed into the
``gluing data'' $\mathbb G$ in \rf{confblnot}.
While cut systems can be used to label boundary components in $\pa\CM_{g,n}$, one may parameterize
boundary components $\pa_\si\CT_{g,n}$ of the Teichm\"uller space $\CT_{g,n}$ with the help of markings $\si$.
Using the markings allows one to properly take care of the multi-valuedness of the conformal
blocks on $\CM_{g,n}$ \cite{T09}.

The conformal blocks $\CF_{\be,C_q}^{\si}(p)$ are entire
analytic with respect to the variables
$\be_r$, meromorphic in the variables $p_\ga$, $\ga\in\CC_\si$ with poles at the zeros of the Kac determinant, and the
dependence on the gluing parameters $q$
can be analytically continued over $\CT_{g,n}$ \cite{T01,T09}.
When the dependence on $\be$ is not important we will
abbreviate $\CF^\si_q(p):=\CF_{\be,C_q}^\si(p)$.

\subsubsection{Change of pants decomposition}

It turns out that the conformal
blocks $\CF_{q_1}^{\si_1}(p)$ constructed by the gluing construction
in a neighborhood of the asymptotic region of $\CT(C)$ that is determined
by $\si_1$ have an analytic continuation
$(\SA_{\si_1}^{\si_2}\CF)_{q_2}^{\si_1}(p)$ to the asymptotic region of
$\CT(C)$ determined by a second marking $\si_2$.
A fact \cite{T01,T03a,T09}\footnote{A full proof of the statements made here does not
appear in the literature yet. It can, however, be assembled from
building blocks that are published. By using the
groupoid of changes of the markings it is sufficient to
verify the claim for the cases $g=0,n=4$ and $g=1,n=1$, respectively.
For $g=0,n=4$ this was done in \cite{T01}, see also \cite{T}. The
case of $g=1,n=1$ was recently reduced to the case $g=0,n=4$ in
\cite{HJS}.}
 of foundational importance for the subject is that
the analytically continued conformal blocks
$(\SA_{\si_1}^{\si_2}\CF)_{q_2}^{\si_1}(p)$
can be represented as a linear combination
of the conformal blocks $\CF_{q_2}^{\si_2}(p)$ which takes
the form
\begin{equation}\label{fustrsf}
(\SA_{\si_1}^{\si_2}\CF)_{q_2}^{\si_1}(p)\,=\,
\int d\mu(p')\;V_{\si_2\si_1}(p|p')\,
\CF_{q_2}^{\si_2}(p')\,.
\end{equation}
The changes from one pants decomposition to another
generate the modular groupoid (\cite{MS},
see also \cite{T09} for nonrational cases). Having a representation of the modular
groupoid via \rf{fustrsf} makes the space of conformal blocks a
representation of the mapping class group via
\begin{equation}\label{modtrsf}
(\SA^{m.\si}_{\si}\CF)_{q}^{\si}(p)\,=\,\int
d\mu(p')\;V_{m.\si,\si}(p,p')\,\CF_{q}^{\si}(p')\,,
\end{equation}
where $m.\si$ is the image of the marking $\si$ under $m\in{\rm MCG}(C)$.

To each marking $\si$
one may associate a Hilbert space $\CH_{\si}\simeq L^2((\BR^+)^{3g-3+n},d\mu)$
of complex valued functions $\psi_\si(p)$ on the space of assignments
$p:\ga\mapsto p_\ga\in \BR$, $\ga\in\CC_\si$ that are square-integrable
w.r.t. $\mu$. The scalar product is defined by means of the same measure $\mu$
that appears in the holomorphic factorization of the full correlation functions
\rf{holofact},
\begin{equation}
\lVert\,\psi\,\rVert^2\,=\,\int d\mu(p)\,|\psi_{\si}(p)|^2\,.
\end{equation}
The integral operators defined in
\rf{fustrsf} and \rf{modtrsf} are unitary w.r.t. this scalar product,
which is equivalent to crossing symmetry and modular invariance of the
physical correlation functions constructed from the
conformal blocks as in \rf{holofact} \cite{T01,T09}.

\subsection{Degenerate fields as probes}\label{probes}

\subsubsection{Insertion of degenerate fields}

An interesting way to probe the
conformal blocks \cite{AGGTV,DGOT}
is to consider insertions of degenerate fields like
\begin{equation}\label{corr+}
\big\langle \;\CO_{n,l}\;\big\rangle_{\hat\BG}\,\equiv\,
\big\langle \;e^{2\al_n\phi(z_n)}\cdots e^{2\al_1\phi(z_1)}\;
e^{-\frac{1}{b}\phi(y_l)}\cdots e^{-\frac{1}{b}\phi(y_1)}
\;\big\rangle_{\hat\BG}
\end{equation}
The conformal blocks satisfy the null vector decoupling equations
\begin{equation}\label{BPZ3'}
\CD_{y_k}^{\rm BPZ}\cdot \big\langle \;\CO_{n,l}\;\big\rangle\,=\,0\,,\qquad \forall \;k=1,\dots,l\,,
\end{equation}
with differential operators $\CD_{y_k}^{\rm BPZ}$ being for $g=0$ given as
\begin{align}
\CD_{y_k}^{\rm BPZ}\!= b^{2}\frac{\pa^2}{\pa y_k^2}
 +\sum_{r=1}^n
\left(\frac{\De_r}{(y_k-z_r)^2}+\frac{1}{y_k-z_r}\frac{\pa}{\pa z_r}\right)
-\sum_{\substack{k'=1\\k'\neq k}}^l
\left(\frac{3b^{-2}+2}{4(y_k-y_{k'})^2}-\frac{1}{y_k-y_{k'}}
\frac{\pa}{\pa y_{k'}}\right).
\nonumber\end{align}
Let us abbreviate
the notation for the
space of conformal blocks on $C_{g,n}$ to $\CFB(C_{g,n})$
and let  $\CFB'(C_{g,n+l})$ be
the space of conformal blocks on $C_{g,n+l}$ with $l$ vertex operators
$e^{-\frac{1}{b}\phi}$ assigned to the extra punctures $y_1,\dots,y_l$,
respectively.
It follows from \rf{BPZ3'} that  the three point conformal blocks
$\langle e^{-\frac{1}{b}\phi(z_3)}e^{2\al_2\phi(z_2)}e^{2\al_1\phi(z_1)}\rangle$ can only
be nonzero if $\De_{\al_2}=\De_{\al_1\mp 1/2b}$, which is
symbolically expressed in the fusion rules
\beq
[e^{-\frac{1}{b}\phi}]\, [e^{2\al\phi}]\,\sim\, 
[e^{(2\al-1/b)\phi}]+[e^{(2\al+1/b)\phi}]\,.
\label{opedegen}
\eeq
This implies that
$\CFB'(C_{g,n+l})$ is isomorphic to
$\CFB(C_{g,n})\otimes (\BC^2)^{\ot l}$
as a vector space.

\subsubsection{Quantum loop operators}

The key observation to be made is that for $l=2$
there is a canonical embedding
\begin{equation}\label{embed}
\imath_{g,n}:\CFB(C_{g,n})\hookrightarrow \CFB'(C_{g,n+2})\,,
\end{equation}
coming from the fact that the fusion of the two degenerate fields
$V_{-1/2b}$ contains the vacuum representation, and that insertions of the vacuum representation
do no alter the space of conformal blocks (propagation of vacua).
It follows from the existence of the embedding \rf{embed}
that the mapping class group action
on $\CFB'(C_{g,n+2})$ can be projected onto $\CFB(C_{g,n})$.
The mapping class group ${\rm MCG}(C_{g,n+2})$ contains in particular
the monodromies generated by moving the insertion point of one of the vertex operators $e^{-\frac{1}{b}\phi}$
along a closed curve $\ga$ on $C_{g,n}$. The projection of the
action of these elements on  $\CFB'(C_{g,n+2})$ down to
$\CFB(C_{g,n})$ defines operators on
$\CFB(C_{g,n})$. Let us denote the operator associated to a generator $\ga$ of
the fundamental group $\pi_1(C_{g,n})$ by $\SL_\ga$. We will call
$\SL_\ga$ a quantum loop operator.

The conformal blocks $\CF_{q}^\si(p)$ defined above
generate a basis for $\CFB(C_{g,n})$.
This basis is such the operators $\SL_{\ga}$
associated to the curves $\ga\in\CC_\si$ in the cut system
corresponding to $\si$ are represented diagonally,
\begin{equation}\label{Hecke}
\SL_{\ga}\cdot \CF_{q}^\si(p)\,=\,
2\cosh(2\pi p_\ga/\ep_1)\,\CF_{q}^\si(p)\,.
\end{equation}
This means that the operators $\SL_{\ga}$ can be used to
``measure'' the intermediate representation that has been used in the
construction of conformal blocks by summing over complete sets of
vectors from given representations. The parametrization
in terms of the data $\si$ and $p$ is therefore equivalent to
a parametrization in terms of the eigenvalues of the
quantum loop operators $\SL_\ga$, $\ga\in\CC_\si$.

\subsection{Parameterizing conformal blocks with degenerate fields}
\label{liouconf++}

In order to get a parametrization for the space
of solutions to \rf{BPZ3'},
we shall consider representations for the Riemann surface $C_{g,n+l}$ which are
obtained as follows. Let us call a marked point special if it will be the insertion point of
a degenerate field, non-special otherwise. We may the consider representations for $C_{g,n+l}$
obtained by gluing surfaces $T_\nu$,
$\nu=1,\dots,2g-2+n$, of genus zero with
$l_\nu$ special marked points and exactly three non-special ones.
For each surface $T_\nu$ we may then pick a pants decomposition which is such that each pair
of pants contains at most one special marked point.
We may therefore view the markings $\hat\si$ on $C_{g,n+l}$ that
have pants decomposition of this type as certain refinements of a marking $\si$
on the surface $C_{g,n}$ obtained from $C_{g,n+l}$ by "forgetting" the
insertion points of the degenerate fields.
We will in the following restrict attention to markings of this type.

Conformal blocks can then be defined by the gluing construction. This
defines solutions to
\rf{BPZ3'} denoted as
\begin{equation}\label{confbl++}
\CF_{q,y}^{\hat\si}(p,\de):=
\big\langle \;e^{2\al_n\phi(z_n)}\cdots e^{2\al_1\phi(z_1)}\;
e^{-\frac{1}{b}\phi(y_{l})}\cdots e^{-\frac{1}{b}\phi(y_1)}
\;\big\rangle_{C_{q,y},{\hat{\mathbb G}}\,.}
\end{equation}
These conformal blocks are
parameterized by the data $p$ and $q$ associated to the underlying marking
$\si$ on $C_{g,n}$ in the same way as explained in
Subsection \ref{gluingpants}, together with the following additional data
\begin{itemize}
\item $\de$ is a map which assigns a sign $\de_k$ to each of the
special marked points $y_k$, which determines the change of representation label
according to the fusion rules
\rf{opedegen}. Noting that $p$ determines the choice of representations associated
to the non-special marked points of $T_\nu$ it is easy to see that this allows one to determine
all representations involved in the gluing construction unambigously.
\item $y$ is the collection of gluing parameters involved in the gluing construction
of $T_\nu$ from three-punctured spheres.
\end{itemize}
In the notation on the left hand side of \rf{confbl++} we have displayed
the  gluing data $\hat{\mathbb G}=(\hat\si,p,\de)$ more explicitly.

The conformal blocks \rf{confbl++} form
a {\it complete} set of solutions to the equations \rf{BPZ3'} in the sense
that the solutions associated to a given marking $\hat\si_1$
can be analytically continued to the boundary component
$\pa_{\hat\si_2}\CT_{g,n+l}$ of
the Teichm\"uller space $\CT_{g,n+l}$ which is associated to any
other marking $\hat\si_2$, and that the analytically
continued solutions associated to $\hat\si_1$
can be represented as a linear combination of the solutions
representable as power series in gluing parameters in a
neighborhood of $\pa_{\hat\si_2}\CT_{g,n+l}$.

\subsection{Quantum Hitchin system from the
semiclassical limit of Liouville theory}\label{LclassI}

\subsubsection{Eigenfunctions of Hitchin's Hamiltonians from 
classical conformal blocks}

Let us now consider the limit $\ep_2\ra 0$ of
the conformal blocks \rf{corr+}, keeping $\ep_1$ finite
in the case $g=0$.
This means that $\hbar\ra 0$ while $b\ra\infty$.
The sum over $k'$
in the expression for $\CD_y^{\rm BPZ}$ becomes subleading in this limit.
To leading
order we can factorize the solutions $\langle \,\CO_{n,l}\,\rangle_{\hat\BG}$
to  \rf{BPZ3'} in the
form
\begin{equation}\label{limfactor}
\big\langle \;\CO_{n,l}\;\big\rangle_{\hat\BG}\,=\,
\exp(-b^2\CW({q}))\prod_{k=1}^l
\chi_k(y_k;{q})\,,
\end{equation}
where $\chi_k(y)\equiv\chi_k(y;q)$  are solutions to the equation
\begin{equation}
(\pa_y^2+t(y))\chi_k(y)=0\,,\qquad
t(y)\,=\,\sum_{r=1}^{n}
\left(\frac{\de_r}{(y-z_r)^2}+\frac{H_r}{y-z_r}\right)\,,
\end{equation}
with $\de_r=\lim_{b\ra\infty} b^{-2}\De_r$, and
\begin{equation}\label{HfromCW}
H_r\,=\,-\frac{\pa}{\pa z_r}\CW(q)\,.
\end{equation}
In \rf{limfactor} and \rf{HfromCW}, we are using the notation $q$ for the
collection of variables $(z_1,\dots,z_n)$ which determine the complex 
structure of the underlying Riemann surface $C_{0,n}=\BP^1\setminus
\{z_1,\dots,z_n\}$.

On the right hand side of \rf{limfactor} we may for $l=n-3$ and 
$z_n\ra\infty$
recognize an eigenfunction of the Gaudin Hamiltonians as
explained in Subsection \ref{qSOV}. It does not satisfy the quantization 
conditions as discussed in Subsection \ref{singleval}, in general.

Considering the limit $b\ra\infty$ of the basis elements $\CF_{q}^{\si}(p)$
we are led to the conclusion that the following limit exists,
\begin{equation}\label{WfromF}
\CW_{q}^\si(l)\,\equiv\,\lim_{b\ra\infty}b^{-2}\log\CF_{q}^{\si}(p)
\end{equation}
where the parameters $l=(l_1,\dots,l_{n-3})$ and $p=(p_1,\dots,p_{n-3})$
are related via
\begin{equation}\label{parid}
{l}_r\,=\,\frac{p_r}{4\pi \ep_1}\,.
\end{equation}
and that the monodromy group of the oper $(\pa_y^2+t(y))\chi(y)=0$,
\begin{equation}\label{H-Cfbl-rel}
t(y)\,=\,\sum_{r=1}^{n}
\left(\frac{\de_r}{(y-z_r)^2}+\frac{H_r(l,q)}{y-z_r}\right)\,,\qquad
H_r(l,q)\,=\,-\frac{\pa}{\pa z_r}\CW_{q}^\si(l)\,,
\end{equation}
satisfies \rf{monofix}.
Let us note in particular that the parameterization of the conformal
blocks in terms of eigenvalues of quantum loop operators introduced
in Subsection \ref{probes} turns into the parameterization of the
opers in terms of the traces of their monodromies introduced
in \rf{monofix}.

\subsubsection{Semiclassical limit of the full correlation functions}

Let us now consider the
classical limit $b\ra\infty$ of full correlation functions \rf{LcorrI}.
We may assume that the measure $d\mu(p)$ which appears in the holomorphically 
factorized representation \rf{holofact} is just the usual Lebesque measure,
$d\mu(p)=\prod_{r=1}^{n-3}dp_r$. This is related to the more conventional 
representation in which $d\mu(p)$ is constructed from the product of 
three-point functions by a change of normalization for the conformal 
blocks, see e.g. \cite{AGT} for explicit formulae. 
The leading behavior of the integrand in \rf{holofact}
is $e^{-2b^2{\rm Re}(\CW_q^{\si}(l))}$, as follows from
\rf{WfromF}.
The integral in the holomorphically 
factorized representation \rf{holofact}
of the full correlation functions
will therefore be dominated by a saddle point
$p_{\rm s}=(p_{1,\rm s}
\dots p_{n-3,\rm s})$,
\begin{equation}\label{corrlim0}
\big\langle \;\CO\;\big\rangle\,\sim\,e^{-b^2S_{\rm L}(q)}\,,\qquad
S_{\rm L}(q)\,=\,2{\rm Re}(\CW_q^\si(l_s))\,,
\end{equation}
with $l$ and $p$ related via \rf{parid}, and the value $l_s=l_s(q,\bq)$ 
at the saddle point is determined by
\begin{equation}\label{saddlecond}
\frac{\pa}{\pa l_r}\,{\rm Re}(\CW_q^{\si}(l))\Big|_{l=l_{\rm s}}\,=\,0\,.
\end{equation}
More explicit analysis of the case $n=4$ can be found in \cite{ZZ}.

\subsubsection{Single-valued Gaudin eigenfunctions from 
Liouville correlation functions}

Let us now consider the semiclassical limit of full correlation functions
containing $n-3$ insertions of degenerate fields $e^{-b\phi(y_k,\by_k)}$.
By the same arguments as used before we find that 
\begin{equation}\label{corrlim}
\bigg\langle\!\!\bigg\langle
\;\prod_{r=1}^n e^{2\al_n\phi(z_r,\bz_r)}\,
\prod_{k=1}^{n-3} e^{-b\phi(y_k,\by_k)}\;
\bigg\rangle\!\!\bigg\rangle\sim e^{-b^{2}S_{\rm L}(q)}
\prod_{k=1}^{n-3}\chi_r(y_k,\bar{y}_k)\,,
\end{equation}
where $S_{\rm L}(q)$ was introduced in \rf{corrlim0}.
On the right hand side of \rf{corrlim} we recognize \cite{RT}
the solutions \rf{factorrep} to the eigenvalue equations 
for the Gaudin model in the Separation of Variables representation.
They are automatically single-valued both with respect to 
the variables $y_k$ and $q$ as the correlation
function on the left hand side of \rf{corrlim} has this property.
We see that the distinguished state $|q\rangle$ of the Gaudin model
introduced in Subsection \ref{singleval}
is reproduced in the semiclassical limit of a 
Liouville correlation function.

\subsubsection{Yang's potential from classical conformal blocks?}

Recall that the space of all differential operators of the form $\pa_y^2+t(y)$
parameterizes via the quantum separation of variables the
commutative algebra of differential operators on ${\rm Bun}_G$
generated by $\SH_r-E_r$. This space can be viewed
as a ``deformation'' $\CB_{\ep}$ of the base $\CB$ of the Hitchin fibration.
Within $\CB_{\ep}$ we want to identify isolated
points representing the quantized eigenvalues with the help of 
a function $\CW$ on $\CB_{\ep}$ called Yang's potential.
We are now going to point out that our discussion of the relation
between the semiclassical limit of the Liouville correlation functions
and the complex quantization of the Hitchin system above suggests
that the classical Liouville conformal blocks are natural candidates for 
the Yang's potential associated to the complex quantization of the
Hitchin system as discussed in Section \ref{sec:quantHitchin}.

For the case $g=0$, $C_{0,n}=\BP^1\setminus\{z_1,\dots,z_n\}$ under 
consideration, let ${l}=({l}_1,\dots,{l}_{n-3})$ be the coordinates 
for the space of opers introduced in \rf{monofix} above.
Our discussion of the semiclassical
limit of the complex quantization of the Hitchin system 
in \ref{Semiquant} implies that 
to leading order in $\ep$ we may identify the
Yang's function $\CW(l)$ 
with the prepotential $\CF(a)$, where $l$ and $a$ are 
related via \rf{lfroma}. The standard relation 
$a_s^{\sst D}=\pa_{a_s}\CF(a)$ then allows us 
to reformulate the Bohr-Sommerfeld quantization conditions \rf{BS}
to leading order in $\ep$ in terms of $\CW(l)$. This suggests that
the exact quantization conditions could likewise 
be formulated in terms of a
potential $\CW(l)$, and that they
should include the conditions
\begin{equation}\label{Qcond2}
\frac{\pa}{\pa {l}_r}{\rm Re}(\CW({l}))
\Big|_{{l}={l}({\mathbf k})}\,=\,\pi{\ep} k_r\,,
\quad{\rm for}\quad r=1,\dots,n-3\,,
\end{equation}
where ${\mathbf k}=(k_1,\dots,k_{n-3})$ is a given
vector of integers.

In our discussion above we had observed that the quantization condition 
in the case of the distinguished state $|q\rangle$ can be formulated as 
the saddle-point condition \rf{saddlecond}. This invites us to 
identify
\begin{equation}\label{Yang-cfbl}
\CW({l})\,\equiv\,\CW_{q}^\si(p),
\end{equation}
with ${l}$ and $p$ related by \rf{parid}. The saddle-point condition 
\rf{saddlecond} would then correspond to the special case 
${\mathbf k}=(0,\dots,0)$
of \rf{Qcond2}. It remains to be seen if other single-valued 
eigenstates of the Gaudin-Hamiltonians can be characterized
in terms of  
the conditions \rf{Qcond2}.  

\subsubsection{Characterization of Yang's potential in terms of
opers}

At the end of Subsection \ref{opers}
we had defined $H_r({l},q)$ as the accessory
parameters which give the oper a monodromy
characterized by the parameters ${l}$. 
It follows from \rf{Yang-cfbl} and \rf{H-Cfbl-rel} that 
$\CW(l)\equiv\CW({l},q)$ satisfies the equations
\begin{equation}\label{H-Wrel}
H_r({l},q)\,=\,-\frac{\pa}{\pa z_r}\CW({l},q)\,.
\end{equation}
The equations 
\rf{H-Wrel} define $\CW({l},q)$ up to addition of $q$-independent 
functions of $l$.\footnote{This corrects
an inaccurate statement in a previous version of this 
paper that has been pointed out by S. Shatashvili.}

The formulation of the quantization conditions in terms of the
Yang's potential via \rf{Qcond2} will only work for a suitable
choice of the $l$-dependence in $\CW({l},q)$. Such a choice is implied in
the identification \rf{Yang-cfbl} with the classical conformal blocks.
The freedom to add $q$-independent 
functions of the variables $l$ is via \rf{Yang-cfbl}
related to the freedom to multiply
the conformal blocks $\CF_{q}^{\si}(p)$ by functions of the parameters
$p$. The latter freedom is fixed if one requires, 
as has been done above, that the {\it single-valued}
correlation functions \rf{LcorrI} are constructed from the conformal blocks
by an expression of the form
\rf{holofact} with measure $d\mu(p)$ being the standard Lebesque
measure. This amounts to absorbing the three-point functions
into the conformal blocks. We see that the correct choice of 
the $q$-independent functions of the variables $l$ in the 
definition of $\CW({l},q)$ is ultimately determined by the
single-valuedness of the Liouville correlation functions
\rf{LcorrI} which determines the measure $d\mu(p)$, as discussed
e.g. in \cite{T09}. Explicit formulae can easily be found with the help of
\cite{ZZ}.
This single-valuedness is directly
related to the single-valuedness of the eigenfunctions of the 
Gaudin-Hamiltonians via \rf{corrlim}. 

The two 
different formulations of the quantization conditions -- from single-valuedness
of the wave-functions on the one hand, and in terms of $\CW({l},q)$ on the 
other hand --
are unified in the condition of single-valuedness of the 
Liouville correlation functions appearing on the left hand side of 
\rf{corrlim} above. These relations fit into a Langlands-duality 
scheme similar to our diagram \rf{dmaindiag1} 
above, in which the single-valued 
Gaudin eigenvectors would appear in the upper right box, and the points on 
$\CB_\ep$ determined from $\CW$ should be placed into the upper left box.

\subsubsection{Quantization conditions in real quantization ?}

In Subsection \ref{Semiquant} we had also considered the quantization of 
a real slice in the phase space in 
the semiclassical limit. It is suggestive to observe that both in the
real and complex quantization schemes discuseed in Subsection 
\ref{Semiquant} it is the {\it same} function (the prepotential) which
appears in the formulation of the leading semiclassical 
quantization conditions. This suggests that
the quantization conditions in real quantization
can be formulated as the equations
\begin{equation}\label{Qcond1}
\frac{\pa}{\pa {l}_r}\CW({l},q)\Big|_{{l}={l}(\mathbf k)}\,=\,2\pi{\ep} k_r\,,
\quad{\rm for}\quad r=1,\dots,n-3\,,
\end{equation}
where ${\mathbf k}=(k_1,\dots,k_{n-3})$.
The critical point(s) of $\CW(l({\mathbf k}),q)$ give the eigenvalues
$E_r$ of the Hitchin Hamiltonians via
\begin{equation}
E_r\,=\,H_r(l({\mathbf k}),q)\,.
\end{equation}
As partially discussed in Section \ref{sec:quantHitchin}
we'll need futher investigations to properly define the eigenvalue problem
in the real quantization and to check if it can be reformulated in the
form \rf{Qcond1}.

\subsubsection{Further remarks}\label{rems}

The identification of Yang's potential with the semiclassical limit
of conformal blocks can also be arrived at by combining the discussion of
\cite{NS} with the observations of \cite{AGT}. It is proposed in 
\cite{NS} that the
Yang's potential is obtained
from Nekrasov's partition function $\CZ(a,\ep_1,\ep_2;q)$ 
in the limit $\ep_2\ra 0$.
One of the main observations made in \cite{AGT} is the coincidence of the
Nekrasov partition functions for the theories of interest 
with Liouville conformal
blocks. This holds in particular in the case of the $N=2^*$-theory discussed in \cite{NS}
for which the Nekrasov partition function coincides according to \cite{AGT,FL}
with the  Liouville conformal blocks on the one-punctured torus.

The observations discussed above appear to be deeply related
to the recent work of Nekrasov and Shatashvili \cite{NS2}.

\subsection{Degenerate fields as heavy sources}

We shall now consider more general Liouville conformal blocks
of the form
\begin{equation}\label{corr++}
\big\langle \;\CO_{n,m,l}\;\big\rangle\,\equiv\,
\bigg\langle \;\prod_{s=1}^n e^{2\al_s\phi(z_s)}
\prod_{r=1}^me^{-b\phi(w_r)}\prod_{k=1}^l e^{-\frac{1}{b}\phi(y_k)}
\;\bigg\rangle_{\hat{\BG}}
\end{equation}
The conformal blocks
\rf{confbl++}
satisfy the null vector decoupling equations
\begin{equation}\label{BPZ1}
\CD_{y_q}^{\rm BPZ}\cdot\big\langle \;\CO_{n,m,l}\;\big\rangle\,=0\,,\qquad
\tilde\CD_{w_r}^{\rm BPZ}\cdot\big\langle \;\CO_{n,m,l}\;\big\rangle\,=0\,,
\end{equation}
where for $g=0$
\begin{align}
\CD_{y_k}^{\rm BPZ}\,=\,&\, b^2\frac{\pa^2}{\pa y^2}+\sum_{s=1}^n
\left(\frac{\De_s}{(y_k-z_s)^2}+\frac{1}{y_k-z_s}\frac{\pa}{\pa z_s}\right)\\
&
-\sum_{r=1}^m
\left(\frac{3b^2+2}{4(y_k-w_r)^2}-\frac{1}{y_k-w_r}\frac{\pa}{\pa w_r}\right)
-\sum_{\substack{k'=1\\k'\neq k}}^l
\left(\frac{3b^{-2}+2}{4(y_k-y_{k'})^2}-\frac{1}{y_k-y_{k'}}
\frac{\pa}{\pa y_{k'}}\right)\,,\nn\\
\tilde\CD_{w_r}^{\rm BPZ}\,=\,& \frac{1}{b^{2}}\frac{\pa^2}{\pa w_r^2}+\sum_{s=1}^n
\left(\frac{\De_r}{(w_r-z_s)^2}+\frac{1}{w_r-z_s}\frac{\pa}{\pa z_s}\right)
\\
& -\sum_{k=1}^l
\left(\frac{3b^{-2}+2}{4(w_r-y_k)^2}-\frac{1}{w_r-y_k}
\frac{\pa}{\pa y_{k}}\right)-\sum_{\substack{r'=1\\r'\neq r}}^m
\left(\frac{3b^2+2}{4(w_r-w_{r'})^2}-\frac{1}{w_r-w_{r'}}
\frac{\pa}{\pa w_r}\right)
\,.
\nonumber\end{align}
The equations \rf{BPZ1} imply the fusion rules
\beq
[V_{-b/2}]\cdot [V_{\alpha}]= [V_{\alpha-b/2}]+[V_{\alpha-b/2}]\,.
\label{furu2}
\eeq
Bases for the space of conformal blocks of the type \rf{corr++}
can be parameterized in a similar way as described in
Subsection \ref{liouconf++}.


As above in Subsection \ref{probes} we may now consider
the insertions of the degenerate fields $e^{-\frac{1}{b}\phi(y_q)}$
as probes.
The key observation to be made is that the monodromy of
$e^{-\frac{1}{b}\phi(y)}$ around any of the degenerate fields
$e^{-{b}\phi(w_k)}$ is minus the identity matrix, therefore
projectively trivial. This can easily be verified with the help of the
well-known expressions for the fusion and braiding matrices of the degenerate field
$e^{-\frac{1}{b}\phi(y)}$ as recollected e.g. in \cite[Appendix B]{DGOT}.
The procedure explained in Subsection
\ref{probes} can therefore be used to construct an operator $\SL_{\ga}$
acting on the
space of conformal blocks \rf{corr++}
for each generator $\ga$ of the
fundamental group $\pi_1(C_{g,n})$. This operator is insensitive to the
insertions of $e^{-b\phi(w_k)}$, and "measures" via a formula analogous to
\rf{Hecke} the intermediate dimensions $p$ used in  the gluing construction
of the conformal blocks only.

\subsection{Isomonodromic deformations from the
semiclassical limit of Liouville theory}

Let us now consider the limit $\ep_1\ra 0$ keeping $\ep_2$ fixed,
which corresponds to $\hbar\ra 0$ and $b\ra 0$.
Analyzing the differential equations satisfied
by $\langle \,\CO_{n,m,l}\,\rangle$ in this limit we find that
\begin{itemize}
\item[(i)] the following limits exist
\begin{align}
& \CW_{q,w}^{\hat\si}(p,\de)\,\equiv\,
\lim_{b\ra 0}b^{2}\log \big\langle \;\CO_{n,m}\;
\big\rangle_{\hat{\mathbb G}}\\
& \Psi(y)\,\equiv\,\lim_{b\ra\0}
\big[{\big\langle \;\CO_{n,m}\;\big\rangle}_{\hat\si}\big]^{-1}
\big\langle \;\CO_{n,m,l}\;
\big\rangle_{\hat{\mathbb G}}
\,,
\end{align}
\item[(ii)] $\Psi(y)$ factorizes as
\begin{equation}
\Psi(y)\,=\,\prod_{k=1}^{l}\chi_k(y_k)\,,
\end{equation}
where $\chi_k(y_k)$ satisfy
an equation of the form
$(\pa_y^2+t(y))\chi_k(y)=0$ with
\begin{equation}\label{texp}
t(y)\,=\,\sum_{s=1}^{n}
\left(\frac{\de_s}{(y-z_s)^2}+\frac{H_s}{y-z_s}\right)
-\sum_{r=1}^{m}
\left(\frac{3}{4(y-w_r)^2}-\frac{\kappa_r}{y-w_r}\right)\,,
\end{equation}
\item[(iii)] the residues $H_s$ and $\kappa_k$ are constrained by
the relations \rf{BPZclass}.
\item[(iv)] the residues
$H_s=H_s(p,\de|q,w)$ and $\kappa_r=\kappa_r(p,\de|q,w)$ of $t(y)$
introduced in \rf{texp} are related to $\CW_{\be}^{\hat\si}(p,\de|q,w)$ as
\begin{equation}\label{Ekappapotential}
E_s\,=\,-\frac{\pa}{\pa z_s}\CW_{q,w}^{\hat\si}(p,\de)\,,\qquad
\kappa_r\,=\,-\frac{\pa}{\pa w_r}\CW_{q,w}^{\hat\si}(p,\de)\,.
\end{equation}
\end{itemize}
In the case $m=n-3$ we may note
that the equations \rf{BPZclass} coincide with the equations
\rf{nullclass} and that \rf{Ekappapotential} are the
the relations defining the isomonodromic tau-function.

\section{Liouville theory as a quantum theory of the space of local systems}
\label{sec:quantlocsys}
\setcounter{equation}{0}

\subsection{Overview}

The results of the previous sections have demonstrated that Liouville theory has
many relations to the moduli spaces of local systems --- it deforms key geometrical
structures of these moduli spaces. We now want to show that the main features
of Liouville theory can be understood in terms of the {\it quantization} of
real slices in $\CM_{\rm H}$.

It is very important that the structure of $\CM_{\rm H}\simeq
\Homsl{C}$ as a
complex algebraic variety has a natural
deformation that is realized
within the quantization of its real slices. The ring $\CO$
of regular functions on
$\Homsl{C}$ is generated from the traces of holonomies ${\rm tr}(\rho(\ga))$.
It is natural that the algebra $\CO_b$ of
quantized observables should be generated from the quantum
operators $\SH_\ga$ associated to the classical observables ${\rm tr}(\rho(\ga))$.
A natural integrable structure is obtained by choosing a maximal set
of non-intersecting closed curves $\ga_r$,  $r=1,\dots,3g-3+n$.
The corresponding observables $\SL_r\equiv\SL_{\ga_r}$ commute,
$[\SL_r,\SL_s]=0$ for all $r,s=1,\dots,3g-3+n$,
so that the subalgebra $\CI\subset\CO_b$ generated by the $\SL_r$ represents the integrable
structure of the quantum theory of $\Homsl{C}$.

However, non-compactness of the moduli spaces $\Locsl{C}$ implies that the elements of the
algebra $\CO_b$ can not be realized by bounded operators on a Hilbert space $\CH$. It is
therefore important to consider the maximal common domain of definition for the elements of $\CO_b$
within $\CH=\CH(C_{g,n})$. This defines a natural analog $\CS_{\CO_b}$ of the Schwartz-space of smooth,
rapidly decreasing functions on the real line. The common eigenstates of the Hamiltonians
$\SL_r$ are elements of the hermitian dual $\CS_{\CO_b}^{\dagger}$ of $\CS_{\CO_b}^{}$.
Let us denote by $\langle \,p\,|$ the element of $\CS_{\CO_b}^{\dagger}$ which satisfies
\begin{equation}\label{p-eval}
\langle\,p\,|\,\SL_r\,=\,2\cosh(2\pi b p_r/\ep_1)\langle\,p\,|\,,\qquad\forall \;r=1,\dots,3g-3+n.
\end{equation}
The spectrum of the operators $\SL_r$ is exhausted by considering $p_r\in\BR^+$.
It will be important for us to note that the eigenstates $\langle\, p\,|$
can be meromorphically continued to arbitrary {\it complex} values of $p$, in the
sense that $\psi(p)=\langle \,p\,|\,\psi\,\rangle$ can be meromorphically continued w.r.t.
$p$ for all $\psi\in\CS_{\CO_b}$. The wave-functions $\psi(p)$ give a concrete
representation for the elements of $\CS_{{\CO_b}}$.

The action of ${\CO_b}$ on the space $\CS_{\CO_b}$ can be represented as the action of
a ring of finite difference operators on the wave-functions $\psi(p)$.
This furnishes a concrete realization of the quantization of the ring of
regular functions on $\Homsl{C}$ as a noncommutative ring of difference operators
acting on $\CS_{\CO_b}$.

The conformal blocks of Liouville theory are found to be wave-functions of
certain
states $|C\rangle$ associated to the Riemann surface $C$. The conformal blocks
$\CF_{q}^{\si}(p)$, for example,
are nothing but the wave-functions $\langle\, p\,|\,C_q\,\rangle$
of  states $|C_q\rangle$ associated to a family of surfaces $C_q$ with complex structure
parameterized by $q=(q_1,\dots,q_{3g-3+n})$ in the representation introduces above.
Using this dictionary it is possible to see that the Liouville loop operators
$\SL_\ga$ introduced in Subsection \ref{probes}
are mapped precisely to the difference operators
which represent the Hamiltonians $\SL_\ga$ on the wave-functions $\psi(p)$.
Parameterizing conformal blocks in terms of the eigenvalues of the Liouville loop
operators corresponds to labeling the eigenstates $\langle\,p\,|$ by their
eigenvalues, \rf{p-eval}.

The Liouville correlation functions \rf{LcorrI}
represent the norm squared of $|\,C_q\,\rangle$,
\begin{equation}\label{norm-L}
\langle\,C_q\,|\,C_q\,\rangle\,=\,\big\langle\!\big\langle
\;e^{2\al_n\phi(z_n,\bz_n)}\cdots e^{2\al_1\phi(z_1,\bz_1)}\;
\big\rangle\!\big\rangle_{C_q,}
\end{equation}
and the holomorphic factorization \rf{holofact} is the representation
of the scalar product on $\CH$ in the representation where the
operators $\SL_r$, $r=1,\dots,3g-3+n$ are diagonal.

\subsection{Fock-Goncharov coordinates}

Let $\tau$ be a triangulation of the surface $C$ such that
all vertices coincide with marked points on $C$. An edge $e$
of $\tau$ separates two triangles defining a quadrilateral $Q_e$ 
with corners being the marked points $P_1,\dots,P_4$. For
a given local system $(\CE,\nabla')$, let
us choose four
sections $s_i$, $i=1,2,3,4$ that are holomorphic in $Q_e$, 
obey the flatness condition
\begin{equation}
\nabla' s_i\,=\,\left(\frac{\pa}{\pa y}+M(y)\right)s_i\,=\,0\,,
\end{equation}
and are eigenvectors of the monodromy around $P_i$.
Out of the sections $s_i$ form \cite{FG,GNR}
\begin{equation}
\CX_e^\tau\,:=\,
\frac{(s_1\wedge s_2)(s_3\wedge s_4)}{(s_2\wedge s_3)(s_4\wedge s_1)}\,,
\end{equation}
where all sections are evaluated at a common point $P\in Q_e$. It is not
hard to see that $\CX_e^\tau$ does not depend on the choice of $P$.

The Poisson structure is particularly simple in terms of these
coordinates,
\begin{equation}\label{poisson}
\{\,\CX_e^\tau\,,\,\CX_{e'}^\tau\,\}
\,=\,\langle e, e'\rangle\,
\CX_{e'}^\tau\,{\CX_e^\tau},
\end{equation}
where $\langle e, e'\rangle$ is the number of faces $e$ and $e'$
have in common, counted with a sign.

A real slice $\Homslr{C}$ in $\Homsl{C}$ can  be defined by the
conditions $\CX_e^*=\CX_e^{}$.
Recall that the real slice
$\Homslr{C}$ decomposes into different connected components, and that
one of these components is canonically isomorphic to the
Teichm\"uller space of deformations of $C$. This
component is characterized by the property that the
functions $\CX_e$ are all positive.

\subsection{Holonomy variables}

Assume given a
path $\varpi_\ga$ on the fat graph homotopic to a simple closed
curve $\ga$ on $C_{g,n}$. Let the edges be labelled $e_i$, $i=1,\dots,r$
according to the order in which they appear on $\varpi_\ga$,
and define $\si_i$ to be $1$ if the path turns left
at the vertex
that connects edges $e_i$ and $e_{i+1}$, and to be equal to $-1$
otherwise. Consider the following matrix,
\begin{equation}\label{fuchsgen}
{\rm X}_{\ga}\;=\;{\rm V}^{\si_r}{\rm E}(z_{e_r})\dots {\rm V}^{\si_1}
{\rm E}(z_{e_1}),
\end{equation}
where $z_e=\log X_e$, and
the matrices ${\rm E}(z)$ and ${\rm V}$ are defined respectively by
\begin{equation}
{\rm E}(z)\;=\;\bigg(\begin{array}{cc} 0 & +e^{+\frac{z}{2}}\\
-e^{-\frac{z}{2}} & 0 \end{array}\bigg),\qquad
V\;=\;\bigg(\;\begin{array}{cc} 1 & 1 \\ -1 & 0 \end{array}\,\bigg).
\end{equation}
Taking the trace of ${\rm X}_\ga$ one
gets the hyperbolic length of the closed geodesic
isotopic to $\ga$ via
\cite{Fo97}
\begin{equation}\label{glength}
L_{\ga}\,\equiv\,2\cosh\big(\fr{1}{2}l_{\ga}\big)\;=\;|{\rm tr}({\rm X}_{\ga})|.
\end{equation}

We may observe that the classical expression for $L_{\ga}\equiv
2\cosh\frac{1}{2}l_\ga$ as given by formula \ref{glength}
is a linear combination of monomials
in the variables $u_e^{\pm1}\equiv
e^{\pm\frac{z_e}{2}}$ of a very particular form,
\begin{equation}\label{Lclass}
L_{\ga}\,
=\,
\sum_{\nu\in\CF}\,C_{\tau,\ga}(\nu)\,\prod_e u_e^{\nu_e}
\end{equation}
where the summation is taken over a finite set $\CF$ of
vectors $\nu\in\BZ^{3g-3+2n}$
with components $\nu_e$. The coefficients
$C_{\tau,\ga}(\nu)$ are positive
integers.

It is proven in \cite[Theorem 12.3]{FG} that the products of traces of monodromies of
finite laminations form a basis for the vector spaces of regular functions on $\Homsl{C}$.
These objects are therefore natural generators for the algebra $\CO$
of observables of interest.

\subsection{Quantization of the Teichm\"uller component}


The simplicity of the Poisson brackets (\ref{poisson}) makes part of the
quantization quite simple. To each edge $e$ of a triangulation
of a Riemann surface $C_{g,n}$ associate a quantum operator
$\sz_e$ corresponding to the classical phase space function
$z_e=\log\CX_e$. Canonical quantization of the Poisson brackets (\ref{poisson})
yields an algebra  $\CA_\tau$ with generators $\sz_e$ and relations
\begin{equation}\label{comm}
[\,\su_e\,,\,\su_{e'}\,]\,=\,{2}{\pi i}
b^2 \langle e, e'\rangle
\,,
\end{equation}
The algebra $\CA_\tau$
has a center with generators
$\sfc_a$, $a=1,\ldots, n$ defined by $\sfc_a=\sum_{e\in E_a}\sz_e$,
where $E_a$ is the set of edges in the triangulation that emanates from the
$a^{\rm th}$ boundary component.
The representations of $\CA_\tau$ that we are going to consider
will therefore be such that the generators $\sfc_a$ are
represented as the operators of multiplication by real
positive numbers ${l_a/2}$. Geometrically one may interpret
$l_a$ as the geodesic length of the
$a^{\rm th}$ boundary component
\cite{Fo97}.
The vector $l=(l_1,\dots,l_n)$ of lengths of the boundary components
will figure as a label of the representation
of the algebra $\CA_\tau$.

Recall furthermore that the variables $\CX_e$ are positive
for the Teichm\"uller component.
The scalar product of the quantum theory
should realize the
phase space functions
$z_e=\log\CX_e$ as
self-adjoint operators $\sz_e$,
$\sz_e^{\dagger}=\sz_e^{}$.
By choosing a maximal set of commuting generators for the algebra $A_\tau$
one may naturally define a Schr\"odinger
type representation of the algebra $\CA_\tau$ in terms of multiplication
and differentiation operators. It is realized on the
Hilbert space $\CH_\tau\simeq L^2(\BR^{3g-3+n})$.

Less trivial is the fact
that one can define on $\CH_\tau$ a projective unitary
representation of the mapping class group ${\rm MCG}(C_{g,n})$.
It is generated by unitary operators $\SW_\tau(m):
\CH_\tau\ra\CH_\tau$, $m\in{\rm MCG}(C_{g,n})$ constructed in
\cite{CF99,Ka98,T05}.

The resulting quantum theory
does not depend on the underlying triangulation in an essential
way. This follows from the existence of a family of unitary
operators $\SU_{\tau_2,\tau_1}$ that satisfy
\begin{equation}
\SU_{\tau_2\tau_1}^{-1}\cdot\SW_{\tau_1}^{}(m)\cdot\SU_{\tau_2\tau_1}^{-1}
\,=\,\SW_{\tau_2}^{}(m)\,.
\end{equation}
The operators $\SU_{\tau_2,\tau_1}$
describe the change of
representation when passing from the quantum theory associated
to triangulation $\tau_1$ to the one associated to $\tau_2$
\cite{CF99,Ka98,T05,FG09}. They allow us to identify $\CH_{\tau_2}\simeq\CH_{\tau_1}=:\CH(C_{g,n})$.

\subsection{Quantizing regular functions on $\Homsl{C}$}

\subsubsection{Quantizing traces of holonomies}

In order to define a set of generators for the quantized algebra
$\CO_b$ of observables one needs
to define the
length operators $\SL_{\tau,\ga}$ associated to simple closed curves $\ga$.
The operators $\SL_{\tau,\ga}$
should be representable as Laurent polynomials in the variables
$\su_e=e^{\pm\frac{1}{2}\sz_e}$
with positive coefficients that reproduce the
expressions \rf{Lclass} in the
classical limit.
It is important
to ensure that the definition of the operators $ \SL_{\tau,\ga}$
is independent of the triangulation in the sense that
\[
\SU_{\tau_2\tau_1}^{-1}\cdot\SL_{\tau_1,\ga}^{}\cdot\SU_{\tau_2\tau_1}^{-1}
\,=\,\SL_{\tau_2,\ga}^{}\,,
\]
where $\SU_{\tau_2\tau_1}$ is the unitary operator
relating the representation associated to
triangulation $\tau_1$ to the one associated to $\tau_2$.
This ensures that the collection of length
operators $\SL_{\tau,\ga}$ associated to the different
triangulations $\tau$ ultimately defines an operator $\SL_{\ga}$
that is {\it independent} of the triangulation.
A general construction of length operators which fulfils this
requirement was given in \cite{T05}. This
construction coincides with the earlier constructions in
\cite{CF99,CF00} whenever both can be applied.

\subsubsection{The Length Representation}

It can be shown that the length operators associated
to non-intersecting simple closed curves
commute with each other. This together with the
self-adjointness of the length operators
allows one to introduce bases of eigenfunctions
for the length operators.

One gets one such basis for each marking $\si$
of $C_{g,n}$. A key result for the connection
between quantum Liouville and quantum Teichm\"uller
theory is that for each marking $\si$ there exists
a basis for $\CH_{g,n}\equiv\CH(C_{g,n})$ spanned by $\langle\,l\,|$,
$l=(l_1,\dots,l_{3g-3+n})$ which
obeys the factorization rules of conformal field theory
\cite{T05}. This means in particular that
for any pair $\si_2$, $\si_1$ of markings one can always decompose the
unitary transformation $\SV_{\si_2\si_1}$ which relates
the representation corresponding to
marking $\si_1$ to the one corresponding to $\si_2$
as a product of operators which represent
the elementary fusion, braiding and
modular transformation moves introduced in \cite{MS}. The
unitary transformation $\SV_{\si_2\si_1}$ can be represented
as an integral operator of the form
\begin{equation}\label{changepants}
\psi_{\si_2}(l_2)\,=\,\int
d\mu(l_1)\;V_{\si_2\si_1}(l_2,l_1)\,\psi_{\si_1}(l_1)
\,.
\end{equation}
The explicit expressions for the kernel
$V_{\si_2\si_1}(l_2,l_1)$ are known for the
cases where $\si_2$ and $\si_1$ differ by one of the elementary
moves.

With the help of \rf{changepants} we may describe the unitary operators
representing the action of the mapping class group as integral operators of the form
\begin{equation}\label{MCGaction}
\psi_{\si}(l_2)\,=\,\int
d\mu(l_1)\;V_{m.\si,\si}(l_2,l_1)\,\psi_{\si}(l_1)\,,
\end{equation}
where $m.\si$ is the image of the marking $\si$ under $m\in{\rm MCG}(C)$, and we
are taking advantage of the fact that the length representations for $\CH_{g,n}$ associated to
markings $\si$ and $m.\si$ are canonically isomorphic.

\subsection{K\"ahler quantization of the Teichm\"uller component}\label{Kaehlerquant}

\subsubsection{Quantization of local observables}

In analogy to the coherent state representation of quantum
mechanics it is natural to consider a quantization scheme in
which states are represented by holomorphic multi-valued wave-functions\footnote{More precisely sections of a projective line bundle on $\CT_{g,n}$.}
\begin{equation}\label{cohwavefct}
\Psi(q)\,=\,\langle\,q\,|\,\Psi\,\rangle\,,\qquad q=(q_1,\dots,q_{3g-3+n})\,,
\end{equation}
in which the operators $\sq_r$ corresponding to the observables $q_r$ introduced in Section \ref{sec:isomon}
are represented
as multiplication operators, and
the operators $\SH_r$ associated to the conjugate "momenta"
$H_r$  should be represented by the differential operators
$b^{2}\pa_{q_r}$ in such a representation,
\begin{equation}\label{K-quant}
\sq_r\Psi(q)\,=\,q_r\Psi(q)\,,\qquad\SH_r\Psi(q)\,=\,{b^{2}}\,\frac{\pa}{\pa q_r}\,\Psi(q)\,.
\end{equation}
The state $\langle\,q\,|$ introduced in \rf{cohwavefct} is thereby identified as an analog of a coherent state
(eigenstate of the "creation operators" ${\mathsf q}_i$) in quantum mechanics.

Formulae \rf{K-quant} turn the
space of holomorphic wave-functions obtained in the
K\"ahler quantization of the Teichm\"uller
spaces into a module over the ring of holomorphic differential operators
on $\CT_{g,n}$.
Let $P$ be the projective connection $\pa_y^2+t(y)$, and let
the difference $P-P_S$ w.r.t. a reference projective connection $P_S$ be expanded as
\[
t(y)-t_S(y)\,=\,\sum_{r=1}^{3g-3+n} \vartheta_r(y)\,H_r\,,
\]
We may then represent the corresponding quantum operator obtained in the K\"ahler quantization
of the Teichm\"uller spaces as
\begin{equation}
\ST(y)-b^{-2}t_S(y)\,=\,\sum_{r=1}^{3g-3+n} \vartheta_r(y)\,
\frac{\pa}{\pa {q}_r}\,.
\end{equation}
The operator $\ST(y)$ may be called the "quantum energy-momentum tensor". For $g=0$ we'll find the
following operator as the counterpart of the classical energy-momentum
tensor $b^{-2}t_{\vf}$,
\begin{equation}\label{STdef}
\ST(y)\,=\,\sum_{r=1}^{n-1}\left(\frac{\De_{r}}{(y-z_r)^2}+
\frac{1}{y-z_r}
\frac{\pa}{\pa z_r}\right)\,,
\end{equation}
where, as before, $z_{n-1}=1$ and $z_{n-2}=0$.
We have introduced the quantum conformal dimensions $\De_{r}$ which are related to the $\de_r$ by $\de_r=b^2\De_r+\CO(b^2)$.
This should be compared with the Virasoro Ward identities \rf{cfWard}.
Comparison of \rf{STdef} and \rf{cfWard}
indicates that the $\CD$-module structure on $\CM_{0,n}$ produced by the
the K\"ahler
quantization of $\CT_{0,n}$ can be identified with the
$\CD$-module structure on the space of
Virasoro conformal blocks.

\subsubsection{Relation between Length Representation and K\"ahler Quantization}
\label{sec:length-Kahler}

The relation between length representation
and the K\"ahler quantization
is described by means of the wave functions
\begin{equation}
\Psi_l^{\si}(q)\,\equiv\,{}_\si^{}\!\langle\, q\,|\, l\,\rangle
\,.
\end{equation}
The following characterization of these matrix elements was obtained in
\cite{T}:
\begin{equation}\label{bl=wf}
\Psi_l^{\si}(q)\,=\,\CF_{q}^{\si}(p)\,,
\end{equation}
where $\CF_{q}^{\si}(p)$ is the Liouville conformal block
associated to a marking $\si$ with fixed intermediate
dimensions given by the parameters $p_\ga$, $\ga\in\CC_{\si}$.
These parameters are related to the lengths $c_a$ of
the boundary components and to the lengths $l_\ga$
around the curves defining the pants
decomposition respectively as
\begin{equation}
\be_s\,=\,\frac{Q}{2}+i\frac{c_s}{4\pi \ep_1}\,,\qquad
p_\ga\,=\,\frac{l_\ga}{4\pi \ep_1}\,,
\label{mal-cl}
\end{equation}
where $s=1,\dots,n$ and $\ga\in\CC_{\si}$.

Let me quickly recall the argument which lead to the identification
\rf{bl=wf}.
It is based on the
observation that the
wave-function $\Psi_l^\si(q)\,\equiv\langle \,q\,|\,l\,\rangle$
can be characterized as the unique solution of
the following Riemann-Hilbert type problem:
\begin{itemize}
\item The mapping class group element $m$ acts on the wave-functions $\Psi(z)$ in the K\"ahler quantization
in the natural way as a deck transformation.
This means if $\SU_m$ is the operator representing an element $m$ of the mapping class group, we should have
$(\SU_m\Psi)(z)\equiv \Psi(m.z)$, with $\Psi(m.z)$ being the analytic
continuation of $\Psi(z)$ along the path associated to $m$. We may, on the other hand,
describe the action of $\SU_m$ on $\Psi_l^\si(z)$ by means of \rf{MCGaction}.
The consistency of these two descriptions implies that the monodromy action $\Psi_{l_2}^{\si}(m.z)$
can be represented as
\[
\Psi_{l_2}^{\si}(m.z)\,=\,\int
d\mu(l_1)\;V_{m.\si,\si}(l_2,l_1)\,\Psi_{l_1}^{\si}(z)
\,.
\]
\item The asymptotic behavior of $\Psi_l^\si(z)$ can be determined by
quantizing the classical relation
\[
q_\ga H_{\ga}\,\sim\,\left(\frac{l_\ga}{4\pi}\right)^2-\frac{1}{4}\,,
\]
which is valid to
leading order in the limit $l_\ga\ra0$
if $q_\ga$ is the gluing parameter
that vanishes when $l_\ga\ra 0$,
and $H_\ga$ is the corresponding accessory parameter.
We refer to \cite{T} for more details and references.
\end{itemize}
This defines a Riemann-Hilbert type problem which characterizes the left hand side of \rf{bl=wf}
uniquely. It remains to show that the  right hand side of \rf{bl=wf} is a solution to this Riemann Hilbert
problem. This was done in \cite{T01,T03a}.

\subsection{Intermediate summary}

It may be helpful to summarize the main arguments in a schematic
form.
On the one hand, we have seen that the K\"ahler quantization,
which can be understood as the quantization of the
holomorphic infinitesimal structure of $\CT_{g,n}$, produces
the action of the the ring of holomorphic differential operators
on $\CT_{g,n}$ realized on the wave-functions of the
quantum Teichm\"uller theory, in other words
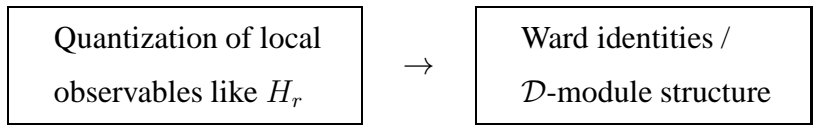
\begin{equation*}
\boxed{\quad
\begin{aligned}
& \text{Quantization of local}\\
& \text{observables like $H_r$}
\end{aligned}
\quad}\quad
\rightarrow \quad
\boxed{\quad
\begin{aligned}
& \text{Ward identities / }\\
& \text{$\CD$-module structure}
\end{aligned}
\quad}
\end{equation*}
The complex structure used here is the one from
Teichm\"uller theory.
The canonical quantization of $\Homslr{C}^0$, on the other
hand, yields
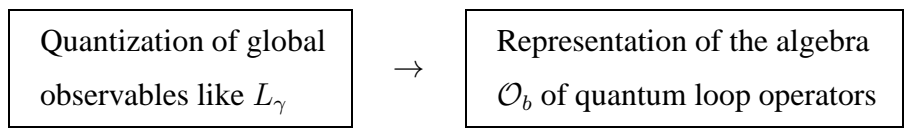
\begin{equation*}
\boxed{\;\;
\begin{aligned}
& \text{Quantization of global}\\
& \text{observables like ${L}_\ga$}
\end{aligned}
\;\;}\quad
\rightarrow \quad
\boxed{\;\;
\begin{aligned}
& \text{Representation of the algebra}\\
& \text{$\CO_b$ of quantum loop operators}
\end{aligned}
\;\;}
\end{equation*}
The realization of the algebra $\CO_b$ deforms the structure
of the ring $\CO$ of algebraic functions on $\Homsl{C}$ in
a natural way. The quantization of the global observables
represents a quantization of
$\Homsl{C}$ in complex structure $J$ with symplectic
form $\Omega_J$.

Different representations for the resulting Hilbert space are
obtained by diagonalizing different maximal subsets
of commuting loop operators. Such subsets are in correspondence
with pants decompositions. The resulting
representation of the groupoid of changes of pants decompostion
(more precisely markings) induces canonically a representation
of the mapping class group via \rf{MCGaction}.

Classically, there is a natural isomorphism between
$\Homslr{C}^0$ and the Teichm\"uller space $\CT(C)$.
Compatibility of canonical quantization of
$\Homslr{C}^0$ and K\"ahler quantization of $\CT_{g,n}$
then defines a Riemann-Hilbert type problem as discussed in
Subsection \ref{sec:length-Kahler}. The Virasoro conformal blocks
are the unique solution of this Riemann-Hilbert type problem.

\subsection{Quantization of the isomonodromic deformation problem}

\subsubsection{Quantization of the local observables}

Let us return to the example of the Garnier system discussed in
Subsection \ref{isomon}. The phase space in question can be identified
with $\Homsl{C_{0,n}}$ via the monodromy map for the
differential operator $\pa_y^2+t(y)$. It will be
useful to start by considering the cases $0<d<n-3$ first. We may then
parameterize $t(y)$ in terms of $H_1,\dots,H_{n-3-d}$ and $\kappa_1,\dots,\kappa_d$,
and the corresponding conjugate
coordinates $z_1,\dots,z_{n-3+d}$, $w_1,\dots,w_d$. The remaining variables
$H_{n-2-d},\dots,H_{n-3}$ are determined
by the constraints \rf{nullclass}, and everything depends on the parameters $z_{n-2-d},\dots,z_{n-3}$.

Contemplating a possible K\"ahler quantization of the Hitchin
moduli space defined by the complex structure $J$ and the
symplectic structure $\varpi_1$, we are lead to
propose a quantization scheme in
which states are represented by holomorphic multi-valued wave-functions
\begin{equation}
\Psi(w,z)\,=\,\langle\,w,z\,|\,\Psi\,\rangle\,,\qquad w=(w_1,\dots,w_{d})\,,\;\;z=(z_1,\dots,z_{n-3-d})\,,
\end{equation}
such that the operators $\sw_r$ corresponding to
the classical observables $w_r$ are represented
as multiplication operators, and
the operators $\sk_r$ associated to the momenta
$\kappa_r$  should be represented by the differential operators
$b^{2}\pa_{w_r}$ in such a representation,
\begin{equation}
\sw_r\Psi(w,z)\,=\,w_r\Psi(w,z)\,,\qquad
\sk_r\Psi(w,z)\,=\,{b^{2}}\,\frac{\pa}{\pa w_r}\,\Psi(w,z)\,.
\end{equation}
The quantum operators $\sz_s$ and $\SH_s$ representing $z_s$ and $H_s$, respectively, should likewise be represented as
\begin{equation}\label{KZ'}
\sz_s\Psi(w,z)\,=\,z_s\Psi(w,z)\,,\qquad
\SH_s\Psi(w,z)\,=\,{b^2}\frac{\partial}{\partial z_s}\Psi(w,z)\,,
\end{equation}
for $s=1,\dots,n-3-d$.
The constraints \rf{nullclass} are quantized as
\begin{align}\label{Hconstr}
\sum_{s=1}^n &\bigg(\frac{b^2\De_s}{(w_r-z_s)^2} +
\frac{1}{w_r-z_s}
\SH_s\bigg)\\
&\qquad\quad+ {b^{4}}\frac{\pa^2}{\pa w_r^2}
+\sum_{\substack{r'=1\\r'\neq r}}^d
\left({b^{2}}\frac{1}{w_r-w_{r'}}
\frac{\pa}{\pa w_{r'}}
-\frac{3+2b^2}{4(w_r-w_{r'})^2}\right)
=0
\,,\nn
\end{align}
for $r=1,\dots,d$. These equations reproduce the equations
\rf{nullclass} or equivalently \rf{BPZclass}
in the limit $b\ra\infty$. The quantum correction proportional to $b^2$
was introduced in the numerator of the last terms 
to ensure commutativity of the operators in \rf{Hconstr}.
Equations \rf{Hconstr} define the Hamiltonians 
$\SH_{n-2-d},\dots,\SH_{n-3}$ as functions of
the remaining
variables.

The wave-functions will depend on $z_{n-2-d},\dots,z_{n-3}$ as parameters. We propose that this dependence
should be expressed by equations of the form
\begin{equation}
\label{KZ''}
{b^2}\frac{\partial}{\partial z_r}\Psi(w,z)\,
\,=\,\SH_r\Psi(w,z)\,,
\end{equation}
with $\SH_{n-2-d},\dots,\SH_{n-3}$ defined by \rf{Hconstr}.
Indeed, let us note that we could equally well have chosen other subsets of $\{H_1,\dots,H_n\}$
and $\{z_1,\dots,z_n\}$ as independent sets of conjugate variables. The consistency with \rf{KZ'}
requires \rf{KZ''}.

The system of equations \rf{Hconstr} is then equivalent to the equations
\begin{align}
\Bigg[\sum_{s=1}^n\bigg(\frac{\De_r}{(w_r-z_s)^2} &+
\frac{1}{w_r-z_s}
\frac{\pa}{\pa z_s}\bigg)\\
&+ b^2\frac{\pa^2}{\pa w_r^2}
-\sum_{\substack{r'=1\\r'\neq r}}^d
\left(\frac{3+2b^2}{4b^2(w_r-w_{r'})^2}-\frac{1}{w_r-w_{r'}}
\frac{\pa}{\pa w_{r'}}\right)\Bigg]\Psi(w,z)
=0
\,,\nn
\end{align}
which are equivalent to the null vector decoupling equations satisfied by
the Liouville conformal blocks \rf{corr++}.

In the case $d=n-3$ we may regard
the second order
differential operators $\SH_r$ as natural
quantization of the Hamiltonian functions of the Garnier system.
The differential equations \rf{KZ''} represent the change of the wave-function
under the change of representation induced by a change of the underlying
complex structure, analogous to the way the KZ equations
were derived by Hitchin in
\cite{Hi90}. We will see later that the equations \rf{KZ''} are indeed essentially equivalent
to the KZ equations in the SL$(2)$ WZNW model.

\subsubsection{Quantization of the global observables}

In the maximal case $d=n-3$,
it seems natural to identify the space of states with the
space spanned by a complete set of solutions to the
equations \rf{KZ'}. We have previously seen in
Subsection \ref{liouconf++} how to identify
a set of solutions to \rf{KZ'} that is complete in the
sense that
changes of the pants decomposition are realized
by linear transformations from one set of solutions
to another. The conformal blocks $\CF_{q,w}^{\si}(p,\de)$
generate a set of solutions which has simple asymptotic
behavior in the boundary component of $\CT_{0,n}$ corresponding
to the marking $\si$. The analytic continuation of
$\CF_{q,w}^{\si_1}(p,\de)$ into the boundary
component of $\CT_{0,n}$ corresponding
to the marking $\si_2$ can be represented as a linear
combination of the solutions $\CF_{q,w}^{\si_2}(p,\de)$.

There is a natural hermitian
form on this space of solutions that is invariant
under the action of the mapping class (braid) group,
given by the Liouville-correlation functions in a
similar way as in \rf{norm-L}. At the moment
it is not clear to the author if this hermitian form
is positive definite for $d>0$. For $d=0$ it certainly is.

As in Subsection \ref{probes} one can define quantum loop
operators acting on the space of states defined above.
These are realized as difference operators. In the
classical limit $b\ra\infty$ we get a distinguished point
in the real slice $\CS_\BR$ in $\Homsl{C_{0,n}}$
defined by the extremum of the absolute value squared of
$\CF_{q,w}^{\si}(p,\de)$. This point lies in the
component of $\Homslr{C_{0,n}}$ labeled by the integer
$d$. The quantum theory described above can therefore
be interpreted as a quantization of this component of
$\Homslr{C_{0,n}}$.

We arrive at a very natural interpretation of the
parameterization of the wave-functions
in terms of their
asymptotic behavior at the boundaries of $\CM_{0,n}$.
The ``interactions'' between degrees of freedom in the
isomonodromic deformation system go to zero near
the boundaries of $\CM_{0,n}$. One may therefore classify
the elements of a basis for the space of states in terms
of the asymptotics of the eigenvalues of the quantized
Hamiltonians. The representation of the space of states in
terms of asymptotic eigenvalues coincides with the
representation for the space of conformal blocks in terms
of the eigenvalues of the quantum monodromy
operators. The unitary operators representing the
transition from one pants decomposition to another
are thereby interpreted as analogs of scattering
operators relating ``In''- and ``Out''-representations
of the space of states.

\section{Geometric Langlands correspondence and conformal field theory}
\setcounter{equation}{0}\label{geolangsect}

In this section we will try to explain some of the relevant
features of the conformal field theory approach to the geometric
Langlands correspondence initiated by Beilinson, Drinfeld, Feigin and
Frenkel to physicists, following mostly the review \cite{Fr2}.

\subsection{Geometric Langlands correspondence and quantization of the
Hitchin system}


The correspondence between opers and the Hitchin
eigenvalue equations is part of the
geometric Langlands correspondence, for the case at hand schematically
\begin{equation}\label{Lcorr1}
\boxed{\quad\phantom{\Big|}
{}^{\rm\sst L}_{}{\mathfrak g}-\text{opers}
\;\quad}\quad
\longrightarrow \quad
\boxed{\quad\phantom{\Big|}
\CD-\text{modules on ${\rm Bun}_G$}
\;\quad}
\end{equation}
The $\CD$-modules on ${\rm Bun}_G$ in question
are in the case of $g=0$ generated by the
differential operators $\CD_r=\SH_r-E_r$.
For the case $n=0$, Beilinson and Drinfeld construct
$3g-3$ differential operators $\CH_r$, $r=1,\dots,3g-3$
on the line bundle $K^{\frac{1}{2}}$ on ${\rm Bun}_G$
which are mutually commuting and have the Hitchin Hamiltonians
as their leading symbols.


\subsection{Conformal blocks for the current algebra}\label{cfbl}

\renewcommand{\fw}{w}
\renewcommand{\fv}{v}
\renewcommand{\fe}{e}

\subsubsection{Definition of the conformal blocks}

Let $\hat\fg=\widehat{\fsl}_{2,k}$ be the central extension of the loop algebra
of $\fsl_2$ with level $k$. The generators of $\hat\fg$ will be denoted
$J_n^a$, $n\in\BZ$, $a=-,0,+$,
the relations are
\begin{equation}
\begin{aligned}
&[\,J_n^0\,,\,J_m^0\,]\,=\,\frac{k}{2}n\de_{n+m,0}\,,\\
&[\,J_n^0\,,\,J_m^\pm\,]\,=\,\pm J_{n+m}^{\pm}\,,
\end{aligned}
\qquad
[\,J_n^+\,,\,J_m^-\,]\,=\,2J_{n+m}^0+kn\de_{n+m,0}\,.
\end{equation}
For representations $\CR_r$, $r=1,\dots,n$
of $\hat\fg$ and Riemann surface $C$ with $n$ marked points
at positions $z_1,\dots,z_n$
one defines the conformal blocks as linear
functionals $\CG:\CR_{[n]}\equiv\otimes_{r=1}^n\CR_r\ra\BC$ that
satisfy the invariance property
\begin{equation}\label{cfbldef}
\CG(\eta\cdot \fv)\,=\,0\qquad \forall \fv\in\CR_{[n]},\;\;\forall\eta\in
\fg_{\rm out}^{}
\end{equation}
where $\fg_{\rm out}^{}$ is the Lie algebra of $\fg$-valued meromorphic functions on $C$ with poles only at $z_1,\dots,z_n$.
The action of $\eta$ on
$\otimes_{r=1}^n\CV_r\ra\BC$ is defined
by taking the Laurent expansions of $\eta$ at the points $z_1,\dots z_n$,
w.r.t. local coordinates $t_r$,
\begin{equation}
\eta(t)\,=\,\sum_{k\in\BZ}\sum_{a=1}^{{\rm dim}(\fg)} t_r^k J^a \,\eta_{r,k}^a\;\;\in\;\;
\fg\otimes\BC(\!(t_r)\!)\,,
\end{equation}
to which we may associate the element
\begin{equation}
J[\eta_r]\,:=\,\sum_{k\in\BZ}\sum_{a=1}^{{\rm dim}(\fg)} J^a_k \,\eta_{r,k}^a\;\;\in\;\;\hat\fg_k\,,
\end{equation}
Denoting by $J_r[\eta_r]$ the operator which acts on $\CR_{[n]}$ nontrivially only on the $r$-th tensor factor of $\CR_{[n]}$,
where the action is given by $J[\eta_r]$ we finally get $\eta=\sum_{r=1}^nJ_r[\eta_r]$.
It can be shown that the central extension vanishes
on the image of the Lie algebra $\fg_{\rm out}^{}$ in
$\bigoplus_{r=1}^n \hat\fg_k$, making the definition consistent.

\subsubsection{Twisted conformal blocks}\label{Ward}

In order to get
differential equations for the conformal blocks from the
conformal Ward identities one possible solution is to modify the definition
\rf{cfbldef} by twisting $\fg_{\rm out}^{}$ by an element $\CE$
of ${\rm Bun}_G$, which means to use \rf{cfbldef} with $\fg_{\rm out}$ replaced
by
\begin{equation}
\fg_{\rm out}^{\CE}\,=\,\Gamma\big(C\setminus\{z_1,\dots,z_n\}\,,\,
\fg_\CE^{}\big)\,, \qquad \fg_\CE^{}\,=\,\CE\underset{G}\times\fg\,.
\end{equation}
The space of linear functionals that satisfy the invariance conditions
in \rf{cfbldef} with $\eta\in\fg_{\rm out}^{\CE}$ will
be denoted $\CFB(\CR_{[n]},C,\CE)$.

Concerning the dependence on the choice of $\CE$ one can a priori only say that one has defined
the conformal blocks as a sheaf over ${\rm Bun }_{G}$.
This means that locally over ${\rm Bun}_G$ we assign
to each bundle $\CE$ the vector space $\CFB(\CR_{[n]},C,\CE)$, but the spaces
assigned to "neighboring" bundles $\CE$ and $\CE'$
do not need to have the same dimension.
The key observation to be made here is that the twisting of conformal blocks by elements of ${\rm Bun}_G$
offers a canonical way to define an action of the differential operators on ${\rm Bun}_G$
on the sheaf of conformal blocks.  In mathematical language this
is expressed as the statement that the space of conformal blocks becomes a $\CD$-module.
In physicists terms this can e.g. be expressed more concretely as follows.
Let us consider conformal blocks with $n+1$ marked points $z_0,\dots,z_n$,
where the vacuum representation is assigned to the marked point $z_0$.
Then for each differential operator $\CD_{\eta}$ on ${\rm Bun }_{G}$
there exists an element $J[\eta]\in\hat\fg_k$ 
such that
\begin{equation}\label{D-mod}
\CD_{\eta}^{}\cdot
\bigg\langle\,\Phi_0(\fv_0|z_0) \,\prod_{r=1}^n
\Phi_r(\fv_r|z_r)\,\bigg\rangle_{C_{g,n+1}}^{\CE}\,=\,
\bigg\langle\,\Phi_0(J_\eta\fv_0|z_0) \,\prod_{r=1}^n
\Phi_r(\fv_r|z_r)\,\bigg\rangle_{C_{g,n+1}}^{\CE}\,.
\end{equation}
The point is that \rf{D-mod} is to be read as the {\it definition} of the action of
the differential operator $\CD_{\eta}$ on the conformal blocks.
The construction of the differential operators $\CD_\eta$ in \rf{D-mod}
is nontrivial in general.
In the mathematical literature there is a construction named "localization functor"
which produces the corresponding sheafs of twisted\footnote{That means roughly "taking care of the central extension"}
differential operators on ${\rm Bun}_G$
under rather general conditions.

In general it is not possible to exponentiate the infinitesimal action of
of the affine Lie algebra $\hat\fg$ given by \rf{D-mod} to a projective
representation of the corresponding loop group. This means that in general one can not
define a parallel transport that would allow one to regard the locally defined
spaces of conformal blocks as a vector bundle over ${\rm Bun}_G$.
For the cases of interest, however, it will turn out that the Lie algebra action \rf{D-mod}
can be exponentiated at least locally, away from a certain divisor of singularities
in ${\rm Bun}_G$.

\subsubsection{More concrete representation of twisted conformal blocks}

In the cases where  the Lie algebra action on the vacuum representation $\CR_0$
exponentiates to a projective representation of the corresponding loop group,
one may represent the relation between twisted and untwisted conformal blocks
more concretely, e.g. for $n=1$
\begin{equation}\label{twist}
\big\langle\,\Phi_0(\fv_0|z_0)\, 
\big\rangle^\CE_{C_{g,1}} \,=\,
\big\langle\,\Phi_0\big(e^{J[\eta]}\fv_0|z_0\big)
\,\big\rangle^{}_{C_{g,1}}\,,
\end{equation}
where $J[\eta]=\sum_n\sum_a J_n^a\eta_n^a$.
$e^{J[\eta]}$ is an operator which represents an element of the (centrally extended)
loop group on $\CV_0$. $e^{J[\eta]}$
can be factorized as $e^{J[\eta]}=Ne^{J[\eta_<]}G_{\rm in}$,
where $G_{\rm in}e_0=e_0$ and $N\in\BC$. Note that
${\rm Bun}_G$ can be represented as double quotient,
\begin{equation}\label{bung-unif}
{\rm Bun}_G\,\simeq\,
G_{\rm out}\,\setminus\, G(\!(t_0)\!)\;/\;
G[[t_0]]\,,
\end{equation}
where $G_{\rm out}$ is the group of algebraic maps
$C_{g,n}\!\!\setminus \{z_0\}\ra G$, and $t_0$ is a
local coordinate around $z_0$ vanishing there. The representation
\rf{bung-unif} follows from the fact that any $G$-bundle can be trivialized
on the complement of a disc ${\mathbb D}_0$ cut out of the surface $C$. This means that
the transition function can be represented by means of an element of the
loop group assigned to the boundary of the disc ${\mathbb D}_0$.
The double quotient representation \rf{bung-unif} implies
a similar representation
for the tangent space $T_\CE{\rm Bun}_G$ as
$\fg_{\rm out}\setminus\fg(\!(z_0)\!)/\fg[[z_0]]$. We may therefore represent
tangent vectors from $T_\CE{\rm Bun}_G$ in terms of derivatives w.r.t.
the parameters $\eta_n^a$ introduced
in \rf{twist}, which explains how \rf{D-mod} comes about.

If all representations $\CR_r$, $r=1,\dots,n$ are integrable one may similarly
introduce the twisting via
\begin{equation}\label{twist'}
\bigg\langle\,\prod_{r=1}^n
\Phi_r(\fv_r|z_r)\,\bigg\rangle^\CE_{C_{g,n}} \,=\,
\bigg\langle\,
\prod_{r=1}^n\Phi_r(e^{J[\eta_r]}\fv_r|z_r)\,\bigg\rangle^{}_{C_{g,n}}\,,
\end{equation}
In this case one should replace \rf{bung-unif} by
\begin{equation}\label{bung-unif'}
{\rm Bun}_G\,\simeq\,
G_{\rm out}\,\setminus\, \prod_{r=1}^n G(\!(t_r)\!)\;/\;
\prod_{r=1}^n G[[t_r]]\,,
\end{equation}
where $t_r$ are local coordinates around the points $z_r$. The representation
\rf{bung-unif'} comes from the existence of a trivialization of the bundle
$\CE$ on the complement of the union $\bigcup_{r=1}^n{\mathbb D}_r$ of
small discs around the points $z_r$.

\subsubsection{Conformal blocks vs. functions on subsets of ${\rm Bun}_G$}

It will also be important for our aims that the twisting allows us to express
the values of the conformal blocks $\CG^{\CE}$
on arbitrary vectors $v\in\CR_{[n]}$ in terms of derivatives on ${\rm Bun}_G$.
This means that for each $v\in\CR_{[n]}$ there
exists a differential operator $\CD_{\CE}(v)$ on ${\rm Bun}_G$ such that
\begin{equation}\label{hwv-red}
\CG_{\CE}(\fv)\,=\,\CD_{\CE}(v)\,\CG_{\CE}(\fe_{[n]})\,,
\end{equation}
where $\fe_{[n]}=\fe_n\ot\dots\ot\fe_1$ is the product of highest weight vectors.

Given a holomorphic bundle $\CE$, a neighborhood $\CU$ of $\CE$ in ${\rm Bun}_G$
and a holomorphic function $\CG$ on $\CU$ we may turn \rf{hwv-red} around and use it
to {\it define} a conformal block. This means that large classes of conformal blocks
actually come from (locally defined) functions on ${\rm Bun}_G$. The point is that
the double quotient representation \rf{bung-unif} of ${\rm Bun}_G$ identifies
this space as a locally symmetric space of the loop group, with infinitesimal structure
given by the loop algebra $\fg\ot \BC(\!(t)\!)$. The relation \rf{hwv-red} describes
how a holomorphic function $\CG$ can be described in terms of this infinitesimal symmetry.

This suggests that one can use conformal blocks as a basis for the space of
holomorphic "functions", or rather sections of bundles, on ${\rm Bun}_G$.
One could thereby put conformal field theory in analogy to the harmonic analysis
on locally symmetric spaces.
The issue raised by this point of view is the possibility to
extend these structures globally over ${\rm Bun}_G$ or some compactification thereof,
possibly allowing "controllable" singular behavior at some divisors.

\subsection{Realization of the geometric Langlands correspondence from
conformal field theory}\label{cfblcrit}

The representation theory of $\hat\fg_{k}$ at the critical level $k=-2$ has
remarkable features. The universal enveloping algebra
$\CU_{\rm crit}(\widehat\fsl_{2})\equiv
\CU(\widehat\fsl_{2})/(k+2)$
has a large center generated by the modes $t_n$ of the rescaled energy-momentum tensor
\begin{equation}
t(y)\,=\,-\frac{1}{k+2}T(y)\,=\,\sum_{n\in\BZ}t_ny^{-n-2}\,.
\end{equation}
This means that there exist representations $\pi_t$ in which
all the generalized
Casimir elements $t_n$ are realized as multiples of the identity.
The generating function $t(y)=\sum_{n\in\BZ}t_ny^{-n-2}$
can be used to parameterize
such representations.

One may then attempt to construct the
conformal blocks with insertions from this class of representations,
\begin{equation}
\bigg\langle\,\prod_{r=1}^n
\Phi_{r,t_r}(\fv_r|z_r)\,\bigg\rangle\,,
\end{equation}
where $\Phi_{r,t_r}$ is the vertex operator associated to a representation $\pi_{t_r}$ with
fixed choice of a generating function $t_r(y)$. The key point to observe about such conformal
blocks is that they can be non-vanishing if, and only if, the generating functions $t_r(y)$
are the Laurent expansions near the
marked points $z_r$ of an oper $\pa_y^2+t(y)$ which is {\it globally defined} on the surface
$C$.

The correspondence between this oper  $\pa_y^2+t(y)$ and the {\it space}
of conformal blocks associated to $C$ and the choice of a collection of representations
assigned to the marked points $z_r$,
\begin{equation}\label{Lcorr3}
\boxed{\quad\phantom{\Big|}
{}^{\rm\sst L}_{}{\mathfrak g}-\text{opers}
\;\quad}\quad
\longrightarrow \quad
\boxed{\quad\phantom{\Big|}
\text{conformal blocks of $\hat\fg_{\rm crit}$}\;
\quad}
\end{equation}
is the origin of the geometric Langlands correspondence
in the approach of Beilinson and Drinfeld.
It remains to remember that
spaces of conformal blocks canonically represent $\CD$-modules to arrive at \rf{Lcorr1}.
The differential equations following from \rf{D-mod} include in particular the
eigenvalue equations for the quantized Hitchin Hamiltonians.
For $g=0$ one finds that the  eigenvalues $E_r$ are given
given by the residues of the oper $\pa_y^2+t(y)$ at $z_r$.

\subsubsection{Hecke action} \label{Heckesect}

\renewcommand{\bg}{${\rm Bun}_G$}

There is a class of natural operations on the $\CD$-modules on \bg\
called Hecke functors.
We refer to \cite{Fr2}
for more discussion of the Hecke functors and their
realization on spaces of conformal blocks at the critical level.
For the moment let us only remark that in the cases where the
$\CD$-modules are produced by the conformal
blocks of $\hat\fg_k$ at the critical level $k=-2$
one may describe the Hecke functors as the
modification of the conformal blocks
by the insertion of certain
representations with rather
special properties. We will later (in Subsection \ref{Hecke+})
discuss natural
analogs of the Hecke functors on the spaces of
conformal blocks for $\hat\fg_k$ at the noncritical level.

Restricting to $\fg=\fsl_2$ for simplicity, the
representations in question are labeled by half-integers
$j$ and denoted $\CW_j$.
As representations of the affine algebra $\hat\fg_{\rm crit}$
these representations are just the
vacuum representation $\CR_0$, but they come equipped with a $2j+1$-dimensional
"multiplicity"-space $V_j$ which is a module for the Lie algebra $\fsl_2$,
\begin{equation}\label{Repfactor}
\CW_j\,\simeq\CR_0\otimes V_j\,.
\end{equation}
The Lie algebra $\fsl_2$ that $V_j$ is a module of has no direct relation
with the $\fsl_2$-subalgebra of the affine algebra $\widehat{\fsl}_{2,k}$
that we started from. It is identified as $\fsl_2$-representation
by its categorical properties, in particular by its behavior under
taking tensor products.
In the case that one is considering a general affine algebra
$\hat\fg_{\rm crit}$
one finds similarly
\begin{equation}\label{Repfactor2}
\CW_\lambda\,\simeq\CR_0\otimes V_\lambda\,.
\end{equation}
with $V_{\lambda}$ being a module of the Langlands dual Lie algebra
${}^L\fg$ to $\fg$.

One may then consider conformal blocks with the representations $\CW_j$ inserted,
\begin{equation}
\bigg\langle\,\Xi_j(\fv|y)\prod_{r=1}^n
\Phi_r(\fv_r|z_r)\,\bigg\rangle_{\CE}\,.
\end{equation}
The special properties of the representations $\CW_j$ imply that the spaces of conformal
blocks with and without insertion of $\CW_j$ are related
as
\begin{equation}\label{Hecke1}
\CFB_{\CE}^{}(\CW_j\ot\CR_{[n]})\,\simeq\,V_j\ot
\CFB_{\CE}^{}(\CR_{[n]})\,.
\end{equation}
The crucial Hecke eigenvalue property of the geometric Langlands correspondence
can loosely speaking be described as the statement that under a variation of the
insertion point $y$ of $\CW_j$ the local isomorphisms \rf{Hecke1}
glue together to generate a local system $E$. If $\CFB(\CR_{[n]},C,\CE,P)$
is the space of conformal blocks associated  to a given oper $P=\pa_y^2+t(y)$
one gets the local system corresponding to the monodromy
representation of $\pa_y^2+t(y)$. The local system $E$ associated to the
oper $\pa_y^2+t(y)$ therefore plays a role analogous to
an eigenvalue. This is roughly what is called the Hecke eigenvalue
property in the context of the geometric Langlands correspondence.



\section{Quantum geometric Langlands correspondence}\label{sec:WZNW}
\setcounter{equation}{0}
\subsection{The Knizhnik-Zamolodchikov equations}

In the case of noncritical level $k\neq -2$ we can use the
Sugawara construction to realize the generators $L_n$, $n\in\BZ$
of the Virasoro algebra
within the universal enveloping algebra $\CU(\hat\fg_k)$.
Recall that the Virasoro
algebra uniformizes infinitesimally the moduli space $\CM_{g,n}$
of Riemann surfaces
in a way that is similar to the way the current algebra uniformizes
${\rm Bun}_G$, as expressed in \rf{VirUni} above.
In the same
way as described in Subsection \ref{Ward} for the case of the current algebra
one may use \rf{VirUni} in order to define a (twisted) action
of the differential operators on $\CM_{g,n}$ on the spaces of conformal
blocks.

The fact that the Virasoro generators $L_n$ can be expressed as bilinear
expressions in the generators $J_n^a$ implies relations between
the differential operators representing the
action of vector fields on $\CM_{g,n}$ and ${\rm Bun}_G$,
respectively, which take the form of the Knizhnik-Zamolodchikov-Bernard (KZB) equations, schematically
\begin{equation}\label{KZ}
(k+2)\frac{\pa}{\pa z_r}\Phi(x|z)\,=\,\SH_r\,\Phi(x|z)\,.
\end{equation}
These equations allow us to ``parallel transport'' conformal blocks
over $\CM_{g,n}$. Any given solution to the equations \rf{KZ} in open
subsets of ${\rm Bun}_G\times\CM_{g,n}$
defines
a conformal block according to our discussion in Subsection
\ref{cfbl}.
We will see, however, that the equations \rf{KZ} have for
fixed point in $\CM_{g,n}$ regular singularities in ${\rm Bun}_G$.
This is related to the fact that the action of $\hat\fg_k$ on the
spaces of conformal blocks defined in \rf{D-mod} does not exponentiate to
a group action in general. However, away from the singularities of the
equation \rf{KZ} it is certainly possible to integrate
the equations \rf{KZ}
in order to extend local solutions to solutions defined on some
covering space of ${\rm Bun}_G\times\CM_{g,n}\setminus{\CS}$,
where $\CS$ is a certain divisor of singularities.

\subsection{Conformal blocks for genus zero}

\subsubsection{Twisting parameters in genus zero}

We will discuss conformal blocks for the ${\rm SL}(2)$-WZNW model
in $g=0$ denoted as
\begin{equation}\label{WZWbl}
\CG(x|z)\,\equiv\,
\big\langle\,\Phi^{j_n}(x_n|z_n)\dots\Phi^{j_1}(x_1|z_1)\,\big\rangle\,.
\end{equation}
The parameters $x_r$ represent a non-minimal twisting of the
conformal blocks as in \rf{twist'}.
In the cases where the representation $\CR_r$
has a highest weight vector $\fe_r$
we may introduce the dependence on the variables $x_r$ via
$\Phi^{j_r}(x_r|z_r)\equiv\Phi^{j_r}(e^{x_rJ_0^-}\fe_r|z_r)$.
The parameters $x_r$ represent the choice of parabolic structures near
the marked points $z_r$.
As vector bundles on a surface of
genus zero are always trivial, we can take the coordinates
$x_r$ to parameterize an open dense subset of  ${\rm Bun}_G(C_{0,n})$.
The current algebra Ward identities now take the familiar form
\begin{equation}
\big\langle\,J^a(t)\,
\Phi^{j_n}(x_n|z_n)\dots\Phi^{j_1}(x_1|z_1)\,\big\rangle
= \sum_{r=1}^{n}\frac{\CJ_r^a}{t-z_r}
\big\langle\,\Phi^{j_n}(x_n|z_n)\dots\Phi^{j_1}(x_1|z_1)\,\big\rangle\,,
\end{equation}
where $\CJ_r^a$ are the differential operators defined in \rf{CJdef}.
The conformal blocks \rf{WZWbl} satisfy the KZ equations \rf{KZ}
with differential operators $\SH_r$ being explicitly given in \rf{Hdef}.

\subsubsection{More general classes of representations}
So far we had assumed that the representations $\CR_r$ of the
current algebra are all of highest weight type. It is worth
noting that the formalism easily allows one to cover representations
of principal or complementary series type, too. Let, for example, 
$\CR_r$ be a
representation of $\widehat{\fsl}_{2,k}$
induced from a principal series representation
of ${\rm SL}(2,R)$. We may assume that the zero mode sub-algebra
$\fsl_2\subset \widehat{\fsl}_{2,k}$ generated by the
$J_0^a$ is realized on functions $f(x_r)\in\CS_r$
by the differential operators $\CJ_r^a$
defined in \rf{CJdef}, with $\CS_r$ being the
Schwartz space of smooth functions on $\BR$ with rapid decay.
The dual space of distributions $\CS_r^{\dagger}$
contains the delta-distributions $\de_x$ with support
at $x$.  In this case we should
identify $\Phi^j(x|z)$ with
$\Phi^j(\de_x|z)$, with $\Phi^j(v|z)$ being the vertex operator
associated to a vector $v\in\CR_j^\dagger$, where $\CR_j^\dagger$ is the hermitian dual
of $\CR_j$.

Correlation functions as considered in
\rf{WZWbl} above are then to be understood as distributions
on a Schwartz space of functions in $n$ variables
$x_1, \dots, x_n$. The type of representation one wants
to consider will determine the precise space of solutions
of the KZ equations that may be relevant for
physical applications. It may, in general, contain distributional solutions supported
on subspaces of ${\rm Bun}_G(C)$.

\subsubsection{Singularities}
In the case $g=0$ it is possible to analyze the singularities of the
differential equations \rf{KZ} which prevent one to extend a local
solution unambigously over ${\rm Bun}_G(C_{0,n})$ in detail.
In the case $n=4$, for example, one may recall the singularity
at $x=z$ found in Subsection \ref{SLCgaudin}. This is the simplest
example of a phenomenon that has also been discussed in the context of
the geometric Langlands correspondence, where it figures under the
name of the ``global nilpotent cone'',
see \cite[Section 9.5]{Fr2} for a discussion and further references.
The global nilpotent cone is the locus in ${\rm Bun}_G(C)$ where all
Hitchin Hamiltonians can vanish. Noting that the leading symbol
of the differential operators $\SH_r$ in the KZ equations coincides with
the Hitchin-Hamiltonians \cite{Hi90}, we are led to
identify  the singularity at $x=z$
exhibited in \rf{CDrep} with the
global nilpotent cone in the example $g=0$, $n=4$.

\subsubsection{Whittaker model}

By means of (formal) Fourier transformation \
$\mu^{j+1}\int dx_r \,e^{\mu_rx_r}$
one can pass to a representation in which the current $J^-(t)$ is represented
diagonally,
\begin{equation}
\big\langle\,J^-(t)\,
\tilde\Phi^{j_n}(\mu_n|z_n)\dots\tilde\Phi^{j_1}(\mu_1|z_1)\,\big\rangle
= \sum_{r=1}^{n}\frac{\mu_r}{t-z_r}
\big\langle\,\tilde\Phi^{j_n}(\mu_n|z_n)
\dots\tilde\Phi^{j_1}(\mu_1|z_1)\,\big\rangle\,.
\end{equation}
This representation will be called the Whittaker model.

The precise
definition of the Fourier-transformation
is delicate since the dependence of the conformal blocks on the variable
$x_r$ is multivalued in general. One would need to choose an appropriate
branch. We plan to discuss this important issue in more detail elsewhere.

This sublety does not affect the relation between the
{\it differential equations} characterizing the conformal blocks in the
two representations.
The conformal blocks must in particular
satisfy the Knizhnik-Zamolodchikov (KZ) equations \rf{KZ}
with differential operators $\SH_r$ represented via \rf{Hdef'} and \rf{Dmu}.
The subtleties coming from additional singularities in the
dependence on the variables $x_r$ will have counterparts in this
representation as well. However, as will be explained below, there
will now be a neat way to handle these singularities in this
representation.

\subsubsection{Gluing construction}

We are interested in the class of solutions that
are properly factorizable in the sense that they have
power series representations in terms of the gluing parameters
defined by a pants decomposition of the surface $C_{g,n}$.
We will in the following construct sets of proberly factorizable
solutions that are complete in a suitable sense.
It is possible to construct such solutions
by means of a gluing construction
which is analogous to the one discussed in Subsection \ref{glue}
for the Virasoro algebra. However, in order to get sufficiently large
families of solutions one also needs to consider representations of
principal series type, as discussed in
a related case in \cite{T99}. We plan to
discuss this important point in more detail elswehere.

\subsection{Solutions to the KZ equations from solutions to
null vector decoupling equations} \label{BPZ->KZ}

In what follows we will describe a construction of
a sufficiently large set of factorizable
solutions to the KZ-equations \rf{KZ} from the solutions to
the BPZ-equations \rf{BPZ3'}. In order to formulate it, we shall again take adavantage of the
fact that projective invariance allows us to reconstruct the conformal 
blocks introduced in \rf{WZWbl} from their limits when $z_n\ra\infty$, $x_n\ra\infty$.
The Fourier-transformation with respect to the remaining $n-1$ variables
$x_1,\dots,x_{n-1}$ will be denoted as $\tilde\CG(\mu|z)$, $\mu=(\mu_1,\dots,\mu_{n-1})$,
$z=(z_1,\dots,z_{n-1})$ in the following. The main claim is that the ansatz
\begin{equation}\label{main}
\begin{aligned}
\tilde\CG(\mu|z)\,=\, u\,\delta\big({\textstyle\sum_{i=1}^{n-1}\mu_i}\big)
\, \Theta_n(y|z)
\, & \CF(y|z)\,,
\end{aligned}
\end{equation}
yields a solution to the KZ-equations \rf{KZ} from any given solution
$\CF(y|z)$ to the BPZ-equations \rf{BPZ3'}.
The function $\Theta_n(y|z)$
which appears here is defined as
\begin{align}
\Theta_n^{}(y|z)\,=\,
\prod_{r<s\leq n-1} z_{rs}^{\frac{1}{2b^2}}
\prod_{k<l\leq n-3} y_{kl}^{\frac{1}{2b^2}}
\prod_{r=1}^{n-1}\prod_{k=1}^{n-3}(z_r-y_k)^{-\frac{1}{2b^2}}.
\label{thetan}
\end{align}
The claim
will hold provided that the respective variables
are related as follows:
\begin{enumerate}
\item The variables $\mu_1,\dots,\mu_{n-1}$ are related
to $y_1,\dots,y_{n-3},u$ via
\begin{equation}\label{magic}
\sum_{r=1}^{n-1} \frac{\mu_r}{t-z_r} =
u\frac{\prod_{k=1}^{n-3}(t-y_k)}{\prod_{r=1}^{n-1} (t-z_r)}
\, .
\end{equation}
In particular, since $\sum_{r=1}^{n-1} \mu_r=0$, we have $u=\sum_{r=1}^{n-1}
\mu_r z_r$.
\item The Liouville parameter $b$
is identified with the $\H$ parameter $b^2=-(k+2)^{-1}$.
\item The
Liouville momenta are given by
\begin{equation}\label{alpha-j}\a_r\equiv\a(j_r):=b(j_r+1)+\hib.
\end{equation}
\end{enumerate}
The fact that \rf{main} solves the KZ equations \rf{KZ}
is a simple generalization of Sklyanin's
observation described in Subsection \ref{qSOV} \cite{Sto,RT}.
\begin{rem}\label{globalinvrem}
Comparing with \cite{RT} one should note
that the formulae in this paper yield the formulae
above in the limit $z_n\ra\infty$ and $x_n\ra\infty$. It is interesting to 
note that the resulting formulae look very similar except that we have
only $n-3$ variables $y_k$ here rather than $n-2$ in \cite{RT}.
In order to understand the relation between the two 
representations note that the solutions
constructed in \cite{RT} automatically satisfy the constraints
of invariance under the global ${\rm SL}(2)$. This follows indirectly
from the proof of the main result in \cite{RT}. To see how this works
one may start by considering the case $n=3$. In this
case one may note that the condition
$\sum_{r=1}^{n-1}{\rm D}_r^+\tilde\CG(\mu|z)=0$ is a second order
differential equation on the variables $\mu_r$, which is true
as a consequence of the fact that the corresponding Liouville
conformal block satisfies a BPZ null vector decoupling equation.
The case of arbitrary $n$ can be reduced to $n=3$ by means of 
the factorization argument 
used in \cite{RT}.

We here prefer the formulation above as the 
relation with the degrees of freedom of the  Gaudin-model
is more directly visible in this way.
\end{rem}

\subsubsection{Bases for the space of conformal blocks from the gluing construction}

We may then {\it define} a family of
$\widehat{\fsl}_{2,k}$-conformal blocks by means of the
formula
\begin{align}\label{LiouWZW}
\big\langle\,  \Phi^{j_n} &
(\mu_n|z_n)\dots\Phi^{j_1}(\mu_1|z_1)\,\big\rangle_{\hat{\mathbb G}}= \\[.5ex]
\,&= \delta({\textstyle\sum_{i=1}^{n}\mu_i})
\,  u\,\Theta_{n}(y|z)\,
\big\langle \;e^{2\al_n\phi(z_n)}\cdots e^{2\al_1\phi(z_1)}\;
e^{-\frac{1}{b}\phi(y_{n-2})}\cdots e^{-\frac{1}{b}\phi(y_1)}
\;\big\rangle_{\hat{\mathbb G}},
\nn\end{align}
where the conformal blocks on the right hand side have been
defined in Subsection \ref{liouconf++}.

We are looking for properly factorizable solutions, i.e. solutions that
have a simple behavior at the boundary component of Teichm\"uller
space corresponding to a chosen marking $\si$. Consider e.g.
a degeneration where $z_2-z_1=\CO(\ep)$ with $\ep\ra 0$.
Considering formula \rf{magic} for values of $t$ such that
$t-z_1=\CO(\ep)$, we may note that the left hand side is of
order $\CO(\ep^{-1})$, while the right hand side would be
of order $\CO(\ep^{-2})$ unless there is an index $j$ such
that $y_j-z_1=\CO(\ep)$. Considering a degeneration of
$\BP^1\setminus\{z_1,\dots,z_n\}$
into pairs of pants connected by thin tubes
we'd similarly find that each pair of pants will contain
exactly one of the insertion points $y_j$
of the degenerate fields $e^{-\frac{1}{b}\phi(y_j)}$
in \rf{LiouWZW}. Specializing the parameterization of
the solutions to the BPZ-equations introduced in
Subsection \ref{liouconf++} accordingly allows us to get a
parameterization for a
complete set of solutions
to the KZ-equations \rf{KZ}.

\subsubsection{Critical level limit}

We want to explain how formula \rf{main}
reproduces the geometric Langlands correspondence in the
limit $b^2\ra\infty$ corresponding to $k\ra -2$.
We may, on the one hand, note that in the limit $k\ra-2$
we may solve the KZ equations in the form
\begin{equation}
\tilde\CG(\mu|z)\,\sim\,\exp(-b^2S(z))\Psi(\mu|z)(1+\CO(b^{-2}))\,,
\end{equation}
provided that $\Psi(x|z)$ is a solution to the Gaudin eigenvalue
equations $\SH_r\Psi=E_r\Psi$ with $E_r$ given in terms of
$S(z)$ by $E_r=-\pa_{z_r}S(z)$. The system of these eigenvalue equations
represents the $\CD$-module on the right hand side of \rf{Lcorr1}.

Considering the right hand side of \rf{main}, on the other hand,
we may use the discussion of the semiclassical limit of
Liouville conformal blocks in Subsection \ref{LclassI}. It shows how the
opers on the right hand side of \rf{Lcorr1} are reproduced.

\subsection{Modular duality}\label{moddual}

\subsubsection{A dual WZNW model from Liouville theory}

An interesting consequence pointed out in \cite{GNN} of the duality of Liouville theory under
$b\ra b^{-1}$
is that one can build a second, dual WZNW model from Liouville theory
by replacing \rf{LiouWZW} by
\begin{align}\label{LiouWZW'}
\big\langle\,  \tilde\Phi^{j_n} &
(\mu_n|z_n)\dots\tilde
\Phi^{j_1}(\mu_1|z_1)\,\big\rangle_{\mathbb G}^{\rm\sst dual}= \\[.5ex]
\,&= \delta({\textstyle\sum_{i=1}^{n}\mu_i})
\,  u\,\tilde\Theta_{n}(y|z)\,
\big\langle \;e^{2\al_n\phi(z_n)}\cdots e^{2\al_1\phi(z_1)}\;
e^{-{b}\phi(y_{n-2})}\cdots e^{-{b}\phi(y_1)}
\;\big\rangle_{\mathbb G}\,,
\nn\end{align}
where $\tilde{\Theta}_n$ is obtained from the definition \rf{thetan} by replacing $b\ra 1/b$ on the left hand side,
and the parameters $j_r$ are related to the $\al_r$ via
\begin{equation}\label{alpha-j'}\a_r=b^{-1}(j_r+1)+\frac{b}{2}.
\end{equation}
The conformal blocks $\check\CG(\mu,z)\equiv\langle\,  \Phi^{j_n}
(\mu_n|z_n)\dots\Phi^{j_1}(\mu_1|z_1)\rangle_{\mathbb G}^{\rm\sst dual}$
on the left hand side satisfy KZ equations of the form
\begin{equation}
(\check{k}+2)\frac{\pa}{\pa q_r}\,\check\CG(\mu,z)
\,=\,\SH_r\,\check\CG(\mu,z)\,,
\end{equation}
which are the KZ equations for the SL$(2)$-WZNW model with level $\check{k}$ related to $k$
via
\begin{equation}
\check{k}+2\,=\,\frac{1}{k+2}\,=\,-b^2\,.
\end{equation}
The limit $b\ra\infty$  corresponds to the classical limit of the
dual SL$(2)$-WZNW model.

\subsubsection{Local systems from the classical limit of WZNW conformal blocks}\label{classlim}

Let us consider the classical limit where
${k}\ra \infty$ corresponding to $b\ra 0$ in
the WZNW model.
Let us consider, in particular, conformal blocks like
\begin{equation}\label{degins}
\CG(x,u|y,z)\,:=
\big\langle\,  \Phi_{(2,1)}^+(x|y)\,\Phi^{j_n}
(u_n|z_n)\dots\Phi^{j_1}(u_1|z_1)\,\big\rangle_{\mathbb G}\,.
\end{equation}
The null vector decoupling equation
for the degenerate field $\Phi^{+}_{(2,1)}(x|y)$
is simply
\begin{equation}
\pa_x^2\,\Phi^{+}_{(2,1)}(x|y)\,=\,0\,,
\end{equation}
which means that $\Phi^{+}_{(2,1)}(x|y)$ transforms in the
two-dimensional representation of $\fsl_2$.
Let $\CG(x,u|y,z)=\CG_+(u,y,z)+x\CG_-(u,y,z)$, and let
$\FG=(\CG_+,\CG_-)^t$.
The system of KZ equations satisfied by the conformal blocks \rf{degins}
can the be written in the form
\begin{equation}\label{KZdeg}\begin{aligned}
&-\frac{1}{b^2}\,\frac{\pa}{\pa y}\,\FG(u,y,z)
\,=\,\sum_{r=1}^n\,\eta_{aa'}\frac{\si^a\CJ_r^{a'}}{y-z_r}\,\FG(u,y,z)\,,
\\
&-\frac{1}{b^2}\,\frac{\pa}{\pa z_r}\,\FG(u,y,z)
\,=\,\sum_{\substack{s=1 \\ s\neq r}}^n\,\eta_{aa'}\frac{\CJ_r^a\CJ_s^{a'}}{z_r-z_s}\,\FG(u,y,z)+
\eta_{aa'}\frac{\CJ_r^a\si^{a'}}{z_r-y}\,\FG(u,y,z)\,,
\end{aligned}\end{equation}
with $\si^a$ being the matrices representing $\fsl_2$ in the
two-dimensional representation,
and $\CJ_r^a$ being the differential operators introduced in
\rf{CJdef}.
Let us assume that $j_r=\CO(\ep_1^{-1})$,
which implies that $\CJ_r^a=\CO(\ep_1^{-1})$,
where $\ep_1=\hbar b$.
Note that this corresponds to $\al_r=\CO(b^{-1})$ in
terms of the Liouville parameters.
We can assume that in the limit where $b\ra 0$, $\hbar\ra 0$
with $\ep_2=\hbar/b$ fixed
\begin{equation}
I_r^a\,:=\,\lim_{\ep_2\ra 0}\,{\ep_1}\,\frac{
\CJ_r^a G_{\pm}(u,y,z)}{G_\pm(u,y,z)}\,,
\end{equation}
is independent of $y$ and the choice of component, and define
\begin{equation}
I(y)\,:=\,
\sum_{r=1}^n\,\frac{\eta_{aa'}\si^a I_r^{a'}}{y-z_r}\,.
\end{equation}
The first equation in \rf{KZdeg} then implies that the vector
${\mathbf S}(y)\equiv {\mathbf S}(y|u,z)$,
\begin{equation}
{\mathbf S}
(y):=\,\frac{{\mathbf G}(u,y,z)}{F(u,z)}\,,
\end{equation}
where $F(u,z):=\langle\Phi^{j_n}
(u_n|z_n)\dots\Phi^{j_1}(u_1|z_1)\rangle$
satisfies the equation
\begin{equation}
(\ep_2\pa_y+I(y)){\mathbf S}(y)\,=\,0\,.
\end{equation}
$I(y)$, by definition, depends on $z$. However, the monodromy
of the degenerate field $\Phi_{(2,1)}^+(x|y)$
inserted in \rf{degins} is completely defined
in terms of gluing parameters $\mathbb G$.
It follows that the monodromy of the $\ep_2$-connection
$\ep_2\pa_y+I(y)$ stays unchanged under variations of $z$.
That's the dual way the isomonodromic deformation problem is
recovered from the classical
limit of Liouville theory which is related to the
observations \cite{Re,Ha} identifying
the KZ equations as a formal quantization of the
isomonodromic deformation problem.

\subsection{Insertions of degenerate fields as quantum Hecke functors}
\label{Hecke+}

Consideration of the relation between Liouville theory
and the WZNW-model in cases where the representations $\CR_j$
of $\widehat{\fsl}_{2,k}$
contain null vectors will reveal important further aspects of the
relation with the geometric Langlands correspondence.
Recall that
the Verma modules $\CV_{j,k}$ of the affine algebra $\widehat{\fsl}_{2,k}$
become degenerate whenever the representation of the
zero mode subalgebra $\fsl_2$ has Casimir eigenvalue $j(j+1)$ with
$j=j_{(k,l)}^\ep$, where
\begin{equation}
j_{(k,l)}^+\,=\,\frac{k-1}{2}+\frac{l-1}{2b^2}\,,\qquad
j_{(k,l)}^-\,=\,-\frac{k+1}{2}-\frac{l}{2b^2}\,
\end{equation}
with $m,n=1,2,\dots$.
In the following we are going to explain how the representations
with $j=j_{(1,1)}^-=\frac{k}{2}$ and $j=j_{(1,2)}^+=1/2b^2$
are related to the so-called Hecke functors.

\subsubsection{Bundle modifications in conformal field theory}

In Subsection \ref{cfbl} we have described how to assign spaces ${\SC\SB}(\CR_{[n]},C_{g,n},\CE)$
of conformal blocks to
a Riemann surface $C_{g,n}$, a collection of representations
$\CR_1,\dots,\CR_n$ assigned to the marked points
$z_1,\dots,z_n$ of $C_{g,n}$ and a holomorphic $G$-bundle $\CE$ on $C_{g,n}$.
We now want to discuss how modifications of the bundle lead to
modifications of ${\SC\SB}(\CR_{[n]},C_{g,n},\CE)$.
Modifications of the bundle $\CE$ can be described e.g. by cutting out a small disc ${\mathbb D}_0$
around a point $z_0\in C_{g,n}$ and taking an element $g_0$
of the loop group $LG$ associated to the boundary of ${\mathbb D}_0$ as the new transition function
between ${\mathbb D}_0$ and the rest of $C_{g,n}$.

Our discussion in Subsection \ref{cfbl} suggests a simple realization of such bundle modifications in
conformal field theory: Use the propagation of vacua to
represent a conformal block $\CG\in {\SC\SB}(\CR_{[n]},C_{g,n},\CE)$ by means of
$\hat\CG\in {\SC\SB}(\CR_{[n+1]},C_{g,n+1},\CE')$ with an insertion of the vacuum $e_0$ at the point $z_0$,
and then replace $e_0$
by a "twisted vacuum vector" $\tilde{e}_0$ which is a vacuum vector w.r.t.
the generators $\tilde{J}_n^a$ obtained
from the $J_n^a$ by acting with the automorphism of $\hfsl_{2,k}$ induced by the element $g_0$ of the
loop group which represents the transition function between ${\mathbb D}_0$ and the rest of $C_{g,n}$.
We are thereby lead to define the modified
conformal blocks $\CG'$ as
\begin{equation}
\CG'(v_{[n]})\,=\,\hat\CG(v_{[n]}\ot \tilde{e}_0)\,,
\end{equation}
If, for example, the automorphism
is represented as $\tilde{J}_n^a=\hat{g}_0^{}\,J_n^a\,\hat{g}^{-1}_0$ with $\hat{g}_0$ being an element of the central extension
of the loop group corresponding to the Lie algebra $\hfsl_{2,k}$,
and if the vacuum representation exponentiates to a representation of this Lie group, we recover the description of the twisting of
conformal blocks given in Subsection \ref{cfbl}.

\subsubsection{Hecke modifications}

In order to get more interesting bundle modifications we need to consider
a slightly more general set-up. Instead of considering transition functions taking values
in ${\rm  SL}(2)$ let us consider transition function with
values in ${\rm GL}(2)$.
Let us in particular consider transition functions of the form
\begin{equation}\label{H-twist0}
g_0\,=\,\bigg(\begin{matrix} 1 & 0 \\
0 & t\end{matrix}\bigg)\bigg(\begin{matrix} 1 & u_0 \\
0 & 1\end{matrix}\bigg),
\end{equation}
where $t$ is a local coordinate inside of ${\mathbb D}_0$ vanishing at $z_0$.
Bundle modifications of this form are called Hecke modifications.
The determinant of the modified
bundle vanishes at $z_0$.

On an $\ep_2$-connection, conjugation by the element
$h=\big(\begin{smallmatrix} 1 & 0 \\
0 & t\end{smallmatrix}\big)$ induces the improper gauge
transformation
\begin{equation}
\ep_2\,\pa_{t}+\tilde{I}\,:=\,t^{-\frac{1}{2}}
h\cdot(\ep_2\pa_{t}+{I})\cdot h^{-1}
t^{\frac{1}{2}}\,=\,\ep_2\,\pa_{t}+\Bigg(\,\begin{matrix}
I^0+\frac{\ep_2}{2t} & \;\;\;\;\frac{1}{t}\,I^+\\ tI^- & -I^0-\frac{\ep_2}{2t}
\end{matrix}\,\Bigg) \,.
\end{equation}
The factors $t^{\pm\frac{1}{2}}$ were inserted to restore the
${\rm SL}(2)$-form of the $\ep_2$-connection. In terms of the
modes defined by $\tilde{I}^a(t)=\sum_{n}t^{-n-1}\tilde{I}_n^a$ and
${I}^a(t)=\sum_{n}t^{-n-1}{I}_n^a$ this is equivalent to
\begin{equation}\label{specflowcl}
I_n^\pm\,\ra\,\tilde{I}_n^\pm:=I_{n\mp 1}^\pm\,,\qquad I_n^0\,\ra\,
\tilde{I}_n^0:=I_n^0+
\frac{\ep_2}{2}\de_{n,0}\,.
\end{equation}
There is an essentially unique counterpart of this transformation called
spectral flow for the centrally
extended Lie algebra $\widehat\fsl_{2,k}$,
\begin{equation}\label{specflow}
J_n^\pm\,\ra\,\tilde{J}_n^{\pm}\equiv
J_{n\mp 1}^\pm\,,\qquad J_n^0\,\ra\,\tilde{J}_n^{0}\equiv
J_n^0-\frac{k}{2}\de_{n,0}\,.
\end{equation}
The spectral flow \rf{specflow} reduces to \rf{specflowcl} in the classical limit
$k\ra\infty$ as considered in Subsection \ref{classlim}.
The Hecke-modified conformal blocks can then be represented in
terms of ordinary
conformal blocks which have at the point $z_0$ a vector $\tilde{e}_0$
with the modified
vacuum property
\begin{equation}\label{twvac}
\tilde J_n^{\pm}\,\tilde{e}_0\,=\,J_{n\mp 1}^\pm \,\tilde{e}_0\,=\,0\,,
\qquad \tilde J_n^0\,\tilde{e}_0\,=\,J_{n}^0\,\tilde{e}_0
\,=\,0\,,
\qquad n\geq 0\,.
\end{equation}
There is no vector $\tilde{e}_0$ with such properties in the vacuum
representation, but there is a distinguished representation  $\CR_{k/2}$  of
$\widehat\fsl_{2,k}$ which has a highest
weight vector $\tilde{e}_0:=e_{k/2}$ that satisfies \rf{twvac}.
The representation is obtained as the quotient of the
Verma module $\CV_{k/2,k}$ by the submodule generated by the null-vector
$J_{-1}^+\,e_{k/2}$ at level $1$.

Inserting the representation $\CR_{k/2}$ at $z_0$ is in the vertex operator notation
represented by inserting the vertex operator $\Phi^{\frac{k}{2}}(u_0|z_0)$. It depends on the extra variable
$u_0$ which parameterizes a choice of a parabolic subgroup at $z_0$. 

In order to eliminate this
dependence and in order to strengthen analogies
with the definitions of the Hecke operators in other
circumstances (see e.g. \cite{Fr05} for a discussion) let us consider
\begin{equation}
\Xi(w)\,:=\Xi(0|w)\,,\qquad
\Xi(u|w):=(\CI_{\frac{k}{2}}\Phi^{\frac{k}{2}})(u|w)\,,
\end{equation}
where $\CI_j$ is the $\fsl_2$-intertwining operator
\begin{equation}
(\CI_j\Phi^{j})(u|w)\,=\,-\frac{2j+1}{\pi}\int du'\;|u-u'|^{-2j-2}
\Phi^{j}(u'|w)\,.
\end{equation}
It is worth noting that the vertex operator $\Xi(u|w)$
transforms under the $\fsl_2$-subalgebra
as a representation with $j'=-1-\frac{k}{2}=1/2b^2=j_{(1,2)}^+$
which vanishes at the
critical level. In this case the definition of $\Xi(w)$ simplifies to
\begin{equation}\label{pushforward}
\Xi(w)\,\equiv\,\int dx \;\Phi^{\frac{k}{2}}(x|w)\,\equiv\,\tilde{\Phi}^{\frac{k}{2}}(0|w)\,.
\end{equation}
The representation
corresponding to the operator
$\Xi(w)$ will become a multiple of the vacuum representation
at the critical level, as is necessary to make contact with the discussion
in Subsection \ref{cfblcrit}.

\subsubsection{${\rm GL}(2)$-twisted conformal blocks}

An ${\rm SL}(2)$-bundle $\CE$ can be represented in many ways
by means of a ${\rm GL}(2)$-bundle $\hat\CE$ with fixed determinant
$\CO(D)$, where
$D$ is an effective divisor of degree $d$.
Let $w_1,\dots,w_d$ be the points of the divisor $D$, and let ${\mathbb D}_1,\dots,{\mathbb D}_d$ be small
discs around the points $w_1,\dots,w_d$, respectively. If $\CE$ is represented by the
transition functions $g_k$ at $\pa{\mathbb D}_k$, $k=1,\dots,d$, we may represent
$\hat\CE$ by the transition functions $\hat g_k$ defined by
\begin{equation}\label{H-twist}
\hat g_k\,=\,g_k\bigg(\begin{matrix} 1 & 0 \\
0 & t_k\end{matrix}\bigg)\bigg(\begin{matrix} 1 & u_k \\
0 & 1\end{matrix}\bigg).
\end{equation}

We are lead to consider a natural family of generalizations
of the space
of conformal blocks ${\SC\SB}(\CR_{[n]},C_{g,n},\CE)$
which will be denoted
${\SC\SB}^{[d]}(\CR_{[n]},C_{g,n},\hat\CE)$. It is defined as the space of
linear functionals  $\CG$ on $\CR_{[n]}$
which can be represented in terms of conformal blocks
$\CG'\in{\SC\SB}(\CR_{[n+d]},C_{g,n+d},\hat\CE)$ of the form
\begin{equation}
\CG(v_{[n]})\,=\,\CG'\big(v_{[n]}\ot \tilde{e}_0^{\ot d}\big)\,.
\end{equation}
It seems reasonable to regard the elements of
${\SC\SB}^{[d]}(\CR_{[n]},C_{g,n},\hat\CE)$ as
natural generalizations of the twisted conformal blocks if the
twisting by elements $\CE$ of ${\rm Bun}_{\rm SL(2)}(C)$ is generalized to
twisting by elements $\hat\CE$ of ${\rm Bun}_{\rm SL(2)}^{\CO(D)}(C)$,
which are ${\rm SL}(2)$-bundles represented by ${\rm GL}(2)$-bundles
$\hat\CE$ with $\det(\hat\CE)\simeq \CO(D)$.

Of particular interest will also be the conformal blocks that are
obtained by inserting the vertex operators $\Xi(w)$, like
\begin{equation}
\big\langle\!\big\langle\,
\Phi^{j_n}(x_n|z_n)\dots\Phi^{j_1}(x_1|z_1)\,\big\rangle\!\big\rangle\,:=\,
\big\langle\,\Xi(w_1)\dots\Xi(w_d)\,
\Phi^{j_n}(x_n|z_n)\dots\Phi^{j_1}(x_1|z_1)\,\big\rangle\,
\end{equation}
These linear functionals can of course not be canonically identified with the conformal blocks
$\langle\,
\Phi^{j_n}(x_n|z_n)\dots\Phi^{j_1}(x_1|z_1)\,\rangle$
for noncritical level,
but the fact that $\Xi(w)$ becomes proportional to the vacuum for $k=-2$ will imply
that they become proportional to the conformal blocks
$\langle\,
\Phi^{j_n}(x_n|z_n)\dots\Phi^{j_1}(x_1|z_1)\,\rangle$
{\it at} the critical level, as will be shown below.

\subsubsection{Representation of Hecke modifications
in terms of Liouville conformal blocks}

Let us consider conformal blocks for the ${\rm SL}(2)$-WZNW model
with $d$ insertions of $\Phi^{\frac{k}{2}}(u|w)$
\begin{equation}\label{modconfbl}
\Phi(u,x|w,z)\,\equiv\,
\big\langle\,\Phi^{\frac{k}{2}}(u_1|w_1)\dots\Phi^{\frac{k}{2}}(u_d|w_d)\,
\Phi^{j_n}(x_n|z_n)\dots\Phi^{j_1}(x_1|z_1)\,\big\rangle\,.
\end{equation}
After Fourier transformation to the $\mu$-representation we get
\begin{equation}\label{modmurep}
\tilde\Phi(\nu,\mu|w,z)\,\equiv\,
\big\langle\,\tilde\Phi^{\frac{k}{2}}(\nu_1|w_1)\dots\tilde
\Phi^{\frac{k}{2}}(\nu_d|w_d)\,
\tilde\Phi^{j_n}(\mu_n|z_n)\dots\tilde\Phi^{j_1}(\mu_1|z_1)\,\big\rangle\,.
\end{equation}
Note that in the case $j=k/2$
the formula \rf{alpha-j} gives
$\alpha(k/2)=Q$. The Virasoro representation with $\alpha=Q$
has conformal weight zero, it therefore corresponds to the
vacuum representation.
The transformed conformal block \rf{modmurep} may therefore be represented in terms of Liouville conformal blocks as
\begin{align}\label{LiouWZW''}
\bigg\langle\,  \prod_{r=1}^d &\,\tilde\Phi^{\frac{k}{2}}(\nu_r|w_r) \prod_{s=1}^n\tilde\Phi^{j_s}
(\mu_s|z_s)
\,\bigg\rangle_{\mathbb G}\,= \\[-.5ex]
\,& = \delta({\textstyle\sum_{s=1}^{n}\mu_s+\sum_{r=1}^{d}\nu_r})
\,  u\,\Theta_{n+d}(y|z)\,
\bigg\langle \;\prod_{s=1}^n e^{2\al_s\phi(z_s)}\prod_{k=1}^{n+d-2}
e^{-\frac{1}{b}\phi(y_{k})}
\;\bigg\rangle_{\mathbb G},
\nn\end{align}
where $u=\sum_{i=1}^{n}
\mu_s z_s+\sum_{r=1}^{d}\nu_rw_r$ and
\begin{equation}\label{magic''}
\sum_{s=1}^n \frac{\mu_s}{t-z_s} +\sum_{r=1}^d
\frac{\nu_r}{t-w_r} \,=\,
u\frac{\prod_{j=1}^{n+d-2}(t-y_j)}{\prod_{s=1}^n (t-z_s)
\prod_{r=1}^d (t-w_r)}
\, .
\end{equation}
We see that an additional insertion of $\Phi^{\frac{k}{2}}(\nu|w)$
produces an extra degenerate field $e^{-\frac{1}{b}\phi}(y)$, but
without producing any other insertion as would be the case for
$\Phi^{j}(u|w)$ with $j\neq k/2$.
It follows in particular from the
fusion rules \rf{opedegen}
that the spaces of conformal blocks with and
without an insertion of $\Phi^{\frac{k}{2}}(\nu|w)$ are related as
\begin{equation}\label{Hecke2}
\CFB(\CR_{k/2}\ot\CR_{[n]})\,\simeq\,\BC^2\ot
\CFB(\CR_{[n]})\,.
\end{equation}
The isomorphism \rf{Hecke2} is not canonical. A useful way to
describe it uses the markings
introduced
in Subsection \ref{liouconf++}. We will get something more canonical
in the case of the Hecke functors at
the critical level.

In the case where $d$ is even one may on the one hand use the fact that
the vacuum representation appears in the fusion rules
$[e^{-\frac{1}{b}\phi}]\, [e^{-\frac{1}{b}\phi}]\,\sim\, 
[1]+[e^{-\frac{2}{b}\phi}]$. 
Subspaces of the space of conformal blocks of the form \rf{modconfbl}
are therefore naturally isomorphic to the original 
space of conformal blocks
with $d=0$. We may, on the other hand, regard the conformal 
blocks with $d$ insertions of fields $\Phi^{\frac{k}{2}}(u_k|w_k)$
as conformal blocks associated to a bundle $\hat\CE$ obtained from 
an original bundle $\CE$ by means of $d$ Hecke modifications.
These facts can be used to represent
at least a part of the dependence of the conformal blocks on the
twisting bundle in terms of the 
variables $(u_1,\dots,u_d)$ and $(w_1,\dots,w_d)$ introduced in 
\rf{modconfbl}.

\subsubsection{Representation of Hecke vertex operators
in terms of Liouville conformal blocks}

In order to describe conformal blocks with Hecke vertex
operators $\Xi(w)$ it suffices to set
$\nu_r=0$ for $r=1,\dots,d$ in \rf{LiouWZW''}, as follows from
\rf{pushforward}.
Note  that setting $\nu_r=0$ in \rf{magic''}
means that the expression on the right hand side does not have
a pole at $t=z_r$, which is only possible if one of the variables $y_a$
coincides with $z_r$ so that the apparent pole on the right hand side is canceled.
Noting that $\Theta_{n+d}$ simplifies to $\Theta_n$ in this
case we arrive at the formula
\begin{align}\label{LiouWZW'''}
\bigg\langle\,  \prod_{k=1}^d \Xi(w_k) &\prod_{s=1}^n\tilde\Phi^{j_s}
(\mu_s|z_s)
\,\bigg\rangle_{\mathbb G}\,= \\[-.5ex]
\,&= \delta({\textstyle\sum_{s=1}^{n}\mu_s})
\,  u\,\Theta_{n}(y|z)\,
\bigg\langle \;\prod_{s=1}^n e^{2\al_s\phi(z_s)}\prod_{r=1}^{n-2}
e^{-\frac{1}{b}\phi(y_{r})}\prod_{k=1}^{d}
e^{-\frac{1}{b}\phi(w_{k})}
\;\bigg\rangle_{\mathbb G},
\nn\end{align}
where $u=\sum_{i=1}^{n}
\mu_s z_s$ and
\begin{equation}\label{magic'''}
\sum_{s=1}^n \frac{\mu_s}{t-z_s}\,=\,
u\frac{\prod_{r=1}^{n-2}(t-y_r)}{\prod_{s=1}^n (t-z_s)}
\, .
\end{equation}
This means that inserting $\Xi(y)$ into an
SL(2)-WZNW conformal blocks simply maps to the insertion of an
extra degenerate field $e^{-\frac{1}{b}\phi(y)}$ on the Liouville side.

We had previously noted that the $\fsl_2$ representation under
which the vertex operator
$\Xi(w)$ transforms is proportional to
the vacuum representation. This can not be the full
story since insertion of $\Xi(y)$ modifies the space of conformal
blocks as described by \rf{Hecke2}. However, from the discussion
of the semiclassical limit of Liouville conformal blocks in
Subsection \ref{LclassI} it follows that
the insertions of $\Xi(w_k)$ will factor out in this limit,
which leads to the formula
\begin{equation}
\bigg\langle\,  \prod_{k=1}^d \Xi(w_k) \prod_{s=1}^n\Phi^{j_s}
(x_s|z_s)
\,\bigg\rangle_{\mathbb G}\,=\,\prod_{k=1}^d \chi_k(w_k)\,
\bigg\langle\,
\prod_{s=1}^n\Phi^{j_s}
(x_s|z_s)\,,
\,\bigg\rangle_{\mathbb G}
\end{equation}
where $\chi_k(w_k)$ are solutions to the differential equation
$(\pa_w^2+t(w))\chi_k=0$. Which of the two linearly independent
solution of the second order differential equation one gets depends
on the choice of intermediate representation in the gluing construction
of the relevant Liouville conformal blocks.
This phenomenon is closely related
to the Hecke eigenvalue property in the geometric Langlands
correspondence as discussed in Subsection \ref{cfblcrit}.

\subsubsection{Quantum local systems}

Monodromies of an extra insertion $\Xi(y)$ define operators
on the space of conformal blocks as follows.
Elements of the fundamental group
$\pi_1(C_{g,n+d-2})$ are canonically identified with edge paths
on the graph $\Ga_{\hat\si}$. Moving $e^{-\frac{1}{b}\phi(y)}$ along a cycle
$\ga$ representing a generator of the fundamental group
corresponds to moving on a path on the marking graph $\Ga_{\hat\si}$ described
as a sequence of edges such that consecutive edges are connected
at vertices. There
is a standard way described in \cite{AGGTV,DGOT}
to associate to this edge path a composition
of the elementary fusion and braiding moves \cite{MS}. Having returned
to the point we started from, one may use the isomorphism
\rf{Hecke2} to define
a two-by-two matrix $\SM_\ga$ of operators acting on the space
conformal blocks with $n+d-2$ fields inserted.
It is easy to see that the change of the choices involved
in the definition of $\SM_\ga$ will change $\SM_\ga$ by
conjugation with a possibly operator-valued matrix.
Considering the operators $\SM_\ga$ associated to the
generators $\ga$ of the fundamental group up to
conjugation therefore defines a representation of the
fundamental group by operator-valued matrices $\SM_\ga$
whose matrix elements are operators acting on the space
of conformal blocks. Considering cycles $\ga$ which are homotopic
to the curves defining the pants decomposition corresponding to
the marking $\hat\si$ one finds operator-valued matrices that
act diagonally. Taking the trace of $\SM_\ga$ defines operators
on $\CFB(\CR_{[n]},C_{g,n},\CE)$ that up to a phase factor are identical to the
operators on $\CFB(\CR_{[n]},C_{g,n},\CE)$ defined by the construction
described in Subsection \ref{probes}. We have a correspondence
\begin{equation}\label{q-Lcorr}
\boxed{\quad
\begin{aligned}
& \text{Eigenvalues of}\\
& \SM_\ga,\;\;\ga\in\CC_{\si}
\end{aligned}
\quad}\quad
\longrightarrow \quad
\boxed{\quad
\begin{aligned}
& \text{Elements of a}\\
& \text{basis for $\CFB(\CR_{[n]},C,\CE)$}
\end{aligned}
\quad}
\end{equation}
We will call the operator-valued matrices $\SM_\ga$
quantum monodromies, and the representation of the fundamental
group generated by the monodromies of the extra insertion $\Xi(y)$
a quantum local system. Parameterizing the space of conformal blocks
by means of quantum local systems may be seen as
a natural quantum analog of the
geometric Langlands correspondence.

\subsubsection{Critical level limit}

Note that the operator-valued matrices $\SM_\ga$ will turn into the
matrices $\rho(\ga)$ representing the monodromy of the
corresponding oper.
We see that the quantum local systems turn
into the classical local systems representing the opers. The
representation $\CR_{k/2}$ gets identified with the representation
$\CW_{\frac{1}{2}}$ representing the elementary Hecke functor
on spaces of
conformal blocks at the critical level according to the discussion in
Subsection \ref{Heckesect}.
Note furthermore that
the eigenvalues of $\SM_\ga$ are parameterized by the variables
$p_r$, which in the limit $b\ra\infty$ get identified via \rf{parid} with the
coordinates $\eta_r$ for the space ${\rm Op}_{\fsl_2}(C_{g,n})$ of opers.
We conclude that the correspondence \rf{q-Lcorr}
reduces to the geometric Langlands correspondence \rf{Lcorr1}
in this limit. This is part of our motivation for calling
\rf{q-Lcorr} the quantum geometric Langlands correspondence.

\subsubsection{Quantum Drinfeld-Sokolov reduction}

Let us finally point out that the insertion of the fields $
\Phi^{\frac{k}{2}}(u|w)$ representing the Hecke modifications
not only
allows us to raise the number of degenerate fields $e^{-\frac{b}{2}\phi(y)}$
in the Liouville-representation, it also allows us to lower this
number. In order to see how this works, let us consider conformal blocks
like
\begin{align}
& \big\langle\,  \tilde\Phi^{j}(\mu|w)\tilde\Phi^{j_n}
(\mu_n|z_n)\dots\tilde\Phi^{j_1}(\mu_1|z_1)\,\big\rangle_{\mathbb G}= \\[.5ex]
&\;\;= \delta\big(\mu+{\textstyle\sum_{i=1}^{n}\mu_i}\big)
\,  u_{n+1}\,\Theta_{n+1}(y|z)\,
\big\langle \;e^{2\al\phi(w)}\,e^{2\al_n\phi(z_n)}\cdots e^{2\al_1\phi(z_1)}\;
e^{-\frac{1}{b}\phi(y_{n-2})}\cdots e^{-\frac{1}{b}\phi(y_0)}
\;\big\rangle_{\mathbb G,}
\nn\end{align}
where $u_{n+1}=\sum_{r=1}^n
\mu_r z_r+\mu w$ and
\begin{equation}\label{magic+1}
\frac{\mu}{t-w}+\sum_{r=1}^n \frac{\mu_r}{t-z_r} =
u\frac{\prod_{k=0}^{n-2}(t-y_k)}{(t-w)\prod_{r=1}^n (t-z_r)}\, .
\end{equation}
In the limit $\mu\ra 0$ we find from \rf{magic+1} that one of the
$y_r$, in the following
taken to be $y_0$ must approach $w$ to cancel the pole at $t=w$
of the right hand side. It follows that the limit $\mu\ra 0$
can be analyzed using the Liouville OPE
\begin{equation}\label{LiouOPE}\begin{aligned}
e^{-\frac{1}{b}\phi(y_0)}e^{2\al\phi(w)}\,\sim\,&(y_0-w)^{b^{-1}\al}
e^{(2\al-b^{-1})\phi(w)}(1+\CO(y_0-w))\\ & +C(\al)
(y_0-w)^{b^{-1}(Q-\al)}
e^{(2\al+b^{-1})\phi(w)}(1+\CO(y_0-w))\,.
\end{aligned}\end{equation}
In this way it is straightforward to check that \rf{LiouOPE} implies that
\begin{equation}\label{mutozero}
\tilde\Phi^{j}(\mu|w)\,\underset{\mu\ra 0}{\sim}\,
\mu^{j+1}\Phi^{j}_+(w)(1+\CO(\mu))+\mu^{-j}\Phi^{j}_-(w)(1+\CO(\mu))\,.
\end{equation}
The vertex operator $\Phi^{j}_-(w)$ is proportional to $\lim_{x\ra\infty}
x^{-2j}\Phi^j(x|w)$, as is simplest seen by noting that both are
annihilated by $J_0^-$ bearing in mind the representations \rf{CJdef}
and \rf{Dmu}. It has the {\it lowest weight} property
$J^-(y)\Phi^{j}_-(w)\,=\,{\rm regular}$\,.

In the case $j=k/2$ one has $\al=0$. This implies that the
term proportional to $\mu^{-j}$ in \rf{mutozero} would be absent unless
one of the $y_r$, $r=1,\dots,n$
happens to be at $w$. This
is equivalent to the constraint
\begin{equation}\label{support}
J^-(w)\,\equiv\,\sum_{r=1}^n \frac{\mu_r}{w-z_r}\,=\,0\,.
\end{equation}
We conclude that conformal blocks like
$\big\langle\,\Phi^{\frac{k}{2}}_-(w)\,\tilde\Phi^{j_n}
(\mu_n|z_n)\dots\tilde\Phi^{j_1}(\mu_1|z_1)\,\big\rangle$
can be defined as distributions
with support given by \rf{support}. In the resulting
representation by Liouville conformal blocks we'll now find instead of
$e^{2\al\phi(w)}$ one of the degenerate fields $e^{-\frac{1}{b}\phi(y_r)}$
with $y_r=w$ in  \rf{LiouOPE}. In this case
the second term in \rf{LiouOPE} will be proportional
to the identity field. This leads to
a representation of the form
\begin{align*}
\bigg\langle \Phi^{\frac{k}{2}}_-& (w) \,\prod_{r=1}^n\tilde\Phi^{j_r}
(\mu_r|z_r)\bigg\rangle_{\mathbb G} = \\[-1ex]
& =\delta\big({\textstyle\sum_{i=1}^{n}\mu_i}\big) \,
\delta\big({\textstyle\sum_{r=1}^n \frac{\mu_r}{w-z_r}}\big)
\,  u_{n}\,\Theta_{n}(y|z)\,
\bigg\langle \prod_{r=1}^n
e^{2\al_r\phi(z_r)}\prod_{k=1}^{n-3}
e^{-\frac{1}{b}\phi(y_{k})}
\bigg\rangle_{\mathbb G,}
\nn\end{align*}
where $u_{n}=\sum_{r=1}^n
\mu_r z_r$ and
\begin{equation}
\sum_{r=1}^n \frac{\mu_r}{t-z_r} =
u(t-w)\frac{\prod_{k=1}^{n-3}(t-y_k)}{\prod_{r=1}^n (t-z_r)}\, .
\end{equation}
The result is related to earlier work \cite{FZZ,MO}
on the spectral flow in the ${\rm SL}(2)$-WZNW model,
and in particular to the description proposed in \cite{Ri}
for correlation functions in the ${\rm SL}(2)$-WZNW model
with winding number violation.
The most important lesson for our purposes is the
fact that the insertion of
$\tilde\Phi^{\frac{k}{2}}(0|w)$ represents imposing the
constraint \rf{support} which is equivalent to $J^-(w)=0$.
Imposing this constraint effectively removes one
of the degenerate fields $e^{-\frac{1}{b}\phi(y_k)}$
from the
representation in terms of Liouville conformal blocks.
The conformal blocks with maximal number of
insertions of
$\tilde\Phi^{\frac{k}{2}}(0|w)$ are proportional to the Liouville
conformal blocks {\it without} degenerate fields.

\subsection{Generalization of the Geometric Langlands Correspondence ---\\
--- from opers to more general local systems}

It was proposed by Beilinson and Drinfeld (see \cite[Section 9.6]{Fr2}
for a discussion) to view the correspondence above
as special case of a correspondence
\begin{equation}\label{Lcorr2}
\boxed{\quad\phantom{\Big|}
{}^{\rm\sst L}_{}{G}-\text{local systems}
\;\quad}\quad
\longrightarrow \quad
\boxed{\quad\phantom{\Big|}
\CD-\text{modules on ${\rm Bun}_G$}
\;\quad}
\end{equation}
In order to realize an example for this generalized version of the
geometric Langlands correspondence, let us consider
instead of \rf{LiouWZW} the following family of conformal blocks,
\begin{equation}\label{genWZWbl}
\bigg\langle\,   \prod_{s=1}^n\Phi^{j_s}
(x_s|z_s)
\prod_{r=1}^m \Phi^{+}_{(2,2)}(u_r|w_r)\,
\bigg\rangle_{\mathbb G}
\end{equation}
where $\Phi^{+}_{(2,2)}$ is the field corresponding to the
degenerate representation corresponding to
$j=j^+_{(2,2)}=\frac{1}{2}(1+b^{-2})$.

\subsubsection{Critical level limit of KZ equations}

In the critical level limit $b\ra\infty$ we may note that
\begin{equation}
j^+_{(2,2)}\,=\,\frac{1}{2}(1+b^{-2})\;\ra\;\frac{1}{2}\,.
\end{equation}
This implies that the null vector decoupling equation
for the degenerate field $\Phi^{+}_{(2,2)}(x|w)$
simplifies in the critical level limit to
\begin{equation}\label{nulldec}
\pa_x^2\,\Phi^{+}_{(2,2)}(x|w)\,=\,0\,.
\end{equation}
Representing the two-dimensional space of solutions of \rf{nulldec}
as $\BC^2$ allows us to represent
the conformal blocks \rf{genWZWbl} in terms of
a vector-valued function ${\mathbf G}(z,w|x)\in(\BC^{2})^{\otimes m}$
as explained in Subsection \ref{moddual}.
In the critical level limit, the KZ equations produce the pair
of eigenvalue equations
\begin{equation}\label{doubleKZ}
\SH_s{\mathbf G}\,=\,E_s{\mathbf G}\,,\qquad
\sk_r{\mathbf G}\,=\,\kappa_r{\mathbf G}\,,
\end{equation}
where
\begin{equation}\label{gengaudin}\begin{aligned}
&\SH_s\,=\,\sum_{s'\neq s}\frac{\CJ_{s}^a\CJ_{s'}^{a'}}{z_{s}-z_{s'}}\eta_{aa'}+
\sum_{r}\frac{\CJ_{s}^a\si_r^{a'}}{z_{s}-w_r}\eta_{aa'}\,,\\
&\sk_r\,=\,\sum_{s}\frac{\si_s^a\CJ_{r}^{a'}}{w_{r}-z_s}\eta_{aa'}
+\sum_{r'\neq r}\frac{\si_{r}^a\si_{r'}^{a'}}{w_{r}-w_{r'}}\eta_{aa'}\,,
\end{aligned} \end{equation}
with $\si_r^a$ being the $2\times 2$-matrices which represent the action
of $\fsl_2$ on the r-th tensor factor in $(\BC^{2})^{\otimes m}$.
The system of differential equations \rf{gengaudin}
will represent the $\CD$-module to appear on the right hand side of
\rf{Lcorr2}.

\subsubsection{Classical limit of corresponding Liouville conformal blocks}

Let us, on the other hand, analyze the conformal blocks  \rf{genWZWbl}
in the $\mu$-representation obtained by Fourier-transformation over the
variables $x_s$ and $u_r$,
\begin{align}\label{LiouWZWa}
& \bigg\langle\,   \prod_{s=1}^n\tilde\Phi^{j_s}
(\mu_s|z_s)
\prod_{r=1}^m \tilde\Phi^{+}_{(2,2)}(\nu_r|w_r)\,
\bigg\rangle_{\hat{\mathbb G}}\,= \\[-.5ex]
\,&\qquad= \delta({\textstyle\sum_{s=1}^{n}\mu_s+\sum_{r=1}^{m}\nu_r})
\,  u\,\Theta_{n+m}(y|z)\,
\bigg\langle \;\prod_{s=1}^n e^{2\al_s\phi(z_s)}\prod_{r=1}^m
e^{-b\phi(w_r)}\!\prod_{q=1}^{n+m-2}
e^{-\frac{1}{b}\phi(y_{q})}
\;\bigg\rangle_{\hat{\mathbb G}},
\nn\end{align}
where $u=\sum_{i=1}^{n}
\mu_s z_s+\sum_{r=1}^{m}\nu_rw_r$ and
\begin{equation}\label{magic'}
\sum_{s=1}^n \frac{\mu_s}{t-z_s} +\sum_{r=1}^m
\frac{\nu_r}{t-w_r} \,=\,
u\frac{\prod_{j=1}^{n+m-2}(t-y_j)}{\prod_{s=1}^n (t-z_s)
\prod_{r=1}^m (t-w_r)}
\, .
\end{equation}
The null vector decoupling equation
\rf{nulldec} becomes $\mu^2 \tilde\Phi^{+}_{(2,2)}(\mu|w)=0$
after the Fourier-transformation to the $\mu$-representation.
The conformal blocks \rf{LiouWZWa} must therefore be distributions
supported at $\nu_r=0$. Formula \rf{magic'} implies that
$m$ of the variables $y_q$, here taken as $y_{n-1},\dots,y_{n+m-2}$,
must equal one of $w_1,\dots,w_m$, respectively.
The expectation values of the remaining fields $e^{-\frac{1}{b}\phi(y_{q})}$
factor out in this limit, producing a factor
$
\prod_{q=1}^{n-2}\chi_q(y_q)
$, with functions $\chi_q(y)$ that satisfy $(\pa_y^2+t(y))\chi_q(y)=0$
with $t(y)$ of the form
\begin{equation}\label{texp'}
t(y)\,=\,\sum_{s=1}^{n}
\left(\frac{\de_s}{(y-z_s)^2}+\frac{E_s(p,z)}{y-z_s}\right)
-\sum_{r=1}^{m}
\left(\frac{3}{4(y-w_r)^2}-\frac{\kappa_r(p,z)}{y-z_r}\right)\,.
\end{equation}
The local system associated to this differential equation
will appear on the left hand side of
\rf{Lcorr2}.

\subsubsection{The correspondence}

We arrive at another interesting example for the geometric Langlands
correspondence as the correspondence between the local systems
corresponding to the differential equation
$(\pa_y^2+t(y))\chi_q(y)=0$
with $t(y)$ of the form \rf{texp'}
and the system of differential equations on
${\rm Bun}_G(C_{0,n})$. Note that for $m=n-3$ the number of
parameters in $t(y)$ coincides with the dimension of
${\rm Loc}_{{\rm PSL}(2,\BC)}(C_{0,n})$.

This example exemplifies the
abstract construction sketched in \cite[Section 9.6]{Fr2}.
It was noted there that the generalization beyond the case of
opers requires introduction of additional parameters, which
here are represented by the variables $w_r$. It was conjectured
in this reference that the resulting system of
differential equations is in a suitable sense independent of
the choices of $w_r$. In this regard we may observe that the
dependence on $w_r$ is controlled by the relation with
Liouville semiclassical blocks in the following way: The function
\begin{equation}
\CW_{\hat\BG}(z,w)\,=\,\lim_{b\ra\infty}b^{-2}\log
\bigg\langle \;\prod_{s=1}^n e^{2\al_s\phi(z_s)}\prod_{r=1}^m
e^{-b\phi(w_r)}
\;\bigg\rangle_{\hat\BG}
\end{equation}
is a potential for $E_s=E_s(p,\de|z,w)$ and $\kappa_r=
\kappa_r(p,\de|z,w)$ in the
sense
that
\begin{equation}
E_s\,=\,-\frac{\pa}{\pa z_s}\CW_{\hat\BG}(z,w)\,,\qquad
\kappa_r\,=\,-\frac{\pa}{\pa w_r}\CW_{\hat\BG}(z,w)\,.
\end{equation}
The knowledge of $\CW_{\hat\BG}$ in principle allows us to compute how the parameters $\kappa_r$
in the differential equations \rf{doubleKZ}
have to be varied if one modifies the positions $w_r$ of the
additional singularities, keeping the local system fixed.

\section{Concluding remarks}

\setcounter{equation}{0}

\subsection{Relation with gauge theory}

We believe that the results of this paper can help understanding the
relation between gauge theory and Liouville theory
suggested in \cite{NW} more precisely. They may thereby contribute
to uncovering
the deeper reasons for the correspondence between instanton partition
functions and Liouville conformal blocks
proposed in \cite{AGT}.

In this regard let us note that the gauge theory set-up
considered in \cite{NW} produces a
Hilbert space $\CH_{\ep_1\ep_2}$ of open strings which has a
representation in terms of holomorphic sections of
a line bundle on the space of opers. Locally these
sections should be representable as holomorphic functions
of the accessory parameters. There is no natural structure
of noncommutative algebra on this space coming from
quantization of a symplectic form on the space of
opers. There are, however, two commuting
actions on $\CH_{\ep_1\ep_2}$
of quantized algebras of functions on
${\rm Loc}_{\rm SL(2)}(C)$. The deformation parameters
can naturally be identified with $\ep_1$ and $\ep_2$, respectively.
This is what strongly
suggests that the Hilbert space of open strings produced
by gauge theory can be identified with the space of
Liouville conformal blocks \cite{NW}.

The discussion in Subsection \ref{Kahlerreal} suggests that the space of holomorphic
sections of the line bundle produced
by the gauge-theory set-up of \cite{NW} should
be seen as a sort of  "momentum-representation" which is dual to
the K\"ahler-quantization of Teichm\"uller
space discussed here, in the sense that one works in
a representation in which the conjugate momenta
(the accessory parameters) of the Teichm\"uller moduli
are diagonalized. Although such a quantization scheme
remains to be developed in detail, we hope that these
observations may help to clarify the relation between the
Hilbert space of open strings coming from gauge theory
and the space of conformal blocks in Liouville theory.

In any case, in order to understand the conjecture of \cite{AGT}
along such
lines one should ultimately work in a third representation,
which is the representation in which the
a maximal set of commuting global observables
(length operators) is diagonal. As pointed out in \cite{DGOT},
one would thereby naturally explain the form that
the gauge theory loop operator expectation values take
according to \cite{Pe}, as discussed and generalized in \cite{AGGTV,DGOT}.

It is furthermore intriguing to note \cite{BT} that the conformal Ward 
identities have a counterpart in the context of the 
gauge-theoretical instanton counting: 
Variations of gauge coupling constants are described by means of 
insertions of ${\rm tr}(\phi^2)$. This observation should 
be compared to the fact that the Hamiltonians $H_r$ obtained from
the Higgs field $\theta$ via \rf{qfromphi} and \rf{H_rdef} end up being 
the generators of infinitesimal variations of the moduli of $C$
in our approach.

\newcommand{\Td}{\ST\so\sd\sa}

\subsection{Generalization to higher rank}

Of obvious interest is the generalization of this picture when
$\fsl_2$ is replaced by a  Lie algebras $\fg$ of higher rank. We may
anticipate the following picture.

The natural higher rank analogs of the Liouville theory are the conformal
Toda theories denoted $\Td_{k}(\fg)$. The conformal symmetry of Liouville theory
is extended to symmetry under the W-algebra $W_k({\mathfrak g})$, where $h^{\vee}$ is the dual Coxeter number.
Let us also consider the Toda theory $\Td_{\check{k}}(\Lfg)$ where $\Lfg$ is the
Langlands dual Lie algebra $\Lfg$ with a Cartan matrix that is
transpose of the Cartan matrix of $\fg$, while
$\check{k}$ is related to $k$ via
\begin{equation}\label{parrel}
(k+h^{\vee})r^{\vee}\,=\,(\check{k}+\check{h}^{\vee})^{-1}\,.
\end{equation}
$r^{\vee}$ is the lacing number of $\fg$, the maximal number of edges connecting two
nodes of the Dynkin diagram.
It was proven in \cite{FF1} that the W-algebras $W_{\check{k}}(\Lfg)$
and $W_k({\mathfrak g})$ are isomorphic,
\begin{equation}
W_k({\mathfrak g})\;\simeq\;W_{\check{k}}(\Lfg)\,.
\end{equation}
It follows that the
Toda theories $\Td_{k}^{}(\fg)$ and $\Td_{\check{k}}(\Lfg)$
are dual to each other in the sense that the conformal
blocks in the two theories coincide. This naturally suggests
the conjecture \cite{T09} that there exist modular functors
associated to  $\Td_{k}^{}(\fg)$ and $\Td_{\check{k}}(\Lfg)$, respectively,
which are dual to each other if the levels are related by \rf{parrel}.

Let us now assume that there is a way to construct
the conformal blocks  in ${\SW\SZ\SN\SW}_k(\fg)$ from those of $\Td_k(\fg)$,
generalizing what was described above for the case $\fg=\fsl_2$. As in the $\fg=\fsl_2$-case
discussed in this paper, we could
then construct the conformal blocks of two different WZNW models from those
of $W_k({\mathfrak g})$, schematically
\begin{equation}\label{dualWZWToda}
\boxed{\quad\phantom{\Big|}{\SW\SZ\SN\SW}_{\check{k}}(\Lfg)\;
\quad}\quad
\longleftarrow \quad
\boxed{\quad\phantom{\Big|}\Td_b(\fg)\;
\quad}
\quad
\longrightarrow \quad
\boxed{\quad\phantom{\Big|}{\SW\SZ\SN\SW}_{{k}}(\fg)\;
\quad}
\end{equation}
For each of the WZNW models there are two different limits one may consider, leading
to diagrams such as
\begin{equation}\label{DdiagA}
\begin{array}{ccccc}
& & \text{$G$-Hitchin system} & & \\
& {\rm (A)}_{\ep_2}\nearrow & & \nwarrow {\rm (B)}_{\ep_1}& \\
\text{$\fg$-Isomonodromic} & & & & \text{$\fg$-Beilinson-} \\
\text{deformations} & & & & \text{Drinfeld system} \\
& {\rm (C)}_{\ep_1}\nwarrow & & \nearrow {\rm (D)}_{\ep_2} & \\
& & \text{$G$-WZNW-model} &  &
\end{array}
\end{equation}
and on the other hand
\begin{equation}\label{DdiagB}
\begin{array}{ccccc}
& & \text{${}^LG$-Hitchin system} & & \\
& {\rm (A)}_{\ep_1}\nearrow & & \nwarrow {\rm (B)}_{\ep_2}& \\
\text{$\Lfg$-Isomonodromic} & & & & \text{$\Lfg$-Beilinson-} \\
\text{deformations} & & & & \text{Drinfeld system} \\
& {\rm (C)}_{\ep_2}\nwarrow & & \nearrow {\rm (D)}_{\ep_1} & \\
& & \text{${}^LG$-WZNW-model} &  &
\end{array}
\end{equation}
This would again lead to two possible ways to describe the same limit in the
conformal Toda theory $\Td_b(\fg)$. Extrapolating from case $\fg=\fsl_2$
we would expect that a good part of the geometric Langlands correspondence can
be understood in this way.

In the $\fsl_2$-case we had discussed the relations
between the $\fsl_2$-Toda (Liouville) theory and the quantization of the
Teichm\"uller spaces. It seems worth pointing out that
higher rank analogs of the quantum Teichm\"uller spaces have
been defined in \cite{FG09}. A relation between modular
duality and Langlands duality that
fits perfectly into the picture proposed above was pointed
out in \cite{FG09}. Proving the modular
functor conjecture \cite{FG09} for the higher
quantized Teichm\"uller theories would be an important step
towards the higher rank generalization of the
quantum geometric Langlands correspondence.

\end{document}